%% file: ms.tex
\documentclass[12pt, oldfontcommands]{memoir}

\setsecnumdepth{subsection}
\maxtocdepth{subsection}
\usepackage[utf8]{inputenc}

\usepackage[a-1a]{pdfx}
\usepackage[british,UKenglish,USenglish,english,american]{babel}
\usepackage[T1]{fontenc}

\usepackage{bigints}
\usepackage{pdfpages}
\usepackage{csquotes}
\usepackage{multirow}
\usepackage{ marvosym } 

\usepackage[normalem]{ulem}
\usepackage{verbatim}
\usepackage{amsmath}
\usepackage{amssymb}
\usepackage{bbold, dsfont}
\usepackage{alphalph, relsize}
\usepackage[singlelinecheck=off, justification=centerlast]{caption}
\usepackage[singlelinecheck=off, justification=centerlast]{subcaption} 
\usepackage{epsfig, wrapfig}
\usepackage[makeroom]{cancel}
\date{}
\usepackage{geometry, graphicx,wrapfig}
\usepackage{colortbl}

\usepackage{xcolor,calc,graphicx,soul}
\makeatletter
\newlength\dlf@normtxtw
\newsavebox{\feline@chapter}
\newcommand\feline@chapter@marker[1][4cm]{%
  \sbox\feline@chapter{%
    \resizebox{!}{#1}{\fboxsep=1pt%
      \colorbox{gray}{\color{white}\bfseries\sffamily\thechapter}%
    }}%
  \rotatebox{90}{%
    \resizebox{%
      \heightof{\usebox{\feline@chapter}}+\depthof{\usebox{\feline@chapter}}}%
    {!}{\scshape\so\@chapapp}}\quad%
  \raisebox{\depthof{\usebox{\feline@chapter}}}{\usebox{\feline@chapter}}%
}
\newcommand\feline@chm[1][4cm]{%
  \sbox\feline@chapter{\feline@chapter@marker[#1]}%
  \makebox[0pt][r]{
    \makebox[1cm][r]{\usebox\feline@chapter}%
  }}
\makechapterstyle{daleif1}{

  \renewcommand\printchapternum{\null\hfill\feline@chm[2.5cm]\par}

}
\makeatother
\chapterstyle{daleif1}
\makeatletter
\newcommand{\setwd}[2]{%
  \phantomsection
  #1\def\@currentlabel{\unexpanded{#1}}\label{#2}%
}
\makeatother

\newcommand{\const}{\textrm{const}}

\newcommand{\stuff}{\mathcal{H}_0^2\Omega_{\rm m0}}

\newcommand{\UD}[2]{\ensuremath{^{#1}_{\phantom{#1} #2}}}
\newcommand{\DU}[2]{\ensuremath{_{#1}^{\phantom{#1} #2}}}

\newcommand{\UDDD}[4]{\ensuremath{^{#1}_{\phantom{#1} #2 #3 #4}}}

\newcommand{\calO}{\ensuremath{\mathcal{O}}}
\newcommand{\calS}{\ensuremath{\mathcal{S}}}
\newcommand{\calQ}{\ensuremath{\mathcal{Q}}}

\newcommand{\WXL}{\ensuremath{W_{XL}}}
\newcommand{\WXX}{\ensuremath{W_{XX}}}
\newcommand{\WLL}{\ensuremath{W_{LL}}}
\newcommand{\WLX}{\ensuremath{W_{LX}}}

\newcommand{\bm}[1]{\ensuremath{\boldsymbol{#1}}}

\newcommand{\bwt}{\begin{widetext}}
\newcommand{\ewt}{\end{widetext}}

\newcommand{\beq}{\begin{equation}}
\newcommand{\eeq}{\end{equation}}
\newcommand{\bea}{\begin{eqnarray}}
\newcommand{\eea}{\end{eqnarray}}
\newcommand{\bean}{\begin{eqnarray*}}
\newcommand{\eean}{\end{eqnarray*}}
\newcommand{\bit}{\begin{itemize}}
\newcommand{\eit}{\end{itemize}}
\newcommand{\bfi}{\begin{figure}}
\newcommand{\efi}{\end{figure}}
\newcommand{\bfic}{\begin{figure*}}
\newcommand{\efic}{\end{figure*}}
\newcommand{\bce}{\begin{center}}
\newcommand{\ece}{\end{center}}
\newcommand{\bt}{\begin{table}}
\newcommand{\et}{\end{table}}
\newcommand{\btb}{\begin{tabular}}
\newcommand{\etb}{\end{tabular}}

\newcommand{\calP}{\ensuremath{\mathcal{P}}}

\newcommand{\calM}{\ensuremath{\mathcal{M}}}

\newcommand{\calE}{\ensuremath{\mathcal{S}}}

\def\blankpage{%
      \clearpage%
      \thispagestyle{empty}%
      \addtocounter{page}{-1}%
      \null%
      \clearpage}

\begin{document}

\include{template/front} 
\blankpage{}
\thispagestyle{empty}%
\topskip0pt
\vspace*{\fill}
\begin{center}
  \flushleft
    \textit{\large{``There is a theory which states that if ever anyone discovers exactly what the Universe is for and why it is here, it will instantly disappear and be replaced by something even more bizarre and inexplicable.}}\\
    
\textit{\large{There is another theory which states that this has already happened.''}}\\
  \flushright 
   Douglas Adams
\end{center}

\vspace*{\fill}
\setlength{\baselineskip}{1\baselineskip} 

\newpage

\frontmatter
\input{template/Abstract.tex}
\blankpage
\input{template/polishabstract.tex} 
\blankpage
\input{template/Declaration.tex}
\blankpage
\input{template/Acknowledgement.tex}
\blankpage

\tableofcontents
\clearpage


\mainmatter

\input{template/intro}
\input{template/cosmology}
\input{template/lightprop}
\input{template/paper_bigonlight}
\input{template/paper_nonlinear}
\input{template/conclusion}




\listoffigures

\bibliography{ms}
\bibliographystyle{acm}
\newpage
\thispagestyle{empty}
\end{document}

%% file: template/front.tex




\begin{titlingpage}
\addtocounter{page}{-1}
\begin{center}
\begin{minipage}{\textwidth}
\begin{minipage}{0.3\textwidth}
\centering
\includegraphics[width=10em]{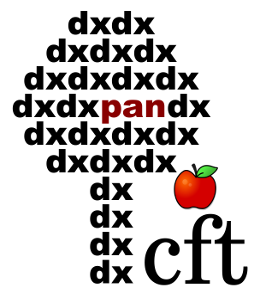}
\end{minipage}
\begin{minipage}{0.7\textwidth}
\centering
\Large{\uppercase{Centrum Fizyki Teoretycznej}}\\
\large{{\sc Polskiej Akademii Nauk}}\\
\end{minipage}
\vspace{0.3truecm}
\end{minipage}
\hbox to \textwidth{\hrulefill}

\vspace{2truecm}

\Large{\bf Michele Grasso}

\vspace{2truecm}

\LARGE{\uppercase{\bf BiGONLight: a new package for computing optical observables in Numerical Relativity}}

\vspace{2truecm}

\centerline{\hbox to 4truecm{\hrulefill}}

\medskip

\large{\it Thesis submitted in fulfillment of the requirements}\\
\large{\it for the degree of Doctor of Philosophy in Physics}\\

\centerline{\hbox to 4truecm{\hrulefill}}

\vspace{2truecm}


\begin{flushright}
\begin{minipage}{0.5\textwidth}
\begin{flushright}
\begin{minipage}{0.9\textwidth}
 \large{Supervisors:} 
\end{minipage}
\end{flushright}
\begin{flushright}
\begin{minipage}{0.7\textwidth}
 \large{Prof. Miko\l{}aj Korzy{\'n}ski} \\ 
\large{Dr. Eleonora Villa} 
\end{minipage}
\end{flushright}
\end{minipage}
\end{flushright}

\vspace{1truecm}

\hbox to \textwidth{\hrulefill}
\large{\sc Warsaw, November 2021}

\end{center}
\end{titlingpage}

%% file: template/Abstract.tex
\section*{Abstract}    
\label{sec:abstract}
\addcontentsline{toc}{chapter}{Abstract}

With the advent of precision cosmology, our theoretical predictions must aspire to the same level of precision as achieved by experimental probes. In this context, numerical simulations including general relativistic effects represent the state-of-the-art method to describe the formation of structures. However, aside from a detailed description of the dynamics, it is necessary to have an equally accurate explanation of the effects of such structures on light propagation and modelling their impacts on measurable quantities. 

The investigation of relativistic effects in the most general way requires a unified treatment of light propagation in cosmology. This goal can be achieved with the new interpretation of the geodesic deviation equation in terms of the bilocal geodesic operators (BGO). The BGO formalism extends the standard formulation, providing a unified framework to describe all possible optical phenomena due to the interaction between light and spacetime curvature. 
	
In my dissertation, I present {\tt BiGONLight}, a {\tt Mathematica} package that applies the BGO formalism to study light propagation in numerical relativity. The package encodes the 3+1 bilocal geodesic operators framework as a collection of {\tt Mathematica} functions. The inputs are the spacetime metric plus the kinematics of the observer and the source in the form of the 3+1 quantities, which may come directly from a numerical simulation or can be provided by the user as analytical components. These data are then used for ray tracing and computing the BGO's in a completely general way, i.e. without relying on symmetries or specific coordinate choices. The primary purpose of the package is the computation of optical observables in arbitrary spacetimes. The uniform theoretical framework of the BGO formalism allows for the extraction of multiple observables within a single computation, while the {\tt Wolfram} language provides a flexible computational framework that makes the package highly adaptable to perform both numerical and analytical studies of light propagation. {\tt BiGONLight} is tested by computing the redshift, angular diameter distance, parallax distance, and redshift drift in well-known cosmological models. We use three different inputs for the metric: two analytical metrics, the homogeneous $\Lambda$CDM model and the inhomogeneous Szekeres model, and 3+1 quantities from a simulated dust Universe. The tests show an excellent agreement with known results.

The characteristics of {\tt BiGONLight} make it a suitable tool for studying the impact of inhomogeneities on light propagation. We investigate various sources of nonlinear general relativistic effects on light propagation induced by inhomogeneous cosmic structures. {\tt BiGONLight} is used to calculate observables computed at different approximations in a plane-parallel inhomogeneous spacetime. The nonlinear effects are evaluated as the fractional difference between the observables obtained at the three different approximations: linear perturbation theory, Newtonian, and post-Newtonian approximations. The inhomogeneities are tuned by varying the model’s free parameters, and their contributions to the observables are obtained by analysing the variations in the fractional differences. Using this method we estimate the Newtonian and post-Newtonian corrections to the linear observables and analyse how these corrections change as we vary the size and magnitude of the inhomogeneities. We also explain the role of the linear initial seed as the dominant post-Newtonian contribution and show that the remaining post-Newtonian nonlinear corrections are less than $1\%$, which is consistent with previous results in the literature.

%% file: template/polishabstract.tex
\section*{Streszczenie} 
 \textbf{Polish translation of the abstract}\\
\addcontentsline{toc}{chapter}{Streszczenie}

Wraz z początkiem kosmologii precyzyjnej, przewidywania teoretyczne powinny zbliżać się do podobnego poziomu precyzji jak eksperymenty. W tym kontekście symulacje numeryczne uwzględniające poprawki związane z ogólną teorią względności stanowią najlepszą obecnie metodę opisu formowania struktury. Jednakże, oprócz dokładnego opisu dynamiki, niezbędne jest także równie dokładne wyjaśnienie oddziaływania tych struktur na propagację światła i precyzyjne modelowanie ich wpływu na wielkości mierzalne. 

Badanie wszyskich relatywistycznych efektów w najbardziej ogólnym sformułowaniu wy\-ma\-ga jednolitego podejścia do problemu propagacji propagacji światła w kosmologii. Można to osiągnąć dzięki nowej interpretacji równania dewiacji geodezyjnych w języku bilokalnych operatorów geodezyjnych (bilocal geodesic operators, BGO). Formalizm BGO jest rozszerzeniem standardowego opisu, wprowadzającym jednolity opis różnych zjawisk optycznych związanych z oddziaływaniem krzywizny czasoprzestrzeni na światło.

W tej rozprawie prezentuję {\tt BiGONLight}, pakiet w języku {\tt Mathematica} implementujący formalizm BGO do badania propagacji światła w numerycznej ogólnej teorii względności. Pakiet implementuje formalizm BGO dla danych w rozkładzie 3+1 jako kolekcję funkcji języka {\tt Mathematica}. Dane wejściowe stanowią metryka czasoprzestrzeni oraz kinematyka obserwatora i źródła, oba w rozkładzie 3+1, które mogą pochodzić bezpośrednio z numerycznej symulacji albo zostać dostarczone bezpośrednio przez użytkownika jako funkcje w jawnej postaci. Dane te służą do śledzenia promieni światła (ray-tracing) i obliczenia bilokalnych operatorów geodezyjnych w najogólniejszy możliwy sposób, bez korzystania z symetrii czasoprzestrzeni lub własności układu współrzędnych. Głównym zadaniem pa\-kie\-tu jest obliczanie obserwabli optycznych w dowolnej czasoprzestrzeni. Formalizm BGO pozwala na wyznaczenie wszystkich obserwabli podczas jednego obliczenia, a język {\tt Wolfram} dostarcza narzędzi numerycznych, dzięki czemu pakiet łatwo nadaje się do zarówno numerycznych, jak i analitycznych badań nad propagacją światła. Pakiet {\tt BiGONLight} został przetestowany przez obliczenie przesunięcia ku czerwieni, odległości kątowej, odległości paralaktycznej i dryfu przesunięcia ku czerwieni w prostych modelach kosmologicznych. Badania przeprowadzamy dla trzech przykładów: dwóch metryk podanych analitycznie, tzn. modelu jednorodnego $\Lambda$CDM bez perturbacji i niejednorodnego modelu z klasy Szekeresa, oraz dla symulowanego wszechświata z pyłem w rozkładzie 3+1. Testy pokazały bardzo dobrą zgodność z analitycznymi wzorami.

Dzięki wyżej wymienionym cechom pakiet {\tt BiGONLight} jest dobrym narzędziem do badania wpływu niejednorodności na propagację światła. Zbadaliśmy źródła efektów relatywistycznych w propagacji światła spowodowane przez niejednorodne struktury. Pakiet {\tt BiGONLight} został użyty do obliczenia obserwabli obliczonych w różnych przybliżeniach w niejednorodnym modelu typu plane-parallel. Efekty nieliniowe obliczone zostały jako względne różnice między trzema przybliżeniami: liniowy rachunkiem perturbacyjnym, przybliżeniem new\-to\-now\-skim oraz post-newtonowskim. Wielkość niejednorodności można regulować zmieniając wolne parametry modelu, a wpływ tych parametrów na obserwable otrzymany został przez a\-na\-li\-zę zmian względnych różnic. W ten sposób szacujemy wielkość newtonowskich oraz post-newtonowskich poprawek do liniowego rzędu rachunku zaburzeń i badamy jak zmieniają się one przy zmianie rozmiaru i amplitudy niejednorodności. Wyjaśniamy także rolę początkowej perturbacji metryki przez potencjał skalarny i pokazujemy, że nieliniowe poprawki są mniejsze niż $1\%$. Wyniki te są zgodne z wcześniejszymi badaniami na ten temat.

%% file: template/Declaration.tex
\section*{Declaration}
\addcontentsline{toc}{chapter}{Declaration}

The work described in this thesis was undertaken between October 2017 and November 2021 while the author was a research student at the Center for Theoretical Physics, Polish Academy of Sciences. The work was done under the scientific supervision of Prof. Miko\l{}aj Korzy{\'n}ski and co-supervised by Dr. Eleonora Villa at Center for Theoretical Physics, Polish Academy of Sciences.
Under the same period, the author completed his coursework at the Institute of Physics, Polish Academy of Sciences. No part of this thesis has been submitted for any other degree at the Center for Theoretical Physics, Polish Academy of Sciences or any other university.
The abstract is translated into Polish by Prof. Miko\l{}aj Korzy{\'n}ski.

This thesis is based on the original results published in the following articles:\\
\begin{itemize}
        \item {\it Chapter}~\ref{chap:bigonlight}: {\textbf{``BiGONLight: light propagation with bilocal operators in Numerical Relativity.''} Authors: Michele Grasso and Eleonora Villa, published in Classical and Quantum Gravity (Class. Quant. Grav.) (\cite{Grasso:2021iwq}).} 
         \item {\it Chapter}~\ref{chap:nonlinearities}: {\textbf{``Isolating nonlinearities of light propagation in inhomogeneous cosmologies.''} Authors: Michele Grasso, Eleonora Villa, Miko\l{}aj Korzy\'nski, and  Sabino Matarrese, published in Physical Review D (Phys. Rev. D) (\cite{Grasso:2021zra}). }
\end{itemize}

\subsection*{Author's contributions}
The work described in Chapters~\ref{chap:bigonlight}-\ref{chap:nonlinearities} was performed in collaboration with the other co-authors (listed above). 
The author's main results are summarised in Sec.~\ref{sec:mainRes}, while a detailed description of the author's contribution is provided in the introductions of Chapters~\ref{chap:bigonlight}-\ref{chap:nonlinearities}.

\subsection*{Other works}
Apart from the two original works mentioned above, the author has contributed to the article \cite{Grasso:2018mei}
\begin{itemize}
     \item {\textbf{``Geometric optics in general relativity using bilocal operators.''} Authors: Michele Grasso, Miko\l{}aj Korzy\'nski, and Julius Serbenta, published in Physical Review D (Phys. Rev. D) (\cite{Grasso:2018mei}). } 
\end{itemize}

In the former work, a new formulation for light propagation in geometric optics by means of the bilocal geodesic operators is presented.
This novel formulation offers a consistent approach to optical phenomena in curved spacetimes, and it may be used to compute observables like angular diameter distance, luminosity distance, magnification, as well as new real-time observables like parallax and redshift drift, all within the same framework.
Although the results in \cite{Grasso:2018mei} are not included as original work of this thesis, the mathematical machinery of the BGO is presented in Chapter~\ref{chap:BGO} as it provides the theoretical framework on which the results in \cite{Grasso:2021iwq, Grasso:2021zra} are based.

%% file: template/Acknowledgement.tex
\section*{Acknowledgements}    
\label{sec:acknowledgement}
\addcontentsline{toc}{chapter}{Acknowledgements}

Searching for the term \textbf{P.h.D.}, the Collins dictionary\footnote{{\color{blue}{https://www.collinsdictionary.com/dictionary/english/phd}}} says :
\begin{displayquote}
\textit{A P.h.D. is a degree awarded to people who have done advanced research into a particular subject. P.h.D. is an abbreviation for ``Doctor of Philosophy''.}
\end{displayquote} 
Similarly, form Oxford dictionary\footnote{{\color{blue}{https://dictionary.cambridge.org/dictionary/english/phd?q=P.h.D}}}
\begin{displayquote}
P.h.D., noun.\\
\textit{Abbreviation for doctor of philosophy: the highest college or university degree.}
\end{displayquote} 
These two definitions make it seem like a solo effort, but in reality, I could never have accomplished this journey alone. Therefore, I would like to thank all those who have supported me during these years.

First and foremost, I would like to express my gratitude to my supervisors Prof. Miko\l{}aj Korzy{\'n}ski and Dr. Eleonora Villa, for their helpful advice, unwavering support, and patience throughout my doctoral studies. They introduced me to the broad and complicated world of research with their vast expertise and wealth of experience and inspired me throughout my studies.
 
I would like to extend my sincere thanks to all other professors of the Center for Theoretical Physics (CFT PAS) for their kind and open-minded attitude and for allowing me to carry out my research in a multidisciplinary and dynamic environment. I would especially like to thank Prof. L. Mankiewicz, Prof. K. Paw\l{}owski, and Prof. M. Bilicki for their help in starting the PhD procedure.
I would also like to thank the administrative and secretarial staff of CFT PAS for their unconditional help and endless patience in facilitating the complicated matters of bureaucracy.

At CTP PAS, I also had the opportunity to meet many more friends and colleagues. I am grateful to the other PhD students in room 304A, Julius, Ishika, Grzegorz, and Suhani, for accompanying me during this intense period and rejoicing with me over my achievements.
Furthermore, I cannot fail to mention ``the boss'', who has been a point of reference for my academic and professional decisions with her sincere and objective advice. \emph{In particular}, thank you for the endless meetings and virtual toasts.

I am immensely grateful to my parents, who have always supported and helped me with every decision I have ever made. Without the support of Orazio and Meletta, as well as Alfredo and Pina, I would never have become who I am today. Thank you for always being there for me.
I would also like to thank those who have become a second family to me: without the delightful distractions from my studies provided by my briscola-enthusiasts friends Mario, Lucia, Silvia, Stefano, Valeria, Fabrizio, Paola and Ruggero, I would not have been able to finish this dissertation. 

Finally, I would like to express my gratitude to Miriam for her unique understanding (over the past 16 years) and for seeing abilities in me that I did not even know I had. This thesis does not represent the end of an academic path but the beginning of a life together.

%% file: template/intro.tex
\chapter{Introduction}
\label{chap:intro}

To answer the most critical questions about the origin, structure, evolution and ultimate destiny of the Universe, cosmologists gather empirical evidence and measurements to create theoretical models that explain real-world observations.  
The evidence used to formulate the current model of the Universe comes mainly from astronomical observations, namely the observation and analysis of signals emitted from distant sources and propagating at the speed of light. The nature of these light-like signals can vary, and they can reveal a range of information: for example, the electromagnetic signals from a star or galaxy can be analysed to draw conclusions about its physical properties (such as temperature, composition, rotational speed, and relative motion), the distance from us, as well as the characteristics of the space(time) through which the signal has passed. Another notable example is gravitational waves, which are direct evidence of the existence of black holes and provide a new window to peer into regions inaccessible with electromagnetic signals, \cite{abbott2016observation}. 
Advances in experimental precision or observations of new physical phenomena lead to improvements in the theoretical model that provides a deeper understanding of our Universe.

Today, this constant exchange between theoretical models and observations has led to a scientific revolution, ushering in the era of \emph{precision cosmology}, \cite{howlett2012cmb, jones2017precision}. The term ``precision'' refers to the targeted accuracy of $1\%$, which is aimed at both experimental observations and theoretical predictions: the next generation of galaxy surveys\footnote{See e.g. {\color{blue}{https://www.skatelescope.org}}, {\color{blue}{https://www.euclid-ec.org}}, {\color{blue}{https://www.lsst.org}}, {\color{blue}{http://litebird.jp/eng/}}, {\color{blue}{https://www.jpl.nasa.gov/missions/spherex}}.} will scan almost the whole extragalactic sky to unprecedented depth and resolution, producing the most detailed map of the Universe ever made. Moreover, this remarkable precision of the experimental data opens up the possibility of measuring small temporal variations in cosmological observables known as \emph{optical drift effects}, \cite{Quercellini:2010zr}. These ``real-time'' effects have the potential to provide new and crucial insights into the structure and evolution of the Universe, \cite{Quercellini:2008ty, Ding:2009xs, Quartin:2009xr,rasanen, PhysRevLett.121.021101}.
In terms of theoretical predictions, numerical simulations have made great strides in describing the formation of cosmic structures, ranging from large to small scales and accounting for relativistic effects, \cite{Giblin:2015vwq, Bentivegna:2015flc, adamek2016general, macpherson2017, Macpherson:2018akp, Barrera-Hinojosa:2020gnx}. In addition, cosmological simulations are used to study nonlinear relativistic effects in lensing observations, \cite{Giblin:2017ezj,Beutler:2020evf, Lepori:2020ifz}, and distance measurements, \cite{adamek2014distance, Adamek:2018rru, Macpherson:2021gbh}.

In this view, the fundamental problem is to describe how an observer perceives signals emitted by a distant object in any spacetime. The difficulty of such an analysis is that the observed quantities depend on the curvature of spacetime and the motion of the emitter and the observer \cite{rasanen, Korzynski:2018}. This problem is easily overcome in the new formulation of light propagation in geometric optics that we presented in \cite{Grasso:2018mei}.
All possible effects on light distortions caused by the curvature between the observer and the source are encoded in the bilocal geodesic operators, which are the fundamental quantities of our formalism. In this way, the effects on the light due to curvature and those caused by the kinematics of the observer and the source can be clearly distinguished, \cite{Grasso:2018mei}. 
Once the bilocal geodesic operators are computed, they can be combined with the source and observer motion to obtain all possible optical observables such as magnification, shear, angular distance, and the real-time observables (i.e., parallax, redshift drift and position drift). In this sense, the bilocal geodesic operators formalism provides a unified framework for studying light propagation and the calculation of optical observables, \cite{Grasso:2018mei, Grasso:2021iwq}.

\section{Main results}
\label{sec:mainRes}
This thesis is devoted to the presentation of {\tt BiGONLight}, {\tt Bi}local {\tt G}eodesic {\tt O}perators framework for {\tt N}umerical {\tt Light} propagation, the {\tt Wolfram} package\footnote{{\color{blue}{ https://github.com/MicGrasso/bigonlight.git.}}} I have developed for the study of light propagation in numerical simulations. 
The original results of this thesis were published in the two peer-reviewed articles \cite{Grasso:2021iwq, Grasso:2021zra}, and are summarised as follows:
\begin{description}
\item[The $3+1$ bilocal geodesic operators framework:] numerically generated spacetimes are evolved in full general-relativistic simulations using the ADM formalism, based on the $3+1$ splitting of the Einstein equations. To make {\tt BiGONLight} compatible with such computer-generated spacetimes, I have obtained the expressions of the parallel transport equations, the optical tidal matrix, and the geodesic deviation equations for the bilocal operators in 3+1 form. Once the geodesic connecting observer and emitter is found (as shown in \cite{Vincent:2012kn}), the above equations can be used to perform the parallel transport of a reference frame and obtain the bilocal geodesic operators along that geodesic. The computation is simplified by using the general matrix form of the optical tidal matrix and the bilocal operators projected into the semi-null frame, which I have obtained.
\item[BiGONLight:] the $3+1$ bilocal geodesic operators framework is encoded in the package as a collection of {\tt Mathematica} functions. These functions take as input the ADM quantities directly from a numerical simulation or provided by the user in analytical components to find the bilocal geodesic operators. The bilocal operators are the starting point to obtain all possible optical observables by combining them with the observer and emitter four-velocities and four-accelerations. The package leaves complete control to the user, who can choose the position of the source and the observer anywhere along the null geodesic with any four-velocities and four-accelerations. 
\item[From forward to backwards-integrated bilocal operators:] I have used the properties of the geodesic deviation equation to obtain the transformation between forward-integrated and backwards-integrated bilocal geodesic operators. Forward-integrated bilocal geodesics operators can be helpful in cosmological simulations to study the properties of light propagation on-the-fly with the simulation of spacetime.
However, observables are obtained using backwards-integrated bilocal geodesic operators since they express observations performed by the observer receiving the light emitted by a source in the past. The explicit transformations between these two methods enlarge the range of applicability of the package.
\item[Tests using three cosmological models:] the accuracy of the package is tested by computing redshift, angular diameter distance, parallax distance, and redshift drift in well-known cosmological models. Three different kinds of inputs are provided: analytical metric components of a homogeneous $\Lambda$CDM model, analytical metric components of the inhomogeneous Szekeres model (as presented in \cite{Meures:2011ke, Meures:2011gp}), numerical data of a uniform dust Universe (EdS) evolved with the {\tt Einstein Toolkit} and {\tt FLRWSolver}, \cite{loffler2012einstein, macpherson2017}.
\item[Isolating nonlinear effects of light propagation:] we present a detailed analysis of the different ways in which inhomogeneities contribute to nonlinearities in cosmological observables. In this study, I have applied {\tt BiGONLight} to compute observables calculated at different approximations in a plane-parallel inhomogeneous spacetime. The nonlinear effects are evaluated as the fractional difference between observables obtained at the three different approximations linear perturbation theory, Newtonian, and post-Newtonian approximations. The inhomogeneities are tuned by varying the model's free parameters, and their contributions to the observables are obtained by analysing the variations in the fractional differences.
\end{description}

\subsection{Structure of the thesis}
The outline of the thesis is the following: the basis of modern cosmology and the description of light propagation with the bilocal geodesic operators are described in Chapters~\ref{chap:cosmology} and~\ref{chap:BGO}, respectively. A detailed presentation of {\tt BiGONLight} and its code tests is given in Chapter~\ref{chap:bigonlight}. Chapter~\ref{chap:nonlinearities} presents the application of the package to study the nonlinear contributions to light propagation in the inhomogeneous plane-symmetric Universe. Finally, the conclusions and a detailed summary of the results of my research are addressed in Chapter~\ref{chap:conclusion}.

\section{Conventions and notations}
This dissertation is a collection of articles, and every effort has been made to make the notation as consistent and clear as possible.
Throughout the text, we assume that the spacetime metric $g_{\mu \nu}$ has signature $(-,+,+,+)$. We also assume the Einstein summation notation $\sum_{\mu} a_{\mu} b^{\mu}\equiv a_{\mu} b^{\mu}$, with the range of the sum depending on the nature of the index $\mu$. 
Greek indices ($\alpha, \beta, ...$) run from 0 to 3, while Latin indices ($i,j, ...$) run from 1 to 3 and refer to spatial coordinates only. Latin indices ($A,B, ...$) run from 1 to 2. Tensors and bitensors expressed in a semi-null frame are denoted using boldface indices: Greek boldface indices ($\boldsymbol{\alpha}, \boldsymbol{\beta}, ...$) run from 0 to 3, Latin boldface indices ($\mathbf{a}, \mathbf{b}, ...$) run from 1 to 3 and capital Latin boldface indices ($\mathbf{A}, \mathbf{B}, ...$) run from 1 to 2. Boldface letters ($\mathbf{X}, \mathbf{Y}, ...$) are also used for vectors in quotient space $\mathcal{P}_{p}$, while objects in $\mathcal{Q}_{p}$ are denoted using square brackets ($[X],\, [Y],\, ...$). Objects defined in tangent spaces $T_{p}\mathcal{M}$ are denoted by standard letters.
Overdotted quantities denote a derivative with respect to conformal time, i.e. $\dot{A}=\frac{d A}{d \eta}$. Quantities with a subscript $0$ are meant to be evaluated at present, whereas the subscript $``{\rm in}"$ indicates the initial time. Quantities with a subscript $\cal S$ (or $\cal O$) are meant to be evaluated at the source (observer) position. 
A description on how to set physical units in {\tt BiGONLight} is presented in Chapter~\ref{chap:bigonlight} (appendix A in \cite{Grasso:2021iwq}).

\begin{table}
\centering
\begin{tabular}{ |p{3cm}||p{8cm}|p{3cm}|  }
 \hline
 \multicolumn{3}{|c|}{\cellcolor{gray!20}List of Acronyms} \\
 \hline
 Acronym & Signification & page\\
 \hline
\ref{acr:ADM}   & Arnowitt-Deser-Misner formalism &   \pageref{acr:ADM}\\ 
\ref{acr:BGO}   & bilocal geodesic operators &   \pageref{acr:BGO}\\
\ref{acr:CMB} & cosmic microwave background &  \pageref{acr:CMB}\\
\ref{acr:EdS}   & Einstein-de Sitter &   \pageref{acr:EdS}\\
\ref{acr:EOS}   & equation of state &   \pageref{acr:EOS}\\
\ref{acr:FLA}   & flat lightcone approximation &   \pageref{acr:FLA}\\
\ref{acr:FLRW}   & Friedmann-Lema\^{i}tre-Robertson-Walker &   \pageref{acr:FLRW}\\
\ref{acr:GDE}   & geodesic deviation equation &   \pageref{acr:GDE}\\
\ref{acr:GR}   & general relativity &   \pageref{acr:GR}\\
\ref{acr:hot}   & higher-order term &   \pageref{acr:hot}\\
\ref{acr:LCDM}   & $\Lambda$-cold dark matter &   \pageref{acr:LCDM}\\
\ref{acr:Lin}   & first-order in perturbation theory &   \pageref{acr:Lin}\\
\ref{acr:LSS}   & large-scale structure &   \pageref{acr:LSS}\\
\ref{acr:LTB}   & Lema\^{i}tre-Tolman-Bondi &   \pageref{acr:LTB}\\
\ref{acr:ODE}   & ordinary differential equations &   \pageref{acr:ODE}\\
\ref{acr:PN}   & post-Newtonian approximation &   \pageref{acr:PN}\\
\ref{acr:PT}   & cosmological perturbation theory &   \pageref{acr:PT}\\
\ref{acr:SNF}   & semi-null frame &   \pageref{acr:SNF}\\
\ref{acr:SnIa}   & type Ia supernov\ae candels &   \pageref{acr:SnIa}\\
 \hline
\end{tabular}
\end{table}

%% file: template/cosmology.tex
\chapter{The standard cosmological model}
\label{chap:cosmology}

\begin{figure*}[ht]
    \centering
    \begin{subfigure}{0.49\linewidth}
        \includegraphics[width=\linewidth]{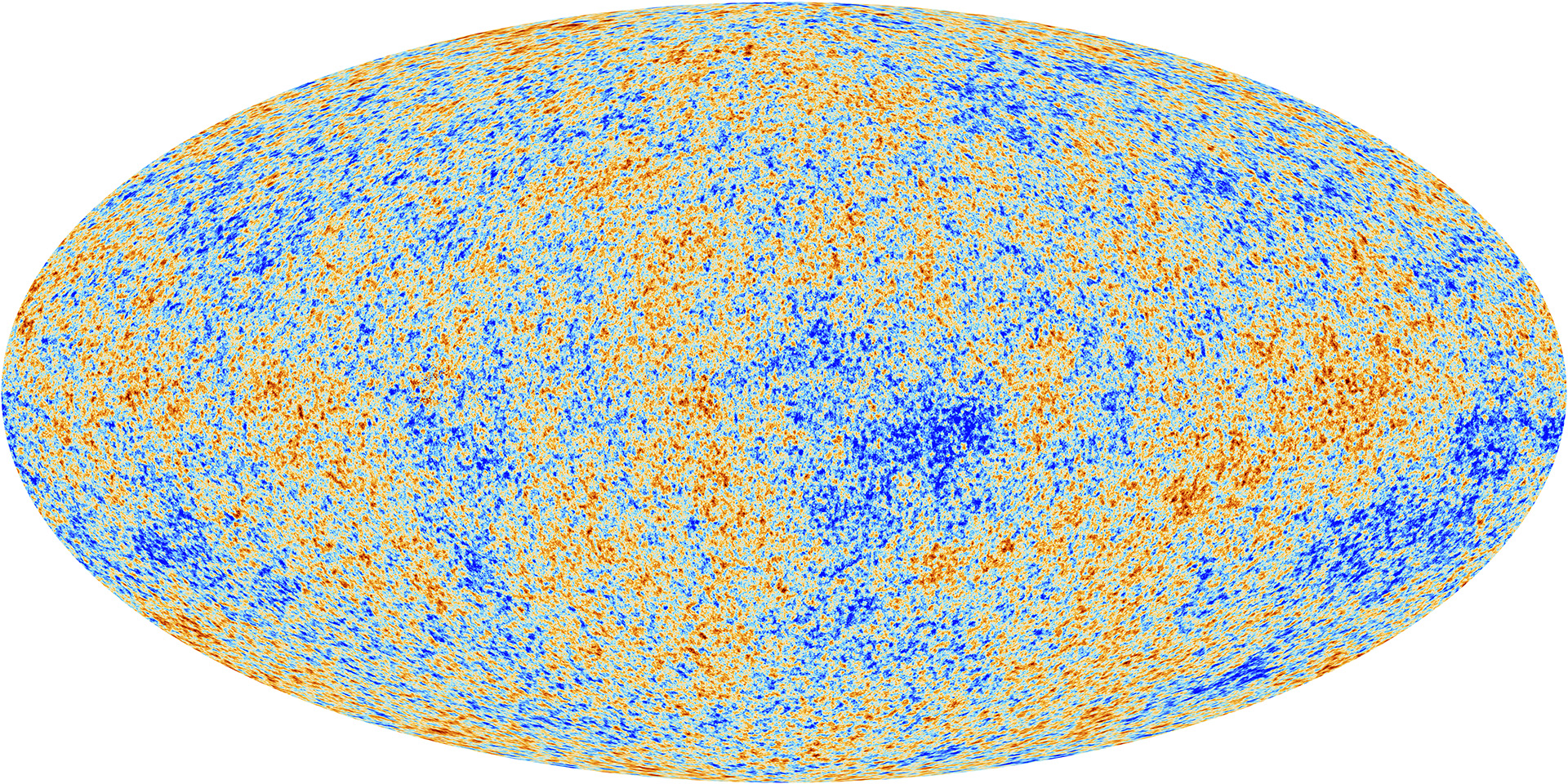}
       \caption{Credits: Planck {\color{blue}{https://www.esa.int/}}}
        \label{fig:CMB}
        \includegraphics[width=\linewidth]{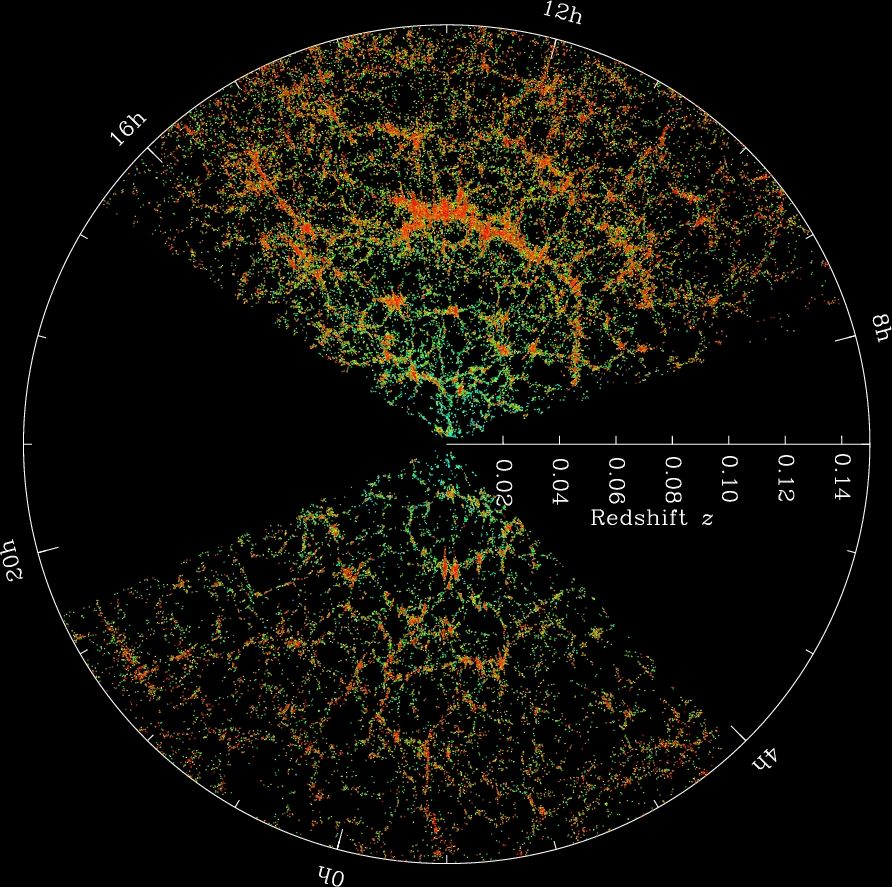}
        \caption{Credits: SDSS {\color{blue}{https://www.sdss.org/}}}
        \label{fig:LSS}
    \end{subfigure}
    \begin{subfigure}{0.5\linewidth}
        \includegraphics[width=\linewidth]{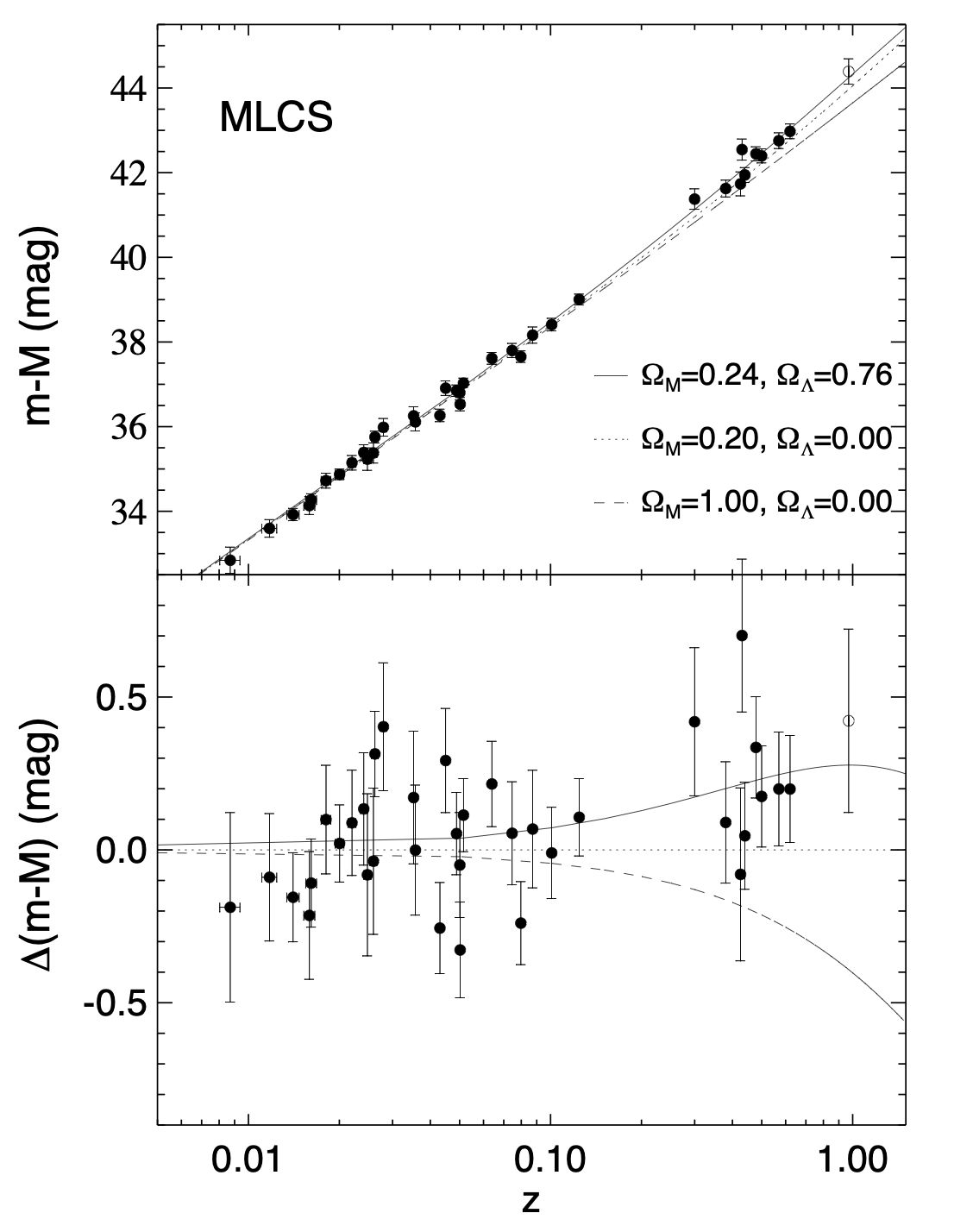}
        \caption{Credits: Riess et.al. 1998, \cite{Riess:1998}. }
        \label{fig:SnIa}
    \end{subfigure}
    \caption{Precise measurements of the CMB (Fig.~\ref{fig:CMB}), the LSS (Fig.~\ref{fig:LSS}), and the SnIa (Fig.~\ref{fig:SnIa}) have been used to constraint the standard model for cosmology.}\label{fig:evidences}
\end{figure*}

The increasing amount of data collected by experimental probes like SDSS, Planck, LSST, SKA and others\footnote{{\color{blue}{https://www.sdss.org}}, {\color{blue}{https://www.cosmos.esa.int/web/planck}}, {\color{blue}{https://www.lsst.org}}, {\color{blue}{http://skatelescope.org/}}} have produced a generally coherent picture of our Universe.
The general purpose of these experiments is to investigate the three main pieces of evidence in cosmology:
%
\begin{itemize}
\item the \emph{cosmic microwave background} (\setwd{CMB}{acr:CMB}): the CMB is the electromagnetic radiation that pervades the Universe as homogeneous and isotropic background noise. It was discovered by A. Penzias and R. Wilson in 1965 \cite{Penzias:1965}, when they observed a thermal black body spectrum with a temperature of $ 3.5 \, {\rm K}$ across the entire sky.
Successive measurements \cite{COBE:1992syq, WMAP:2008lyn, planck2018param, planck2019anl, planck2020CMB} have revealed further details in the CMB structure, showing small fluctuations of $\Delta T/ T \sim 10^{-5}$ around the average temperature of $2.7 \, {\rm K}$. Accurate mapping of the small anisotropies in the temperature of the CMB is of fundamental importance to cosmology, as it provides a clue to the structure of the Universe at very early times, see Fig.~\ref{fig:CMB}. 
 
\item the \emph{large-scale structure} (\setwd{LSS}{acr:LSS}): observations in different wavelength ranges of electromagnetic radiation have revealed a hierarchical organisation of astrophysical objects: massive objects tend to form gravitationally bound structures such as galaxies and galaxy clusters, which organise themselves on cosmological scales into superclusters, filaments, and voids, forming the so-called \emph{cosmic web} \cite{carroll2017book}. Increasingly better galaxy surveys, like \cite{york2000sloan, colless20012df, mccracken2003virmos, vogt2005deep, beasley2002vlba}, have measured the distance and shape of cosmological structures with great precision and produced an accurate three-dimensional map of the LSS of the Universe, see Fig.~\ref{fig:LSS}. Furthermore, since these cosmological structures observed today are the result of the evolution of the tiny CMB anisotropies under the influence of gravity, we expect to find features of the CMB radiation anisotropy in the observed LSS.

\item the \emph{type Ia supernov\ae} (\setwd{SnIa}{acr:SnIa}): the distance-redshift relation is one of the landmarks of modern cosmology. It is also known as ``Hubble law'' and was first derived by G. Lema\^{i}tre in 1927, \cite{lemaitre1927univers}, as a linear relation between the distance of a galaxy $D$ and its recession velocity $v$ ($v = c z$ for small redshifts $z < 1$), i.e. $z= \frac{H_{\rm 0}}{c} D$. The same relation was measured by E. Hubble in 1929, \cite{hubble1988relation}, finding a value of the constant of proportionality $H_{\rm 0}=500 \, {\rm km/s/Mpc}$. This is considered the first evidence for the expansion of the Universe\footnote{This relationship is universal and does not depend on the location of observation. So if we make the same measurement from another galaxy, we get the same relation with the same constant $H_{\rm 0}$. From this, we can conclude that the Universe is expanding isotropically at the expansion rate $H_{\rm 0}$.
}. An accurate measurement of $H_{\rm 0}$ requires a precise estimate of the redshift and the distance of the source: the redshift is obtained directly from spectroscopic analysis, while the distance is derived using indirect methods, such as geometrical relations and/or physical properties of astronomical candles. 
Precise measurements of the distance-redshift relation in cosmology are performed using SnIa standard candles\footnote{Standard candles are astronomical objects whose absolute luminosity $L$ is known, so their distance is determined by measuring the luminosity flux at the observer $F$ and using the relation $D=\sqrt{\frac{L}{4 \pi F}}$ (see Eq.\eqref{eq:D_lum_def}). SnIa are explosions of white dwarf stars characterised by a precise relation between the brightness and the timescale of the explosion.},  see Fig.~\ref{fig:SnIa}, which led not only to a better estimate of the present-day expansion rate $H_{\rm 0}=73.48\pm 1.66\, {\rm km/s/Mpc}$ \cite{Riess:2018uxu}, but also to the first confirmation of the accelerated expansion of the Universe, \cite{Riess:1998, Perlmutter_1999}.
\end{itemize}

This chapter describes the basics of modern cosmology and how the information obtained from the CMB, LSS, and SnIa observations are merged to produce a theoretical model of the Universe.

\section{Cosmological models}
As noted above, observations on cosmological scales suggest that the gravitational attraction of the primordial anisotropies observed in the CMB shaped the Universe's large-scale structure. In general, gravity is the dominant interaction responsible for the formation of structures at all scales (from planets to the LSS), and it represents the fundamental mechanism underlying the formation and evolution of the Universe. Therefore, a model of the Universe must be consistent with the laws of gravity.

\subsection{General relativity}
From a theoretical point of view, the gravitational interaction at the macroscopic level\footnote{Gravity is the weakest of the four fundamental interactions of nature and has a negligible influence on the behaviour of subatomic particles. However, there are events in the cosmos that involve strong gravitational effects at the quantum scale that can only be described by a theory of quantum gravity.} is described by \emph{general relativity} (\setwd{GR}{acr:GR}), which was proposed by A. Einstein in 1915, \cite{Einstein:1915gr}. 
The primary distinction between general relativity and Newtonian gravity is Einstein's interpretation of gravity as a geometric property of spacetime. 
This interpretation was supported by the \emph{equivalence principle}, which states that gravitational acceleration is the same as the acceleration of an inertially moving body and by the fact that there can be no absolute concept of inertia, but only the inertia of masses relative to each other. 
In this perspective, Einstein assumes that free particles move along geodesics in a four-dimensional Riemannian manifold $\mathcal{M}$ whose points represent physical locations in space and time. 
Locally, we can specify a reference frame and label the point in $\mathcal{M}$ by a coordinate system $\{ x^{\mu} \}$. However, the coordinate system $\{ x^{\mu} \}$ is not uniquely defined since the laws of physics must be independent with respect to this choice.
To manifest this ``gauge invariance'', general relativity is formulated using the invariant structures of tensors.
Indeed, at each point $x$ of the manifold $\calM$, we can introduce a tangent space\footnote{Defined as the real vector space that intuitively contains all the possible directions in which one can tangentially pass through $x$.} $T_{x}\calM$ and the cotangent space $T^*_{x}\calM$, being the dual space to $T_{x}\calM$, and define the type $(r, s)$ tensor $\bm{T}$ as the multilinear map 
$$\bm{T}: \underbrace{T^*_{x}\calM \times \cdots \times T^*_{x}\calM}_\text{r} \times \underbrace{T_{x}\calM \times \cdots \times T_{x}\calM}_\text{s} \, \to\, \mathbb{R}\, .$$
Although this invariant coordinate formulation fits well the purposes of general relativity, for practical calculations it is better to express tensors by their components: a type $(r, s)$ tensor may be written as
\begin{equation}
\bm{T} =T\UD{\mu_1 \cdots \mu_r}{\nu_1 \cdots \nu_s} \,  \dfrac{\partial}{\partial x^{\mu_1}} \otimes \cdots \otimes \dfrac{\partial}{\partial x^{\mu_r}} \otimes d x^{\nu_1} \otimes \cdots \otimes d x^{\nu_s}\, 
\end{equation}
where $\dfrac{\partial}{\partial x^{\mu_i}}$ is a basis for the $i$-th tangent space, with $i=1, 2, \cdots r$, and $d x^{\nu_j} $ a basis for the $j$-th cotangent space, with $j=1, 2, \cdots s$. In other words, the type $(r, s)$ tensor $\bm{T}$ associates r vectors $\dfrac{\partial}{\partial x^{\mu_1}}, \cdots , \dfrac{\partial}{\partial x^{\mu_r}}$ and s covectors $d x^{\nu_1} , \cdots, d x^{\nu_s}$ to a scalar $T\UD{\mu_1 \cdots \mu_r}{\nu_1 \cdots \nu_s}$. It is indeed evident that the tensor components $T\UD{\mu_1 \cdots \mu_r}{\nu_1 \cdots \nu_s}$ depend on the choice of coordinates, although the tensor itself $\bm{T}$ is independent. This freedom in the choice of coordinate system can be used to simplify the computation of the tensor components.

The geometry of Riemannian spacetime is encoded in the metric tensor $g_{\mu \nu}$, which is the core object of general relativity. Indeed, the metric is used to define coordinate invariants, such as the squared line element $ds^2=g_{\mu \nu}dx^{\mu}dx^{\nu}$, which expresses the measure of the proper distance between two arbitrarily close events in spacetime $x^{\mu}$ and $x^{\mu}+ dx^{\mu}$.
Moreover, the metric tensor is used to introduce essential structures like the \emph{covariant derivative} (or connection) $\bm{\nabla}$, which is the covariant generalization of the partial derivative that allows to derive and transport tensors along the manifold. The action of the covariant derivative $\nabla_{\sigma}$ on a type $(r, s)$ tensor is expressed as
\begin{equation}
\nabla_{\sigma}T\UD{\mu_1 \cdots \mu_r}{\nu_1 \cdots \nu_s} =\partial_{\sigma} T\UD{\mu_1 \cdots \mu_r}{\nu_1 \cdots \nu_s} + \Sigma_{i}\Gamma^{\mu_{i}}_{\sigma \lambda} T\UD{\mu_1 \cdots \lambda \cdots \mu_r}{\nu_1 \cdots \nu_s} - \Sigma_{j}\Gamma^{\lambda}_{\sigma \nu_{j}} T\UD{\mu_1 \cdots \mu_r}{\nu_1 \cdots \lambda \cdots \nu_s}\, ,
\end{equation}
with $\partial_{\mu}=\dfrac{\partial}{\partial x^{\mu}}$, and $\bm{\Gamma}$ being the \emph{Christoffel symbols} expressing the difference between the covariant and the partial derivative\footnote{It worth mentioning that $\bm{\Gamma}$ are not tensor quantities since they do not transform as tensor. However, the difference $C\UD{\rho}{\mu \nu}=\Gamma^{\rho}_{\mu \nu}-\Gamma^{\rho}_{\nu \mu}$ defines a tensor expressing the torsion of the spacetime. In general relativity we assume that the spacetime is torsion-less, implying that $\Gamma^{\rho}_{\mu \nu}$ are symmetric in the lower indices.} in terms of metric tensor derivatives
\begin{equation}
\Gamma^{\rho}_{\mu \nu}=\dfrac{1}{2} g^{\rho \lambda}\left( \partial_{\mu}g_{\lambda \nu}+\partial_{\nu}g_{\mu \lambda}-\partial_{\lambda}g_{\mu \nu}\right)\, . \label{eq:christoffel_symb}
\end{equation}
One application of the covariant derivative is the \emph{directional derivative} $t^{\sigma}\nabla_{\sigma} v^{\mu}$, namely the derivative of a vector field $v^{\mu}$ in the direction of a given vector $t^{\mu}$. If $t^{\mu}$ is the tangent vector to the curve $C$ in the manifold and $t^{\sigma}\nabla_{\sigma} v^{\mu}=0$ along all points of the curve, we say that the vector $v^{\mu}$ is \emph{parallel transported along} $C$. In the special case where the tangent vector $t^{\mu}$ is parallel transported along itself $t^{\sigma}\nabla_{\sigma} t^{\mu}=0$, we call the curve $C$ a \emph{geodesic}, which in general relativity represents the trajectory of a free particle. In special relativity, the trajectories of free particles are straight lines, but in general relativity, the structure of spacetime is curved by the presence of matter, and this ``bends'' the particle's trajectories.

The equations which make it possible to derive the metric tensor associated with a given distribution of matter are the Einstein's field equations
\begin{equation}
G_{\mu \nu}\equiv R_{\mu \nu}-\dfrac{1}{2} R g_{\mu \nu}=\dfrac{8 \pi G}{c^4}T_{\mu \nu}
\, , \label{eq:einsteinEq}
\end{equation} 
where $G_{\mu \nu}$ is the Einstein tensor representing the curvature of the spacetime, $T_{\mu \nu}$ is the stress-energy tensor representing the energy and momentum of matter and radiation, while $G$ and $c$ are the gravitational constant and the speed of light, respectively. 
The two quantities $R_{\mu \nu} \equiv R\UD{\sigma}{\mu \sigma \nu}$ and $R \equiv g^{\mu \nu} R_{\mu \nu}$ are the Ricci tensor and the Ricci scalar, and they are directly defined from the Riemann tensor
\begin{equation}
R\UD{\rho}{\mu \sigma \nu}=\partial_{\sigma}\Gamma^{\rho}_{\mu \nu} -\partial_{\nu}\Gamma^{\rho}_{\mu \sigma}+\Gamma^{\rho}_{\lambda \sigma}\Gamma^{\lambda}_{\mu \nu}-\Gamma^{\rho}_{\lambda \nu}\Gamma^{\lambda}_{\mu \sigma}\, .
\end{equation}

\subsection{The FLRW class of models}
A valid cosmological model must be based on general relativity, and its spacetime must satisfy the Einstein equations~\eqref{eq:einsteinEq}. Thus, it is a matter of finding the correct stress-energy tensor to describe the correct distribution of energy and momentum in the Universe. 
According to cosmological observations, the Universe seems to be filled with an isotropic and homogeneous mass distribution on large scales, which can be represented as a perfect fluid characterised by an average energy density $\bar{\rho}$ and isotropic pressure $p$. 
The expression for stress-energy tensor of this cosmic fluid is written as
\begin{equation}
T^{\mu \nu}=\left(\bar{\rho} + \dfrac{p}{c^2} \right)u^{\mu}u^{\nu} + p g^{\mu \nu}\, , \label{eq:T_fluid}
\end{equation}
with $u^{\mu}=g^{\mu \nu}u_{\nu}$ the four-velocity vector field of the fluid. In addition, the tensor $T_{\mu \nu}$ satisfies the following conservation rules
\begin{equation}
\nabla_{\mu}T^{\mu \nu} \equiv \partial_{\mu}T^{\mu \nu} +\Gamma^{\mu}_{\mu \lambda}T^{\lambda \nu} + \Gamma^{\nu}_{\mu \lambda}T^{\mu \lambda}=  0\, ,\label{eq:T_conservation}
\end{equation}
whose components are the continuity equation for the energy density $ \bar{\rho}$ and the Euler equation for the fluid. 

Under the same assumptions of isotropy and homogeneity, we may derive the following form of the metric\footnote{It is customary to express the metric components by specifying the line element.}
\begin{equation}
ds^2= - c^2 dt^2+ a(t)^2\left[ \dfrac{d r^2}{1-k r^2}+ r^2 d\theta^2+r^2 \sin(\theta)^2 d\phi^2\right]\, , \label{eq:FLRW_metric}
\end{equation} 
where $(t, r, \theta, \phi)$ are coordinates, $a(t)$ is the scale factor, and $k$ is a constant expressing the spatial curvature: usually the coordinates are rescaled such that $k$ is set to $-1$, $1$, or $0$ for space of constant negative, positive, or null spatial curvature, respectively.  
The metric in Eq.~\eqref{eq:FLRW_metric} is the \emph{Friedmann-Lema\^{i}tre-Robertson-Walker} metric (\setwd{FLRW}{acr:FLRW}) and it represents a class of cosmological models with the dynamics determined by the scale factor $a(t)$.
The time coordinate $t$ is known as the \emph{cosmic time}, and it is the proper time measured in a comoving frame with the observer, i.e. the frame in which the observer's position remains unchanged $(r, \theta, \phi)=\const$. In this frame the four-velocity of a fluid element is simply $u^{\mu}=(1, 0, 0, 0)$ and the stress-energy conservation, Eq.~\eqref{eq:T_conservation}, for the FLRW metric simply reduces to 
\begin{equation}
\dfrac{d}{d t}(\bar{\rho} a^3)=-\dfrac{p}{c^2} \dfrac{d}{d t}(a^3)\, .\label{eq:eos_1}
\end{equation}
This formula expresses the mass-energy conservation, relating the change of the energy density in a element volume $\dfrac{d}{d t}(\bar{\rho} a^3)$ to the pressure acting on that volume $p \dfrac{d}{d t}(a^3)$. In cosmology Eq.~\eqref{eq:eos_1} is usually written as
\begin{equation}
\dfrac{d \log(\bar{\rho})}{d t}=-3 (1+w) \dfrac{d \log(a)}{d t}\, ,\label{eq:FLRW_eos}
\end{equation}
with $w=p/(c^2 \bar{\rho})$. If $w$ is time independent, Eq.~\eqref{eq:FLRW_eos} has solution 
\begin{equation}
\bar{\rho}= \bar{\rho}_0 \left( \dfrac{a_0}{a}\right)^{3 (1+w)}\, ,\label{eq:eos_sol}
\end{equation} 
with $\bar{\rho}(t_{\rm 0})=\bar{\rho}_0$ and $a(t_0)=a_0$ the mass-energy density and the scale factor at present time $t_{\rm 0}$, respectively. The quantity $w=p/(c^2 \bar{\rho})$ gives the \emph{equation of state} (\setwd{EOS}{acr:EOS}) of the cosmic fluid, where some notable examples are:
\begin{itemize}
\item $w=1/3$ for radiation, which gives $\bar{\rho}_{r}\propto a^{- 4}$;
\item $w=0$ for (pressureless) matter, which gives $\bar{\rho}_{\rm m}\propto a^{- 3}$;
\item $w=-1$ for vacuum (dark) energy, which gives $\bar{\rho}_{\Lambda}= \const$.
\end{itemize} 
From this we can conclude that the mass-energy density is the sum of different species $\bar{\rho}(t)=\Sigma_{\rm x} \bar{\rho}_{\rm x}(t) $ and each species scales as a different power of the scale factor according to its EOS $w_{\rm x}$. In cosmology, it is usually assumed that the only species contributing to the mass-energy density are baryonic matter $\bar{\rho}_{\rm b}$, dark matter $\bar{\rho}_{\rm m}$, radiation $\bar{\rho}_{\rm r}$, and dark energy $\bar{\rho}_{\rm \Lambda}=\frac{\Lambda c^2}{8 \pi G}$, i.e. $\bar{\rho}=\bar{\rho}_{\rm b}+\bar{\rho}_{\rm m}+\bar{\rho}_{\rm r}+\bar{\rho}_{\rm \Lambda}$. Similarly, the species contributing to the cosmic pressure are the radiation pressure $p_{\rm r}$ and the dark energy pressure $p_{\rm \Lambda}=-\frac{\Lambda c^4}{8 \pi G}$, since the  (baryonic and dark) matter have vanishing\footnote{The baryonic and dark particles are non-relativistic and therefore their energy density is much larger than their pressure.} pressure $w=0$. More often we refer to ``dust'', or simply matter, to indicate both baryonic and dark matter components as the part of the perfect fluid that has positive mass density and vanishing pressure.
By defining the \emph{Hubble parameter} (also known as expansion rate) $H=\dfrac{1}{a}\dfrac{d a}{d t}$ and the \emph{critical density} as $\rho_{\rm c}(t)=\frac{3 H(t)^2}{8 \pi G}$, we express the mass-energy density by the dimensionless \emph{density parameter} $\Omega(t)=\frac{\bar{\rho}(t)}{\rho_{\rm c}(t)}$. This is particularly convenient in cosmology, since using Eq.~\eqref{eq:eos_sol} one can write the density parameter of each species $\Omega_{\rm x}(t)$ in terms of its value at the present time, $\Omega_{\rm x}(t_0)=\Omega_{\rm x_0}$, i.e.
\begin{equation}
\Omega_{\rm x}=\dfrac{8 \pi G}{3 H_0^2}\bar{\rho}_{\rm x}= \Omega_{\rm x_0} \left(\dfrac{a_0}{a}\right)^{3 (1+w_{\rm x})}\, ,\label{eq:Omega_Omegat0}
\end{equation}
and thus clearly separate the contributions of the different species at present time in the density parameter $\Omega(t)$ as
\begin{equation}
\Omega(t)= (\Omega_{\rm b_0}+\Omega_{\rm m_0})\left(\dfrac{a_0}{a(t)}\right)^{3} +\Omega_{\rm r_0} \left(\dfrac{a_0}{a(t)}\right)^{4} +\Omega_{\rm \Lambda_0}\, . \label{eq:Omega(t)}
\end{equation}

The Einstein field equations in Eq.~\eqref{eq:einsteinEq} for the FLRW metric give the \emph{Friedmann equations} 
\begin{align}
\dfrac{1}{a^2}\left(\dfrac{d a}{d t}\right)^2 & = \dfrac{8 \pi G}{3} \bar{\rho}- \dfrac{k c^2}{a^2} \label{eq:friedmann_1}\\
\dfrac{1}{a}\dfrac{d^2 a}{d t^2} & = -\dfrac{4 \pi G}{3} \left(\bar{\rho}+\dfrac{3 p}{c^2}\right)\, . \label{eq:friedmann_2}
\end{align}
The first Friedmann equation, Eq.~\eqref{eq:friedmann_1}, expressed in terms of the density parameter Eq.~\eqref{eq:Omega(t)} reads
\begin{equation}
H(t)^2 = H_0^2 \left[ \Omega_{\rm r_0} \left(\dfrac{a_0}{a(t)}\right)^{4} + (\Omega_{\rm b_0}+\Omega_{\rm m_0}) \left(\dfrac{a_0}{a(t)}\right)^{3} + \Omega_{\rm k_0} \left(\dfrac{a_0}{a(t)}\right)^{2}+ \Omega_{\rm \Lambda_0} \right] \, , \label{eq:Friedmann_edH}
\end{equation}
with $\Omega_{\rm k_0}= - \dfrac{k c^2}{H_0^2 a_0^2}$. 
Similarly, the second Friedmann equation, Eq.~\eqref{eq:friedmann_2}, in terms of density parameters reads
\begin{equation}
q\equiv -\dfrac{\dfrac{d H}{dt}+H^2}{H^2}=\dfrac{1}{2}\sum_{\rm x}  (1+3 w_{\rm x}) \Omega_{\rm x} \, , \label{eq:Friedmann_edQ}
\end{equation}
where $q$ is the \emph{deceleration parameter}, and we remind that $\Omega_{\rm x}(t)=\frac{8 \pi G}{3 H(t)^2}\bar{\rho}_{\rm x}(t)$ and $w_{\rm x}$ are the density parameter and the EOS parameter for the ${\rm x}^{th}$ specie at time $t$, respectively.

The peculiarity of the FLRW models is that they predict the beginning of the Universe from a singularity point at $t=0$, i.e. $a(0)=0$: this is known as \emph{Big Bang} and corresponds to the origin of the Universe ($13.8$ billion years, according to the standard cosmological model, \cite{ellis2012relativistic, planck2018param}) from extreme conditions of density, pressure, and temperature. From this extreme initial state, the Universe began its adiabatic expansion, becoming less dense and colder, allowing the formation of all elementary particles and gradually, electrons, photons, and baryons. One of the strongest evidence in favour of the Big Bang theory is the CMB, corresponding to the radiation relict formed $379000$ years after the Big Bang in the \emph{recombination epoch}, \cite{ellis2012relativistic}. Before that time, the Universe was indeed a plasma of electrons, protons, and nuclei, in which the photons were constantly scattered. Due to cosmic expansion, conditions in the era of recombination became favourable for the formation of atoms (mainly hydrogen and helium) until the photons decoupled from matter and began to move freely through the expanding Universe: this is observed today as the CMB thermal radiation. Thus, on the one hand, the CMB is an excellent source of information, as it provides a ``snapshot'' of the Universe as it was $379000$ years after the Big Bang, but on the other hand, it also constitutes a limit to cosmological observations with electromagnetic radiation. 

Some notable examples from the class of FLRW models are the \emph{de Sitter} and \emph{Einstein-de Sitter} (\setwd{EdS}{acr:EdS}) models, each representing a (spatially flat) Universe containing only dark energy and only dark matter, respectively. The de Sitter model represents an empty Universe (without matter) containing only dark energy, which determines the expansion rate $H \propto \sqrt{\Lambda}$. It is characterised by an exponentially growing scale factor $a(t)=e^{H t}$, which causes an accelerated expansion of the de Sitter Universe. Since no other mechanism opposes the accelerated expansion, at a certain point, any observer in a de Sitter Universe will start experiencing event horizons, beyond which it is impossible to see or perceive anything.
In contrast, the Einstein-de Sitter model represents a Universe containing only dust (pressureless matter) with a density of $\bar{\rho}_{\rm m}\propto H^2$. In this model, the distance between two comoving observers increases with $t^{2/3}$, but this expansion is balanced by gravitational attraction so that it tends asymptotically to zero as time approaches infinity.
Although these models do not explain current observations, they can be considered reasonable approximations for past epochs of the Universe, \cite{ellis2012relativistic}. Using Friedmann equation Eq.~\eqref{eq:Friedmann_edH}, we can distinguish the following epochs: at very early times, i.e. $a(t)$ small, the Universe was dominated by radiation, since $H^2 \sim H_0^2 \Omega_{\rm r_0} a^{-4}$. After that, followed a matter-dominated era (for a spatially flat Universe), with $H^2 \sim H_0^2 \Omega_{\rm m_0} a^{-3}$, which lasted until dark energy took over, leading to an accelerated expansion of the Universe, $q=\Omega_{\rm m_0}/2 -\Omega_{\rm \Lambda_0} <0$.

\section{The $\Lambda$CDM model}

The Friedmann equations Eqs.~\eqref{eq:Friedmann_edH}-\eqref{eq:Friedmann_edQ}, together with the EOS and Eq.~\eqref{eq:FLRW_eos}, completely define the dynamics and composition of the cosmological model. Therefore, to build a complete picture of the cosmological model that best fits our Universe, one must measure the values of the cosmological parameters $H_0$, $\Omega_{\rm r_0}$, $\Omega_{\rm b_0}$, $\Omega_{\rm m_0}$, $\Omega_{\rm k_0}$, and $\Omega_{\rm \Lambda_0}$ from the observations, and determine the dynamics by solving the Friedmann equations. 
The first important constraint on the total energy-density of the Universe is provided by the Friedmann equation Eq.~\eqref{eq:Friedmann_edH} evaluated at present time, i.e.
\begin{equation}
 \Omega_{\rm r_0} + \Omega_{\rm b_0} + \Omega_{\rm m_0} + \Omega_{\rm k_0} + \Omega_{\rm \Lambda_0} = 1 \, . \label{eq:components}
\end{equation}
Several cosmological parameters can be measured within the same observation, and the specific observation can constrain each parameter differently: their values are obtained by best-fitting the different measurements of the cosmological parameters by the CMB, SnIa, and LSS. The latest results, published in $2018$ by the Planck Collaboration \cite{planck2018param}, depict a cosmological model consistent with a spatially flat Universe $\Omega_{\rm k_0}= 0.001\, \pm 0.002$, and dominated by the dark sector, i.e. dark matter $\Omega_{\rm m_0}=0.315 \, \pm 0.007$ and dark energy $\Omega_{\rm \Lambda}=0.685\, \pm 0.0073$. The baryonic matter represents only few percent of all the energy content of the Universe, i.e. $\Omega_{\rm b_0}=0.022\, \pm 0.0001$, while the radiation component (intended as photons and massless neutrinos) have a negligible effect at present time $\Omega_{\rm r_0}\sim 10^{-5}$. The value of the Hubble constant\footnote{This value of $H_{\rm 0}$ is in $3.7\, \sigma$ tension with the local measurement of $H_{\rm 0}$ from SnIa, $H_{\rm 0}=73.48\, \pm 1.66\, {\rm km/s/Mpc}$ \cite{Riess:2018uxu}.}
 is $H_{\rm 0}=67.4\, \pm 0.5\, {\rm km/s/Mpc}$.
This parametrization defines the standard cosmological model, also dubbed as the \setwd{$\Lambda$CDM}{acr:LCDM} model.

The dynamics of the $\Lambda$CDM model is prescribed by the Friedmann equations Eqs.~\eqref{eq:Friedmann_edH}-\eqref{eq:Friedmann_edQ}, with $ \Omega_{\rm m_0} + \Omega_{\rm \Lambda} = 1$ (here $\Omega_{\rm m_0}$ considers both baryonic and dark matter), and completely encoded in the scale factor. 
Before proceeding to find the expression for the scale factor, let us make some considerations about the coordinates. The FLRW line element, Eq.~\eqref{eq:FLRW_metric}, for the $\Lambda$CDM model simplifies to 
\begin{equation}
ds^2= -c^2 dt^2 + a(t)^2 \delta_{i j} dq^idq^j \, ,
\end{equation}
where $q^i$ are the spatial coordinates in flat space. As for time, we prefer to use the \emph{conformal time} $\eta$ coordinate, which is related to the cosmic time $t$ as $d t = a(\eta) d \eta$. The advantage is that in conformal time the metric further reduces to
\begin{equation}
ds^2=a(\eta)^2\left(-c^2 d\eta^2+ \delta_{i j} dq^idq^j \right)\, ,
\label{eq:conf_ds}
\end{equation} 
simplifying also the calculations. Moreover, the conformal time has the clear physical meaning of particle horizon $c\, \eta$, i.e. the maximum distance ideally travelled by a photon since the beginning of the Universe, \cite{dodelson2003modern, ellis2012relativistic}.
In conformal time $d t = a(\eta) d \eta$, the Hubble parameter transforms as
\begin{equation}
H(t)=\dfrac{\mathcal{H}(\eta)}{a(\eta)}\, , \label{eq:H_to confH}
\end{equation}
where we have defined the conformal Hubble parameter $\mathcal{H}(\eta)=\frac{1}{a(\eta)}\frac{d a(\eta)}{d\eta}$.
The Friedmann equations~\eqref{eq:Friedmann_edH}-\eqref{eq:Friedmann_edQ} for $\Lambda$CDM in conformal time reads
\begin{align}
\mathcal{H}^2 & = \mathcal{H}_0^2 \left[ \dfrac{\Omega_{\rm m_0} }{a} + \Omega_{\rm \Lambda_0} a^2 \right] \label{eq:LCDM_friedmann_H}\\
\dot{\mathcal{H}} & = \mathcal{H}^2-\dfrac{3}{2} \dfrac{\mathcal{H}_0^2 \Omega_{\rm m_0}}{a}\, , \label{eq:LCDM_friedmann_q}
\end{align}
where dotted quantities indicates derivative with respect to conformal time, i.e. $\dot{\mathcal{H}}=\frac{d \mathcal{H}}{d \eta}$, and we have used the standard convention of setting the scale factor today to unit $a_0=1$. The expression for the density parameters in conformal time \cite{Villa:2015ppa} is 
\begin{align}
\Omega_{\rm m} & =\dfrac{8 \pi G a^2}{3 \mathcal{H}^2}\bar{\rho}_{\rm m}=\dfrac{\mathcal{H}^2_0 \Omega_{\rm m_0}}{a \mathcal{H}^2} \\
\Omega_{\rm \Lambda} & =\dfrac{a^2 c^2 \Lambda}{3 \mathcal{H}^2}=\dfrac{a^2 \mathcal{H}^2_0 \Omega_{\rm \Lambda_0}}{\mathcal{H}^2}\, .
\end{align}
The scale factor for the $\Lambda$CDM model can be explicitly found (using the results in \cite{gradshteyn2014table}) by solving Eq.~\eqref{eq:LCDM_friedmann_H}
\begin{equation}
a(\eta)=\frac{\sqrt[3]{\frac{\Omega_{\rm m_0}}{\Omega_{\rm \Lambda}}} \Big(1-{\rm cn}\left(\mathit{y} |\mathit{r} \right)\Big)}{(\sqrt{3}-1)+(\sqrt{3}+1) {\rm cn}\left(\mathit{y} |\mathit{r} \right)}\, ,\label{eq:a_LCDM}
\end{equation}
where ${\rm cn}(\mathit{y}|\mathit{r})$ is the Jacobi elliptic cosine function, with $\mathit{y}=\left(\sqrt[4]{3} \sqrt[6]{\Omega_{\rm \Lambda}} \sqrt[3]{\Omega_{\rm m_0}}\right) \mathcal{H}_0 \eta$, and $\mathit{r}=\sqrt{\frac{\sqrt{3}+2}{4}}$.

The $\Lambda$CDM model presented so far is based on the fundamental assumptions of homogeneity and isotropy of the Universe. However, as has been noted several times, these properties are only satisfied on average and on very large scales, with the transition from clustered structures to a homogeneous distribution beginning on scales\footnote{This scale is also known as \emph{End of Greatness}.} of $\sim 100\, {\rm Mpc}$, \cite{Yadav:2010cc, Scrimgeour:2012wt, Sam:2020}. 
So, while the $\Lambda$CDM model describes the overall dynamics of the Universe, we still need to account for the evolution of structures such as galaxies and galaxy clusters that are visible on smaller scales.

\subsection{Early fluctuations and cosmological perturbation theory}
The features of the CMB map have been thoroughly examined, revealing that in the first moments after the Big Bang, small temperature variations were generated by quantum fluctuations on microscopic scales generating the seeds for galaxies and clusters, \cite{COBE:1992syq, WMAP:2008lyn, planck2020CMB}. Thanks to the inflationary paradigm\footnote{\emph{Inflation} is an epoch of accelerated expansion in the early Universe, and it was introduced to explain the coherence of CMB anisotropies on angular scales larger than the apparent cosmological horizon at recombination.}, \cite{tsujikawa2003introductory}, the early evolution of perturbations is well described on cosmological scales by relativistic \emph{perturbation theory} (\setwd{PT}{acr:PT}), \cite{ kodama1984cosmological}. Indeed, the small primordial fluctuations can be conceived as tiny perturbations $\delta \rho$ over a homogeneous and isotropic density distribution $\bar{\rho}$, so that the real mass-energy density is expanded as $\rho = \bar{\rho}+\delta \rho$ up to linear order. Similarly, the real spacetime is modelled by a perturbed metric $g_{\mu \nu}=\bar{g}_{\mu \nu}+\delta g_{\mu \nu}$, where $\bar{g}_{\mu \nu}$ is the background metric and $\delta g_{\mu \nu}$ is the small linear perturbation. 
It is important to note that $\delta g_{\mu \nu}$ is not uniquely defined due to coordinate gauge freedom: the same physical perturbation can be described by a different tensor perturbation
\begin{equation}
\widetilde{\delta g_{\mu \nu}}=\delta g_{\mu \nu} + \mathcal{L}_{\xi} \bar{g}_{\mu \nu}\, , \label{eq:PT_gauge_trans}
\end{equation}
see e.g. \cite{wald2010general}.
The term $\mathcal{L}_{\xi} \bar{g}_{\mu \nu}\equiv \xi^{\sigma}\partial_{\sigma} \bar{g}_{\mu \nu} + \bar{g}_{\mu \nu} \partial_{\mu} \xi^{\sigma} + \bar{g}_{\mu \nu} \partial_{\nu} \xi^{\sigma}$ is the Lie derivative and it represents the action on $\bar{g}_{\mu \nu}$ of an ``infinitesimal diffeomorphism'' generated by the vector field $\xi^{\mu}$ (see e.g. App. C in \cite{wald2010general}). In other words, the form of the perturbed metric $g_{\mu \nu}$ depends\footnote{For a different approach see \cite{bardeen1980gauge, kodama1984cosmological}.} on the specific choice of the gauge, and the first-order transformation between two different gauge choices is given by Eq.~\eqref{eq:PT_gauge_trans} (see \cite{Villa:2015ppa} for gauge transformations of 3 different gauges up to second-order PT).

\begin{figure}[ht]
    \centering
    \includegraphics[width=0.7\linewidth]{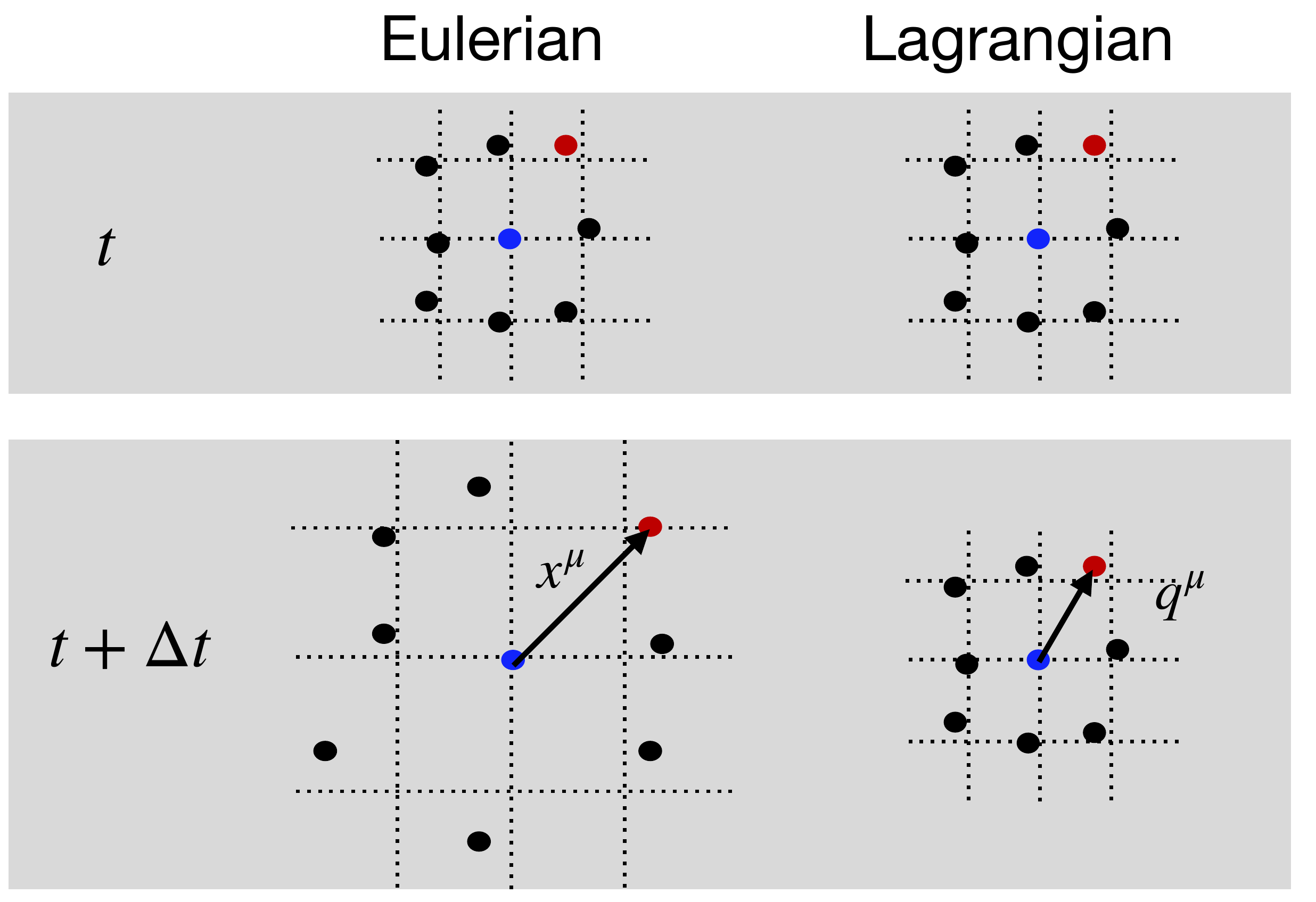}
    \caption{Illustration of the evolution of a group of particles in an expanding Universe in the Eulerian (left) and Lagrangian (right) frames. At the initial time $t$, the particles have a certain position with respect to the uniform grid. Due to the expansion and the gravitational interaction, at a later time $t+\Delta t$ the particles are in a different position with respect to the grid in the Eulerian frame (left), with the new position being marked by the vector $x^{\mu}$. In the Lagrangian frame (right), on the other hand, the positions of the particles $q^{\mu}$ do not change with time.}\label{fig:E_L_frames}
\end{figure}

Considering linear perturbations (\setwd{$\rm Lin$}{acr:Lin}) over the flat FLRW background Eq.~\eqref{eq:conf_ds}, the most general form of the spacetime metric is \cite{Villa:2015ppa}
\begin{align}
g_{0 0}&=-a^2(1+2 \Psi) \nonumber \\
g_{0 i}&=a^2(\partial_i B +\omega_i) \label{eq:metric_PT}\\
g_{i j}&=a^2\{(1-2 \Phi)\delta_{i j}+2 D_{i j} E +\partial_{(i}F_{j)}+\chi_{i j}\}\, , \nonumber 
\end{align}
where $\Psi, \, \Phi,\, B,\, E$ are the scalar modes, $\omega_i,\, F_i$ are the transverse vector modes ($\partial^i \omega_i=\partial^i F_i=0$), and $\chi_{i j}$ is the transverse and tracefree tensor mode ($\partial^i \chi_{i j}=\chi\UD{i}{i}=0$). The operator $D_{i j}=\partial_i \partial_j - 1/3 \delta_{i j} \nabla^2$ is the traceless symmetric double gradient operator, see e.g. \cite{Bertschinger:1993xt}.
The expressions in Eq.~\eqref{eq:metric_PT} define the \emph{scalar-vector-tensor decomposition} of the metric tensor \cite{kodama1984cosmological, bardeen1980gauge}. 
Usually, in the study of structure formation, the linear vector and tensor modes can be neglected, i.e. $\omega_i \sim F_i \sim \chi_{i j}\sim 0$ \cite{matarrese1998relativistic, Bartolo:2005kv}, simplifying the expression of Einstein equations. The reason is that linear vector modes are decaying in time and linear tensor modes, i.e. primordial gravitational waves, are decoupled from the other perturbation modes. Each gauge corresponds to a specific choice of $\Psi, \, \Phi,\, B,\, E$:
\begin{itemize}
	\item the \emph{Lagrangian frame} (or \emph{synchronous-comoving gauge}) is defined by choosing $B=\Psi=0$ and corresponds, in analogy with fluid dynamics, to the reference frame associated with the coordinates comoving with the cosmic flow. In this reference frame, the positions of the fluid particles do not evolve in time, see Fig.~\ref{fig:E_L_frames}.
	\item \emph{Eulerian gauges} are all gauges identified by the (spatial) choice $E=0$. They correspond to a frame associated with the observer measuring the matter stream, i.e. not comoving with the cosmic flow. The expression of $B$ fully specifies the gauge. For example, a common choice in PT is the \emph{Poisson gauge} identified by $B=0$,  \cite{Bertschinger:1993xt}. With this coordinate choice, the positions of the particle of the fluid evolve from their initial positions, see again Fig.~\ref{fig:E_L_frames}.  
\end{itemize}

The form of the first-order line element in the Poisson gauge considering scalar perturbations only is \cite{Villa:2015ppa}
\begin{equation}
ds^2=a(\eta)^2\left[-\left(1+2 \dfrac{\Psi(\eta, q^i)}{c^2}\right) c^2d\eta^2 + \left(1- 2\dfrac{\Phi(\eta, q^i)}{c^2}\right)\delta_{i j} dq^idq^j \right]\, ,
\end{equation}
where the perturbations $\Psi(\eta, q^i)$ and $\Phi(\eta, q^i)$ are obtained by solving\footnote{The equations are solved order by order, with the zeroth-order resembling the Friedmann equations Eqs.~\eqref{eq:LCDM_friedmann_H}-\eqref{eq:LCDM_friedmann_q}.} the Einstein equations, Eq.~\eqref{eq:einsteinEq}, expanded up to linear order, see \cite{Bertschinger:1993xt}. 
Let us start by noting that $\Psi=\Phi=\phi$, as it follows from the trace-free part of the $(i, j)$ components.
At first order, the $(0,0)$ component returns the Poisson equation
\begin{equation}
\nabla^2 \phi(\eta, q^i)-\dfrac{3}{2}\dfrac{\stuff}{a}\delta_{\rm Lin}(\eta, q^i)=0\, , \label{eq:Poisson_eq}
\end{equation}
with $\delta_{\rm Lin}(\eta, q^i)$ the \emph{linear Newtonian density contrast} defined as $\delta_{\rm Lin}=\dfrac{\delta \rho}{\bar{\rho}}+ \calO(\delta \rho^2)$, and $\nabla^2=\partial^i \partial_i$. 
From the other components of Einstein equations, we obtain that the scalar perturbation decomposes\footnote{Actually, the scalar perturbation is composed by growing and decaying modes like $\phi(\eta, q^i)=g_{+}(\eta) \phi_{+}(q^i)+g_{-}(\eta) \phi_{\rm -}(q^i)$. However the decaying modes are quickly suppressed leaving only with growing modes.} into the \emph{present time gravitational potential} $\phi_{\rm 0}(q^i)$ and a time-dependent part containing the \emph{growth factor} $\mathcal{D}(\eta)$ 
\begin{equation}
\phi(\eta, q^i)=\dfrac{\mathcal{D}(\eta)}{a(\eta)} \phi_{\rm 0}(q^i)\, .
\end{equation}
The growth factor $\mathcal{D}(\eta)$ is the growing mode solution for the linear density contrast, i.e. $\delta_{\rm Lin}(\eta, q^i)=\mathcal{D}(\eta) \delta_{\rm Lin}(\eta_0, q^i)$, and it is found by solving the evolution equation for the first-order density contrast \cite{peebles1993principles}
\begin{equation} \label{eqforF}
\ddot{\delta}_{\rm Lin} +\mathcal{H}\dot{\delta}_{\rm Lin}-\frac{3}{2}\mathcal{H}_0^2\Omega_{m_0}\frac{\delta_{\rm Lin}}{a}=0\,.
\end{equation}
The analytical solution for $\mathcal{D}$ is given in \cite{Villa:2015ppa},
\begin{equation}
\mathcal{D} (\eta) = \frac{a}{\frac{5}{2}\Omega_{\rm{m 0}} } \sqrt{1+\frac{\Omega_{\rm{\Lambda}}}{\Omega_{\rm{m 0}}}a^3}\,  {}_2 F_{1} \left( \frac{3}{2}, \frac{5}{6}, \frac{11}{6}, -\frac{\Omega_{\rm{\Lambda 0}}}{\Omega_{\rm{m 0}}}a^3 \right)\,,
\label{eq:grow_mode}
\end{equation}
with ${}_2 F_{1} \left(a,b,c, y\right)$ being the Gaussian (or ordinary) hypergeometric function.
In conclusion at early times 
the small inhomogeneities are described by the density contrast 
\begin{equation}
\delta_{\rm Lin} = \dfrac{2}{3 \stuff}\left( \mathcal{D} \nabla^2 \phi_0 - 3 \mathcal{H}\dot{\mathcal{D}}\phi_0 \right)\, .
\end{equation}

The evolution of inhomogeneities can be modelled by linear PT only if $\delta \ll 1$. Initially, this is the case, but at later times the density fluctuations become larger under the influence of gravity, reaching values of $\delta \sim 10^2$ for filaments and $\delta \sim 10^6$ for galaxies. 
The evolution of the gravitational instability that led from the early linear perturbations to the present-day inhomogeneities is the primary goal of the study of \emph{structure formation}.

\subsection{Analytical approaches to structure formation}

The equations of GR control gravitational instability, but some applications are well described by the \emph{Newtonian approximation}, namely by a weak-field and slow-motion limit of GR. These requirements are indeed satisfied on small scales, where the dimensionless peculiar gravitational potential $\phi_g/c^2$ remains small ( $\phi_g /c^2 \sim 10^{-5}$), and the peculiar velocity is never relativistic. In particular, for a fluctuation of proper scale $L$, the dimensionless peculiar gravitational potential is
\begin{equation}
\dfrac{\phi_g}{c^2} \sim \delta \left(\dfrac{L}{r_{\rm H}}\right)^2\, 
\end{equation}
with $r_{\rm H}= c H^{-1}$ the Hubble radius, implying that $\phi_g/c^2$ remains small even if $\delta \gg 1$. Usually, this legitimises the use of cosmological simulations based on Newtonian dynamics to describe the nonlinear structure growth on small scales. Formally, the Newtonian approach is obtained by perturbing only the time-time component of the FLRW metric Eq.~\eqref{eq:conf_ds} by $2\phi_g /c^2$
\begin{equation}
ds^2=a^2 \left[- \left( 1+2 \dfrac{\phi_g}{c^2} \right) c^2 d \eta^2 + \delta_{i j}d x^i dx^j \right]\, . \label{eq:generic_new}
\end{equation}
The Einstein equations give the Poisson equation Eq.~\eqref{eq:Poisson_eq} again, while the dynamics is described by the stress-energy conservation Eq.~\eqref{eq:T_conservation} with continuity and Euler equations. 
Recently, cosmologists have begun to investigate cosmic dynamics beyond the Newtonian approximation and to search for measurable relativistic effects on cosmic scales. 

Estimating the importance of relativistic corrections in structure formation is of paramount importance in cosmology (see e.g. \cite{bruni2014computing, bertacca2015galaxy, thomas2015fully, Barrera-Hinojosa:2020gnx} and refs. therein), and several approximation techniques have been developed to account for nonlinear GR effects in structure formation: 
to this list belongs the \emph{post-Newtonian approximation} (\setwd{PN}{acr:PN}), which we will encounter in Chapter~\ref{chap:nonlinearities}.
Formally, it is obtained by expanding the equations of GR in inverse powers of the speed of light, where the zero-order is the Newtonian limit. For the application of the PN approach to cosmological perturbations, see \cite{tomita1988post, shibata1995post, carbone2005unified} and \cite{mater} for the formulations of PN cosmology in two different gauges. In the following, we give a brief overview of the PN approximation as presented in \cite{mater}.


The authors use the Lagrangian approach by relating the evolved (Eulerian) position $x^{\mu}$ and the initial (Lagrangian) position $q^{\mu}$ of the fluid particles with the transformation
\begin{align}
x^{\mu}(\eta, q^j)&= q^{\mu} + \calS^{\mu}(\eta, q^j)\, .\label{eq:Euler_to_Lagran}
\end{align}
The vector $\calS^{\mu}(\eta, q^i)$ is the \emph{displacement vector} and represents the difference in matter flow induced by the inhomogeneities. In a homogeneous Universe, the comoving Eulerian coordinate $x^{\mu}$ matches the Lagrangian coordinate $q^{\mu}$. The presence of inhomogeneities locally alters the expansion as the perturbations grow with time. This is encoded by the displacement vector $\calS^{\mu}(\eta, q^i)$, which is the fundamental field describing the evolution of the inhomogeneities \cite{Catelan:1994kt, Catelan:1994ze}. Equivalently, the relation in Eq.~\eqref{eq:Euler_to_Lagran} can be expressed by the Jacobian of the transformation \cite{mater}
\begin{equation}
\mathcal{J}\UD{\mu}{\nu}=\dfrac{\partial x^{\mu}}{\partial q^{\nu}}=\delta\UD{\mu}{\nu}+\calS\UD{\mu}{\nu}\, ,\label{eq:Jacobian_EtoL}
\end{equation}
 where $\calS\UD{\mu}{\nu}=\dfrac{\partial \calS^{\mu}}{\partial q^{\nu}}$ is the deformation tensor. The expression of $\calS\UD{\mu}{\nu}$ is found perturbatively by searching for solutions of the trajectories $x^{\mu}$: this is the key point of the PN approach presented in \cite{mater}. Instead of perturbing over the density fluctuations and the velocity fields, as in the Eulerian approach, the perturbation is performed only in the trajectories. 

The post-Newtonian expression of the Jacobian $\mathcal{J}\UD{\mu}{\nu}$ has the form
\begin{equation}
\mathcal{J}\UD{\mu}{\nu}=\begin{pmatrix}
1+ \dfrac{1}{c} \dfrac{\partial \calS^0}{\partial \eta} && \dfrac{1}{c} \dfrac{\partial \calS^0}{\partial q^{j}} \\
 && \\
v^{i} && \mathcal{J}\UD{i}{j}
\end{pmatrix}\, ,\label{eq:jacobian_PN}
\end{equation}
with $v^i= \delta^{i j}\dfrac{\partial \calS^0}{\partial q^{j}}$ the peculiar velocity, and the spatial deformation $\mathcal{J}\UD{i}{j}=\dfrac{\partial \calS^{i}}{\partial q^{j}}$ being the Newtonian limit of the Jacobian Eq.~\eqref{eq:jacobian_PN}.
In synchronous-comoving gauge the post-Newtonian metric assumes the form
\begin{align}
ds^2=a^2 \left\{- c^2 d \eta^2 + \right.& \left. \gamma_{l k}d q^{l} dq^{k} \right\}=  \nonumber\\
a^2 \left\{- c^2 d \eta^2 + \right.& \left. \left[\left( 1+\dfrac{\chi}{c^2}\right)\delta_{i j}\mathcal{J}\UD{i}{l}\mathcal{J}\UD{j}{k} + \dfrac{1}{c^2} \pi_{l k}\right]d q^{l} dq^{k} \right\}\, . \label{eq:generic_Lagrangian}
\end{align}
The PN scalar and tensor modes $\chi$ and $\pi_{i j}$ are sourced by combinations of the peculiar gravitational field $\phi_{g}$ and the peculiar velocity gradient tensor $\theta\UD{i}{j}=1/2 \gamma^{i l}\dfrac{\partial \gamma_{l j}}{\partial \eta}$. In particular, they are found from 
\begin{align}
\chi = & 2 \mathcal{H}\mathcal{S}^{0}-2 \phi_{g}- \Upsilon \\
D^2 \pi_{i j } = & D_i D_j \Upsilon + \delta_{i j} D^2\Upsilon+2 (\theta\UD{k}{k}\theta_{i j}-\theta_{i k}\theta\UD{k}{j})\, ,
\end{align}
where $D_i$ is the covariant spatial derivative of $\gamma_{i j}$ in the Newtonian limit, while $\Upsilon$ and $\mathcal{S}^0$ are the solutions of
\begin{align}
D^2 \Upsilon = & -\dfrac{1}{2} \left[(\theta\UD{k}{k})^2-\theta\UD{i}{j}\theta\UD{j}{i}\right] \\
D^2 \mathcal{S}^0 = & \theta\UD{k}{k}\, .
\end{align}
Once that $\gamma_{i j}$ is known, one obtains the density contrast $\delta(\eta, q^i)$ from the exact expression of the continuity equation in Lagrangian frame
\begin{equation}
\delta(\eta, q^i)=\left(1+\delta_0(q^i)\right)\sqrt{\dfrac{\gamma_0(q^i)}{\gamma(\eta, q^i)}}-1\, , \label{eq:delta_in_pn}
\end{equation}
where $\gamma={\rm det}(\gamma_{i j})$, and $\delta_0(q^i)=\delta(\eta_0, q^i)$ and $\gamma_0(q^i)=\delta(\eta_0, q^i)$ are respectively the density contrast and the determinant of the spatial metric at present time.
Note that the expression of the density fluctuations in Lagrangian framework Eq.~\eqref{eq:delta_in_pn} is not expanded in $\gamma$. In other words, it is capable of mimicking the  nonlinear behaviour of structure formation\footnote{However, this perturbation technique is limited by the formation of caustic singularities.}.

The PN approximation is one of the many proposed methods to describe nonlinear GR effects in structure formation. 
Other perturbative approaches are: the \emph{post-Friedmann approximation} (see \cite{Milillo:2015cva, Rampf:2016wom} for a different approach, which adapts to cosmology the weak-field post-Minkowskian approximation and reproduces linear-order cosmological perturbation theory at their zeroth-order), the \emph{weak-field approximation}\footnote{The leading order of the last two approximation schemes were shown to be equivalent for a dust Universe in the Poisson gauge in \cite{kopp2014newton}, whereas \cite{mater, carbone2005unified} were constructed on purpose to include second-order perturbation theory at their PN order.} (see \cite{green2011new} for the development of the framework and \cite{Adamek:2013wja} for estimations with the use of N-body simulations for a plane-symmetric Universe), and, more recently, a \emph{two-parameters gauge-invariant approximation} (see \cite{Goldberg:2016lcq}).  

\subsection{Inhomogeneous models}
Another analytical approach to structure formation is to search for exact (i.e. non-perturbative) inhomogeneous solutions to the Einstein equations. These inhomogeneous solutions 
are not assumed to have the symmetries of the FLRW models. However, interesting classes of such solutions typically contain the FLRW models as a limit (in \cite{bolejko2011inhomogeneous} this was used to define suitable classes for representing inhomogeneous cosmological models).
Two noticeable examples are the \emph{Lema\^{i}tre-Tolman-Bondi} (\setwd{LTB}{acr:LTB}) and the \emph{Szekeres} models (see \cite{lemaitre1933univers, tolman1934effect, bondi1947spherically} and \cite{Szekeres:1974ct} for the original articles), which have been extensively studied in cosmology. The LTB is a spherically symmetric solution containing only dust that is inhomogeneously distributed along the radial direction, i.e. the matter is condensed into concentric shells (overdensities) separated by underdense regions. In this model, an observer at the centre of a local underdense region measures a local accelerated expansion caused by the large-scale inhomogeneities. This feature of the LTB models has been investigated as an alternative explanation for the SnIa observations without the need for dark energy, \cite{celerier1999we, alexander2009local}.
Although the LTB solution has been shown to be a valuable toy model\footnote{An interesting review of misleading concepts on LTB models can be found in~\cite{krasinski2012drift}, Sec. 4.} to test possible probes for inhomogeneities and anisotropies at late times, \cite{Quercellini:2010zr}, they cannot be considered as a realistic model of the Universe for its intrinsic symmetries.

A further improvement is represented by the Szekeres models, a class of exact solutions of the Einstein equations that includes both LTB and FLRW solutions as limits.
In his original paper \cite{Szekeres:1974ct}, Szekeres finds all solutions of the form
\begin{equation}
ds_{\rm Sz}^2=-c^2 d t^2 + e^{2 \alpha(t,q_{\rm 1},q_{\rm 2},q_{\rm 3})} {dq^{\rm 1}}^2 + e^{2 \beta(t,q_{\rm 1},q_{\rm 2},q_{\rm 3})} ({dq^{\rm 2}}^2+ {dq^{\rm 3}}^2)
\label{eq:Old_Szekeres_metric}
\end{equation}
Two distinct classes of spacetime metrics can be distinguished: class \textsc{I}, which are a generalization of the  Lema\^{i}tre-Bondi-Tolman model, and class \textsc{II}, which are a generalization of the Kantowski-Sachs and FLRW models. These original solutions were obtained for a pressureless matter (dust) and later extended by Barrow and Stein-Schabes in \cite{Barrow:1984zz} to include a cosmological constant $\Lambda$. 
M. Bruni and N. Meures presented a more recent formulation of the class-\textsc{II} solutions in \cite{Meures:2011ke} that will be later used in Chapter~\ref{chap:bigonlight}. This formulation distinguishes the contribution of inhomogeneities from the FLRW background and allows us to express the spacetime metric in a form more convenient for cosmological applications: the expression of the line element Eq.~\eqref{eq:Old_Szekeres_metric} for this Szekeres model\footnote{We choose here to use our notation instead that of \cite{Meures:2011ke}. The line element (\ref{eq:Szekeres_metric}) is different from the one presented in \cite{Meures:2011ke} because we use conformal time. Of course, this does not affect the results, since it can be easily shown that the two metrics are equivalent under a coordinate transformation.} is rewritten as
\begin{equation}
ds_{\rm Sz}^2=a^2 \left[ -c^2  d \eta^2 + {dq^{\rm 1}}^2 + {dq^{\rm 2}}^2+ Z^2(\eta,q^{\rm 1},q^{\rm 2},q^{\rm 3}) {dq^{\rm 3}}^2 \right]\, .\label{eq:Szekeres_metric}
\end{equation}
As it is shown in \cite{Meures:2011ke}, thanks to the symmetry of the problem, the function $Z(\eta,q^{\rm 1},q^{\rm 2},q^{\rm 3})$ is decomposed as 
\begin{equation}\label{eq:sz_Z}
Z(\eta,q^{\rm 1},q^{\rm 2},q^{\rm 3})=F(\eta,q^{\rm 3})+A(q^{\rm 1},q^{\rm 2},q^{\rm 3})\,,
\end{equation}
where the function $F(\eta, q^{\rm 3})$ satisfies Newton's evolution equation for the first-order density contrast\footnote{This was implicitly shown in Sec.~$5$ of the Szekeres'original paper \cite{Szekeres:1974ct} and later by many other authors such as those of \cite{bonnor:1977pp}. However, it was Goode and Wainwright who explicitly recognized that the relativistic equations for the density fluctuations in Szekeres model are the same as in Newtonian gravity, \cite{Goode:1982pg}. They also provide a new formulation of Szekeres solutions which is much more useful in cosmology and in which the relation to the FLRW solution is clarified.}, Eq.~\eqref{eqforF}. Neglecting the decaying modes, it is therefore possible to factorize $F(\eta, q^{\rm 3})$ without loss of generality as\footnote{The time-dependent-only growing mode is denoted by $f_+$ in \cite{Meures:2011ke} and is given in Eq.~(11b) in a dimensionless time variable $\tau$. To reconcile $\cal D$ in \eqref{eq:grow_mode} and $f_+$, one must: first transform $f_+(\tau)$ into a conformal time $f_{+}(\eta)$ and then normalise so that $f_{+}(\eta_{\rm 0})=1$. The final result is $F(\eta, q^{\rm 3})$ as in \eqref{eq:F_Sz}.}
\begin{equation}
\label{eq:F_Sz}
F(\eta, q^{\rm 3})= {\cal D}(\eta) \beta_+(q^{\rm 3})\, ,
\end{equation}
where we remind that $\mathcal{D}$ is the growing mode solution for the density contrast, Eq.~\eqref{eq:grow_mode}. It follows that  $F(\eta, q^{\rm 3})$ coincides with the linear density contrast and more precisely we have $\delta_{\rm Lin}(\eta, q^{\rm 3})=- {\cal D}(\eta) \beta_+(q^{\rm 3})$ \footnote{The minus sign between $F$ and $\delta$ follows from the fact that in eq. ($A8$) of \cite{Meures:2011ke} the authors set, in full generality,  $\delta_{in}=-\frac{F_{in}}{F_{in}+A}$.}.

The purely spatial function $A(q^{\rm 1},q^{\rm 2},q^{\rm 3})$ sets the spatial distribution of the density contrast 
\begin{equation}
\delta_{\rm Sz}=-\dfrac{F}{F+A}\, . \label{eq:Sz_density}
\end{equation}
From Einstein equations follows that $A$ is decomposed as, \cite{Meures:2011ke}
\begin{equation}
A(q^{\rm 1}, q^{\rm 2}, q^{\rm 3})=1+\beta_{\rm +}(q^{\rm 3}) B \left\{\left[q^{\rm 1}+\omega(q^{\rm 3})\right]^2+\left[q^{\rm 2}+\gamma(q^{\rm 3})\right]^2\right\}\, .
\label{eq:Sz_A}
\end{equation}
with $\omega$ and $\gamma$ being two real functions which reduces to $\omega=\gamma=0$ for the special case of axial symmetry around $q^{\rm 3}$.
The term $B$ in Eq.~\eqref{eq:Sz_A} is a constant and is given by (see App. C in \cite{Grasso:2021zra})
\begin{equation}
B= \frac{5}{4} \stuff \frac{\cal D_{\rm in}}{a_{\rm in}}\, ,
\label{eq:link_on_B}
\end{equation}
where $\mathcal{D}_{\rm in}= a_{\rm in}$ for initial conditions set deeply in the matter-dominated era. As noted before, the function $\beta_{\rm +}$ is the part of $A$ which specifies the spatial distribution of the first-order density contrast, and it can be related to the peculiar gravitational potential $\phi_{\rm 0}$ via the cosmological Poisson equation Eq.~\eqref{eq:Poisson_eq}.

Other examples of exact inhomogeneous cosmological models are \emph{black-hole lattices} (see \cite{linquist1957dynamics, clifton2009archipelagian, clifton2015applications, bentivegna2018black} and \cite{bentivegna2012evolution, bentivegna2013evolution} for numerical investigations), \emph{plane symmetric models} or \emph{wall Universe} (see \cite{diDio2012back, adamek2014distance, Villa:2011vt} for an exact, numerical, and PN analysis of the wall Universe), and \emph{Swiss Cheese} models (\cite{einstein1945influence}).

\subsection{Numerical simulations}
Together with the latter two approaches, numerical simulations in cosmology have become a valuable tool for modelling nonlinear regimes in structure formation. 
The first generation of numerical codes in cosmology used the Newtonian approximation of GR to simulate systems of N self-gravitating identical objects. A pioneering application of these \emph{N-body simulations} was proposed by J. Peebles in 1970, \cite{peebles1970structure}, to model the formation of the Coma cluster.
The steep growth in computational power led to the development of more detailed N-body codes capable of simulating up to $10^{12}$ particles and producing a realistic structure of the cosmic web, \cite{vogelsberger2014introducing, nelson2021illustristng}.
A number of attempts have been made to incorporate relativistic corrections into N-body simulations using some of the perturbation methods listed above, see e.g.  \cite{bonvin2011galaxy, green2012newtonian, bertacca2014observed}, as well as N-body simulations have been used as an input for approximated GR equations, as in \cite{Bruni:2013mua, Adamek:2014xba, Fidler:2017pnb}.
 
Another approach to cosmological simulations is represented by \emph{full-GR numerical codes}, which use numerical methods to directly solve Einstein's equations (see \cite{loffler2012einstein, Bentivegna:2016stg, Mertens:2015ttp, Adamek:2016zes, macpherson2017,east2018comparing, Daverio:2019gql, barrera2020gramses} for the codes used in cosmology, and \cite{Adamek:2020jmr} for the comparison between them).
Numerical solutions of the Einstein equations were used early on to study the dynamics of strong-field gravitational systems, such as the study of the two-body problem for the two ends of a wormhole (Hahn and Lindquist in 1964, \cite{hahn1964two}) and the generation of pure gravitational waves (Eppley in 1977, \cite{eppley1977evolution}), where the highly nonlinear relativistic phenomena dominate. 
Since these early applications, continuous improvements in numerical techniques have shaped \emph{numerical relativity} and enabled many successes in the description of compact astrophysical objects \cite{Pretorius:2005gq, baker2006gravitational, campanelli2006accurate, buonanno2007inspiral, hinder2018eccentric, baiotti2008accurate, chaurasia2018gravitational} and in cosmological dynamics, \cite{Giblin:2015vwq, Bentivegna:2015flc, adamek2016general, macpherson2017, Macpherson:2018akp, Barrera-Hinojosa:2020gnx}.

To simulate full-GR dynamics in numerical relativity, the Einstein equations Eq.~\eqref{eq:einsteinEq} must be reformulated as an initial value problem, clearly separating the temporal and spatial dependence in the equations. This procedure is known as the \emph{$3+1$ splitting of spacetime} (or \emph{\setwd{ADM}{acr:ADM} formalism}) \cite{arnowitt1959, Smarr:1977uf}, and forms the common theoretical framework for most of the numerical codes mentioned so far. 
In what follows, we will discuss some of the key concepts of the $3+1$ approach to GR that we will encounter later in this thesis. For comprehensive references on the 3+1 formalism, see \cite{Alcubierre2008, baumgarte2010numerical, gourgoulhon20123+}.

\begin{figure}[ht]
    \centering
    \includegraphics[width=0.9\linewidth]{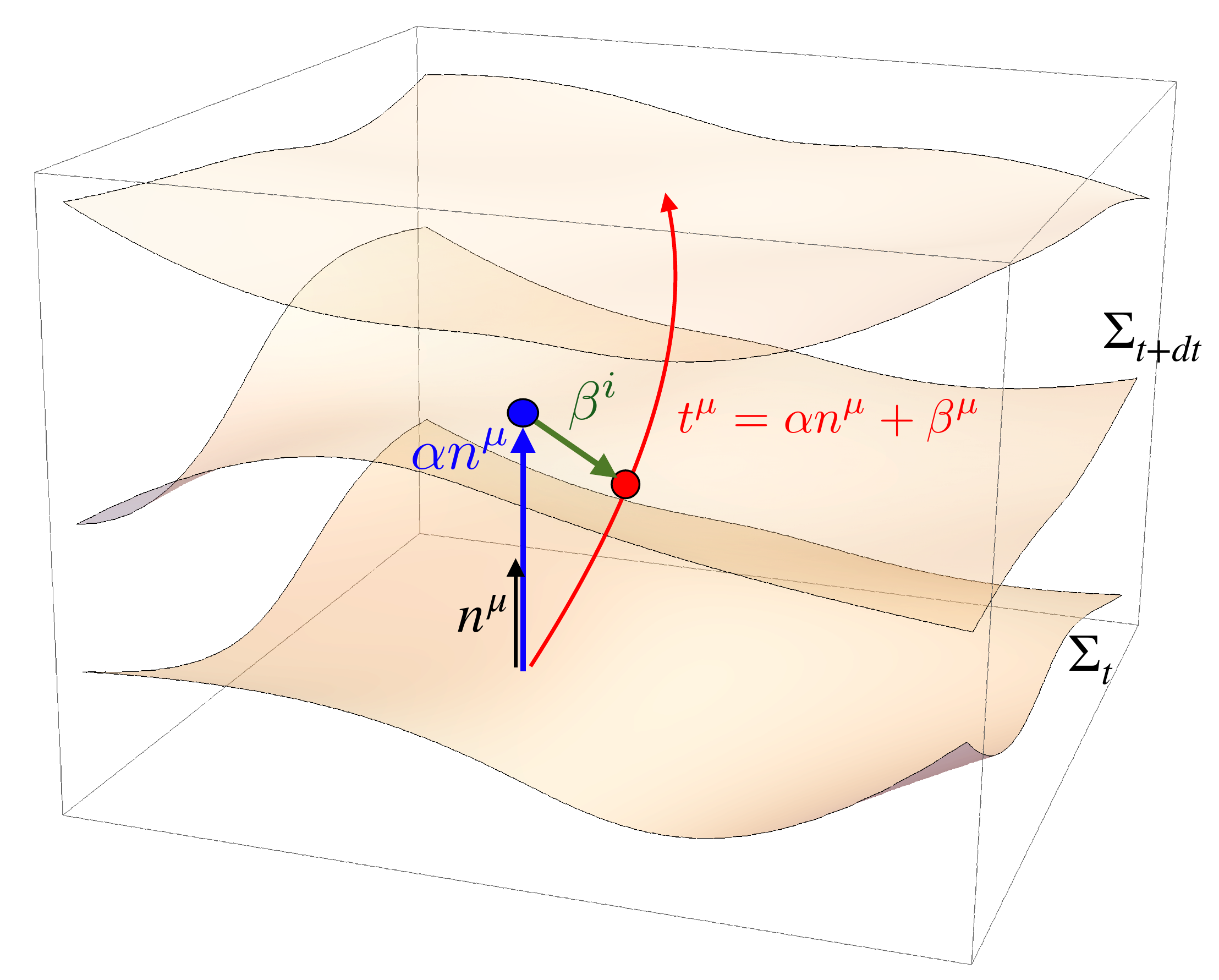}
    \caption{The constant-time hypersurfaces $\Sigma_{t}$ foliating the four-dimensional spacetime $\mathcal{M}, g_{\mu \nu}$, introduce a unit normal vector $n^{\mu}$ orthogonal to $\Sigma_{t}$. The coordinate flow, i.e. the lines of constant spatial coordinates, is given in terms of the lapse $\alpha$ and the shift $\beta^i$ gauge functions.}\label{fig:slice}
\end{figure}

Let us start by considering a manifold $(\mathcal{M}, g_{\mu \nu})$ which is globally foliated by a family of three-dimensional space-like hypersurfaces $\Sigma_t$. We also assume that the foliation is labelled by a monotonic function $t$ such that $t=\const$ on each slice, see Fig.~\ref{fig:slice}. In other words, each slice $\Sigma_t$ is identified as the level (hyper)surface of $t=\const$ and is characterised by the timelike  vector $\nabla^{\mu} t=g^{\mu \nu}\nabla_{\nu} t$ orthogonal to the hypersurface and such that $(\nabla^{\mu}t) (\nabla_{\mu} t)= - \alpha^{-2}$.
 In this view, the function $t$ can be interpreted\footnote{Note that $t$ will not necessary coincide with the proper time of any particular observer.} as a ``global time'' (synchronising all points on $\Sigma_t$), whose flow is represented by $\nabla^{\mu} t$. 
The unit \emph{normal vector} to $\Sigma_t$ $n^{\mu}$ is given as $n^{\mu}=-\alpha \nabla^{\mu}t$ and represents the normalised time-like vector colinear to the time flow $\nabla^{\mu}t$. 
The \emph{lapse function} $\alpha$ gives the flow rate of the proper time  $\tau$ of an ``Eulerian observer'', i.e. an observer moving along the normal vector $n^{\mu}=\frac{d x^{\mu}}{d \tau}$, with respect to the global time $t$ \cite{Alcubierre2008}
\begin{equation}
d\tau=\alpha d t\, . \label{eq:lapse}
\end{equation}
This can be shown by computing the variation of the global time flow $\nabla^{\mu}t$ along $n^{\mu}$
\begin{equation}
n^{\mu}\nabla_{\mu}t=(-\alpha \nabla^{\mu}t)\nabla_{\mu}t=\dfrac{1}{\alpha}\, .
\end{equation}

The adjective ``Eulerian'' refers to the fact that it is the observer who measures how the points on the foliation evolve in time. 
In general, any other observer is called a ``coordinate'' observer and identified by the timelike vector field
\begin{equation}
t^{\mu}=\alpha n^{\mu}+\beta^{\mu}\, ,
\end{equation}
where $\beta^{\mu}$ is the \emph{shift vector} quantifying the displacement on $\Sigma_t$ of the coordinate observer $t^{\mu}$ with respect to the position of the Eulerian observer. The choice of $\alpha=1$ and $\beta^i=0$ for $t^{\mu}$ is equivalent to choosing a Lagrangian observer.
The unit normal vector $n^{\mu}$, together with the spacetime metric $g_{\mu \nu}$ define the induced \emph{spatial metric} on each slice \cite{gourgoulhon20123+}
\begin{equation}
\gamma\UD{\mu}{\nu}=g\UD{\mu}{ \nu} + n^{\mu}n_{\nu}\, , \label{eq:spatial_metric}
\end{equation}
which is linked to the orthogonal projector tensor on $\Sigma_t$ as $\gamma\UD{\mu}{\nu}=g^{\mu \rho}\gamma_{\rho \nu}$.
In the adapted coordinate system $t^\mu$, the components of the normal vector and the metric $g_{\mu \nu}$ are written in terms of $(\alpha, \beta^i, \gamma_{i, j})$ as 
\begin{equation}
\begin{matrix}
n^{\mu}=(\dfrac{1}{\alpha},-\dfrac{\beta^i}{\alpha})\, , && n_{\mu}=(-\alpha,0)
\end{matrix}
\label{eq:nu_nd}
\end{equation}
and
\begin{equation}
\begin{matrix}
g_{\mu \nu}=\begin{pmatrix}
\beta_{i}\beta^i-\alpha^2 &~ \beta_i\\
\beta_j &~ \gamma_{i j}
\end{pmatrix}\, , && g^{\mu \nu}=\begin{pmatrix}
-\alpha^{-2} &~ \alpha^{-2}\beta^i\\
\alpha^{-2} \beta^j &~ \gamma^{i j}-\alpha^{-2} \beta^i \beta^j
\end{pmatrix}\, ,
\end{matrix}
\label{eq:g_3+1}
\end{equation}
where the Latin indices runs from $1$ to $3$, and $\beta_{i}=\gamma_{i j}\beta^j$, \cite{Alcubierre2008}. 

The projection of a generic (r, s) tensor ${}^{(4)} T^{\mu_{\rm 1} \cdots \mu_{\rm r}}_{ \ \ \nu_{\rm 1} \cdots \nu_{\rm s}}$ on the slice is obtained by using the projector $\gamma\UD{\mu}{\nu}$
\begin{equation}
{}^{(3)}T^{\mu_{\rm 1} \cdots \mu_{\rm r}}_{ \ \ \nu_{\rm 1} \cdots \nu_{\rm s}}={}^{(4)} T^{\rho_{\rm 1} \cdots \rho_{\rm r}}_{ \ \ \sigma_{\rm 1} \cdots \sigma_{\rm s}} \gamma\UD{\mu_{\rm 1}}{\rho_{\rm 1}}\cdots\gamma\UD{\mu_{\rm r}}{\rho_{\rm r}} \gamma\UD{\sigma_{\rm 1}}{\nu_{\rm 1}}\cdots\gamma\UD{\sigma_{\rm s}}{\nu_{\rm s}}\, ,
\label{eq:3Dproj}
\end{equation}  
where we have denoted as ${}^{(3)}T^{\mu_{\rm 1} \cdots \mu_{\rm r}}_{ \ \ \nu_{\rm 1} \cdots \nu_{\rm s}}$ the projected tensor, \cite{gourgoulhon20123+}. 
The very same projector is also used to define the operation of \emph{covariant derivative on the slice} $D_{\mu}=\gamma\UD{\sigma}{\mu}\nabla_{\sigma}$: for the $(r, s)$ tensor ${}^{(4)} T^{\mu_{\rm 1} \cdots \mu_{\rm r}}_{ \ \ \nu_{\rm 1} \cdots \nu_{\rm s}}$ (see e.g. \cite{gourgoulhon20123+})
\begin{equation}
D_{\lambda} T^{\mu_{\rm 1} \cdots \mu_{\rm r}}_{ \ \ \nu_{\rm 1} \cdots \nu_{\rm s}}= \gamma\UD{\mu_{\rm 1}}{\rho_{\rm 1}}\cdots\gamma\UD{\mu_{\rm r}}{\rho_{\rm r}} \gamma\UD{\sigma_{\rm 1}}{\nu_{\rm 1}}\cdots\gamma\UD{\sigma_{\rm s}}{\nu_{\rm s}}\, \gamma\UD{\epsilon}{\lambda}\nabla_{\epsilon} T^{\rho_{\rm 1} \cdots \rho_{\rm r}}_{ \ \ \sigma_{\rm 1} \cdots \sigma_{\rm s}} \, ,
\label{eq:3D_covD}
\end{equation}
which is written in terms of the 3D Christoffel symbol ${}^{(3)}\Gamma^{k}_{\ i j}=\frac{1}{2}\gamma^{k l}(\frac{\partial \gamma_{l j}}{\partial x^i}+\frac{\partial \gamma_{i l}}{\partial x^j}-\frac{\partial \gamma_{i j}}{\partial x^l})$.

With the introduction of the spacetime foliation, one must distinguish between the \emph{intrinsic curvature} of the hypersurface and an \emph{extrinsic curvature} in order to fully characterise the curvature of spacetime. The intrinsic curvature is the curvature of the hypersurface and is defined by the three-dimensional Riemann tensor, i.e. the Riemann tensor with respect to the spatial metric $\gamma_{i j}$
\begin{equation}
{}^{(3)}R\UD{k}{i s j}=\partial_{s} {}^{(3)}\Gamma^{k}_{i j} -\partial_{j} {}^{(3)}\Gamma^{k}_{i s}+ {}^{(3)}\Gamma^{k}_{l s} {}^{(3)}\Gamma^{l}_{i j}- {}^{(3)}\Gamma^{k}_{l j} {}^{(3)}\Gamma^{l}_{i s}\, .
\end{equation} 
On the other hand, the \emph{extrinsic curvature} $K_{\mu \nu}$ represents the curvature of the hypersurfaces with respect to the embedding higher-dimensional spacetime $\mathcal{M}$. It can be determined as the covariant variation of the normal vector $\nabla_{\mu} n_{\mu}$ along the slice $\Sigma_{t}$, namely
\begin{equation}
K_{\mu \nu}=-\gamma\UD{\sigma}{\mu}\gamma\UD{\rho}{\nu}\nabla_{\sigma}n_{\rho}\, .
\label{eq:K_def}
\end{equation}
In conclusion, after the introduction of the foliation, the geometry of spacetime $(\calM, \, g_{\mu \nu})$ is completely defined by the four quantities $(\alpha, \, \beta_{i}, \, \gamma_{i j}, \, K_{i j})$.

The projection of the Einstein equations Eq.~\eqref{eq:einsteinEq} decomposes the system into 3 relations \cite{Alcubierre2008, baumgarte2010numerical}:
\begin{itemize}
\item the \emph{Hamiltonian constraint}: \begin{equation}
{}^{(3)}R+(K\UD{i}{i})^2-K_{i j}K^{i j}- \frac{16 \pi G}{c^2} \rho=0\, ,
\end{equation}
\item the \emph{momentum constraint}: \begin{equation}
D_{j}K\UD{j}{i}- D_{i}K\UD{j}{j}- \frac{8 \pi G}{c^3} S_{i}=0\, ,
\end{equation}
\item the \emph{evolution equation}: \begin{align}
\left(\dfrac{\partial}{\partial t}- \mathcal{L}_{\beta}\right)K_{i j}=\alpha & \left[{}^{(3)}R_{i j}-2 K_{i l}K\UD{l}{j}+K_{i j}K\UD{l}{l} \right]- D_i D_j\alpha\\
-&\dfrac{8 \pi G}{c^4}\alpha \left[S_{i j} -\dfrac{1}{2}\gamma_{i j}(S\UD{i}{i} - \rho c^2) \right]\, ,
\end{align}
\end{itemize}
where we have defined the various components of the stress-energy tensor as $\rho c^2\equiv T_{\mu \nu}n^{\mu}n^{\nu}$, $S_{\mu}\equiv -\gamma\UD{\sigma}{\mu} n^{\rho}T_{\sigma \rho}$, and $S_{\mu \nu}\equiv \gamma\UD{\sigma}{\mu}\gamma\UD{\rho}{\nu} T_{\sigma \rho}$.
An additional evolution equation
\begin{equation}
\left(\dfrac{\partial}{\partial t}- \mathcal{L}_{\beta}\right)\gamma_{i j}=- 2 \alpha K_{i j}\, ,
\end{equation}
 is obtained from Eq.~\eqref{eq:K_def}.
The first two are constraints arising from the conservation of energy and momentum, while the last two indicate the evolution of the metric. As mentioned earlier, the covariant formulation of GR leaves us free to choose the gauge in which we perform the computation. The gauge choice in the ADM formalism is represented by a specific choice of the lapse $\alpha$ and the shift $\beta^i$. 
Despite the success of the ADM formalism in transforming the Einstein equations into an initial value problem, the equations obtained are weakly hyperbolic, which prevents the simulations from evolving over a long time and becoming rapidly unstable.
This problem is overcome by the \emph{BSSN formalism}, \cite{shibata1995evolution, baumgarte1998numerical}, which rewrites the ADM equations into a highly hyperbolic form and allows arbitrarily long and stable evolutions of the Einstein equations.

\subsection{Numerical relativity and cosmological observations: the state of the art}

The methods presented above provide a comprehensive general relativistic description of cosmic dynamics. 
Although they allow the inclusion of GR effects in the growth of structures, the key aspect is to describe and evaluate the nonlinear GR effects on cosmological observations.
This requires an equally accurate description of light propagation, which is necessary for tests and comparisons with real data.
These studies are still in the early stages and have been approached in a variety of ways. For example, the distance-redshift relation has been investigated using perturbative methods (see e.g. \cite{DiDio:2016DlPT} for calculations up to second order PT, and  \cite{Sanghai:2017yyn} for PN calculations within a class of cosmological models and the deviation from the homogeneous FLRW), with exact methods (see e.g. \cite{Fanizza:2013GLC} for exact calculations in the geodesic light-cone gauge and \cite{celerier1999we, alexander2009local} for calculations in LTB models), and in various cosmological simulations (see e.g. \cite{adamek2014distance, Adamek:2018rru, Macpherson:2021gbh}). 
Another important example is the estimation of the weak gravitational lensing effect, i.e. the phenomenon of light deflection in the presence of massive objects. Weak lensing on cosmic scales can be used to probe the presence of dark matter and gain insight into the constituents of the Universe, \cite{Stafford:2021uvk}. Furthermore, by comparing the statistical features of the distortion map from galaxy surveys with those obtained from theoretical models, weak lensing can be used to distinguish between different models of modified gravity, \cite{KiDS:2021opn,Euclid:2021icp, Ruan:2021rqv}.
Theoretical estimates of weak lensing observables in the post-Friedmann formalism are presented in \cite{Thomas:2014aga, Gressel:2019jxw}.
Weak lensing map and power spectrum have also been extracted from numerical simulations \cite{Giblin:2017ezj, Lepori:2020ifz}.
According to these initial findings, the codes used to simulate GR dynamics appear to be consistent with Newtonian simulations for predicting weak lensing observables \cite{Thomas:2014aga, Lepori:2020ifz}, although there is a shift in the luminosity distance statistics \cite{Adamek:2018rru}. Moreover, the PN approximation leads to results different from $\Lambda$CDM for certain cosmological models \cite{Sanghai:2017yyn}. However, some efforts still need to be made to adapt the truly GR numerical codes to (observational) cosmology.

The central importance of estimating relativistic effects in cosmological observables requires a theoretical framework that comprehensively describes the propagation of light and all optical effects that result from its interaction with cosmological structures. In the next chapter, we give an overview of the theory of light propagation in geometric optics and introduce the theoretical foundations on which this work is based.
 

%% file: template/lightprop.tex
\chapter{The BGO formalism for light propagation}
\label{chap:BGO}
Cosmologists and astronomers use electromagnetic and gravitational radiation as their major tools for studying the structure and development of the Universe.  
These ''light-like'' signals contain information about the emitting source as well as of the spacetime geometry, the latter derived from the effects induced by gravity.
In the near future, these effects will be measured with unprecedented precision over a wider range of scales and redshift by the next generation of galaxy surveys and CMB experiments\footnote{{\color{blue}{\tt {https://www.skatelescope.org}}}, {\color{blue}{\tt {https://www.euclid-ec.org}}}, {\color{blue}{\tt {https://www.lsst.org}}}, {\color{blue}{\tt {http://litebird.jp/eng/}}}, {\color{blue}{\tt {https://www.jpl.nasa.gov/missions/spherex}}}}. This revolution in cosmology marks also the beginning of the \emph{real-time cosmology} era, \cite{Quercellini:2010zr}, in which it will be possible to measure small temporal changes in cosmological observables, called \emph{optical drift effects}. These real-time effects can provide important and new information about the structure and evolution of the Universe.  
From the point of view of the basic theory of light propagation, a new approach was presented in \cite{Grasso:2018mei}. The key ingredients of this new formulation are the \emph{bilocal geodesic operators} which represent the map from the portion of spacetime occupied by the observer to that occupied by the source and provide the complete description of the distortion of the light rays in between.

This chapter is divided into two parts: in the first part, we review the fundamental equations of light propagation in geometric optics, starting from Maxwell's equations in curved spacetimes. This is standard knowledge, see e.g. \cite{mtw, wald2010general, perlick-lrr}, and serves here as an introduction to the basic concepts of geometric optics in general relativity.
In the second part, we discuss the bilocal geodesic operators formulation of light propagation in geometric optics, based on the results presented in \cite{Grasso:2018mei}. This part provides the theoretical framework for the original results presented in Chapters~\ref{chap:bigonlight} and \ref{chap:nonlinearities}. 

\section{Light propagation in curved spacetime}
\label{sec:Maxwell_to_GeomOpt}

\emph{Light signals}, intended as radiation travelling at the speed $c=299792.5\, {\rm km/s}$, are governed by \emph{Maxwell's equations}
\begin{align}
F_{\mu \nu}&=\nabla_{\mu} A_{\nu}-\nabla_{\nu} A_{\mu} \label{eq:curvF}\\
\nabla_{\nu} F^{\mu \nu}&= 4 \pi J^{\mu} \label{eq:curvME1}\\
\nabla_{[\lambda}F_{\mu \nu]}&=0\, , \label{eq:curvME2}
\end{align}
with $J^{\mu}=(c \rho, \, J^i)$ the four-current of the charge density $\rho$ and current density $J^i$. $F^{\mu \nu}$ is the Faraday tensor (also known as field strength or electromagnetic tensor) defined as the field strength of the four-vector potential $A^{\mu}=(\phi / c, A^i)$
\begin{equation}
F_{\mu \nu}=\nabla_{\mu} A_{\nu}-\nabla_{\nu} A_{\mu}\, , \label{eq:F}
\end{equation}
whose components are the electric $F^{0 i}=-F^{i 0}=E^i$ and magnetic $F^{i j}=\varepsilon^{i j k}B^k$ fields. 
From the very definition of $F_{\mu \nu}$, Eq.~\eqref{eq:F}, we see that the electromagnetic tensor is invariant under the gauge transformation $A_\mu \to A_\mu + \nabla_\mu \chi$, with $\chi$ a scalar function. This constitutes the gauge freedom of electromagnetism. A convenient gauge choice is the Lorentz gauge condition $\nabla_{\mu} A^{\mu}=0$, in which Eq.~\eqref{eq:curvME1} takes the simpler form of a wave equation for the four-vector potential
\begin{equation}
\nabla_{\nu}\nabla^{\nu}A^{\mu}=R\UD{\mu}{\sigma }A^{\sigma}- 4 \pi J^{\mu}\, , \label{eq:curvelectromagn}
\end{equation}
where the Ricci tensor $R\UD{\mu}{\sigma}=R\DU{\sigma}{\mu}=R^{\nu \, \, \, \, \, \mu}_{\,\,  \sigma \nu}$ appear from the commutation of the two covariant derivatives as $\nabla_{\nu}\nabla^{\mu}A^{\nu}=\nabla^{\mu}\nabla_{\nu}A^{\nu}+R^{\nu \, \, \, \, \, \mu}_{\,\,  \sigma \nu}A^{\sigma}$.
The expression in Eq.~\eqref{eq:curvelectromagn} describes the dynamics of electromagnetic potential in a generic spacetime, (see e.g. \cite{mtw, wald2010general, perlick-lrr}).

In the absence of sources $J^{\mu}=0$ Eq.~\eqref{eq:curvelectromagn} gives the propagation equation for the electromagnetic radiation
\begin{equation}
\nabla_{\nu}\nabla^{\nu}A^{\mu} - R\UD{\mu}{\sigma }A^{\sigma}=0\, , \label{eq:curvER}
\end{equation}
which one has to solve to study light propagation in cosmology. 
However, this equation is too general, and needs to be ``adapted'' to apply to astronomical observations.
As described in \cite{mtw}, one can distinguish three characteristic lengths\footnote{These characteristic lengths are  evaluated in a local inertial frame, e.g. the one at rest respect a nearby galaxy.}:
\begin{enumerate}
\item the typical wavelength of the electromagnetic radiation $\lambda$,
\item the typical length over which the amplitude, the polarization and the wavelength vary $L$, 
\item the typical radius of curvature of the spacetime $\mathcal{R}$, defined such that $\mathcal{R}=\mathcal{O}\left( |R\UD{\mu}{\rho \sigma \nu}|^{-1/2}\right)$, with  $|R\UD{\mu}{\rho \sigma \nu}|$ denoting the magnitude of the typical component of the Riemann tensor.
\end{enumerate}
In the range of astrophysical observations, the typical electromagnetic wavelength extends\footnote{For gravitational waves the typical wavelength can be larger, as the one of the first gravitational waves detection $\lambda \sim 10^4 \, {\rm km}$, \cite{abbott2016observation}.} from $\sim 10 \, {\rm m}$ for radio waves emitted by active radio galaxies to the smaller wavelengths of visible, X-rays, and gamma-rays emissions. 
On the other hand, the typical curvature's radius $\mathcal{R}$ of the spacetime where these electromagnetic signals propagate is usually much larger. For instance on cosmological scales we have that\footnote{As an estimate of $|R\UD{\mu}{\rho \sigma \nu}|$ we use the Ricci scalar that in a flat FLRW metric gives $R=6\left(\dfrac{\dot{a}^2+a \ddot{a}}{a^2} \right) \sim \left(\dfrac{\dot{a}}{a}\right)^2=H^2$. In other cases, like e.g. in a Schwarzschild metric, the Ricci scalar is not a good indicator and we need to use a different estimator for $|R\UD{\mu}{\rho \sigma \nu}|$, like the Kretschmann scalar $|R^{\mu \rho \sigma \nu}R_{\mu \rho \sigma \nu}|$.} $\mathcal{R} \sim c / H$, with $H$ the Hubble parameter whose value depends on the cosmological era\footnote{At the recombination era (around $3.7 \times 10^5$ years after the Big Bang, the epoch at which the ionised plasma of electrons and protons first became bound forming neutral hydrogen atoms, which did not scatter the photons but allowed them to travel freely) the Hubble parameter was $H=1255\, \, {\frac{km}{s\, Mpc}}$ corresponding to $\mathcal{R} \sim 0.2 \, {\rm Mpc}$.}: at present time $H_0=67.4\, {\frac{km}{s\, Mpc}}$, see~\cite{planck2018param}, which gives $\mathcal{R} \sim 4.5 \, {\rm Gpc}$. Also on smaller scales the condition $\lambda \ll \mathcal{R}$ remain valid, for instance close to the surface of the Sun, one can calculate that the curvature radius is of the order of $ 10^8\, {\rm km}$. 
We need to go to very strong gravitational regimes, like close to the event horizon $r_s$ of a black hole where $\mathcal{R}\sim r_s$ to have the condition $\lambda \ll \mathcal{R}$ no longer valid for part of the electromagnetic spectrum\footnote{For instance, for a Sun-like black hole we have that $\mathcal{R}\sim r_s\sim 3\, {\rm km}$ which is smaller than the low frequency radio waves, or for a Earth-like black hole $\mathcal{R} \sim 8\, {\rm mm}$ which is smaller than micro waves $\lambda \sim 1\, {\rm cm}$.}.
Except for the last instance, we can treat light propagation within the so called \emph{geometric optics approximation}, which is valid whenever $\lambda$ is much smaller than each of the other scales involved, i.e.
\begin{align}
\lambda \ll L\, &\text{ and }\, \lambda \ll \mathcal{R}\, .
\end{align}
Within this approximation one can look for solutions of Eq.~\eqref{eq:curvER} in the form of a rapidly oscillating wave with a nearly constant amplitude, namely
\begin{equation}
A^{\mu}=C^{\mu} e^{i \theta} \, , \label{eq:em_ansaz}
\end{equation}
where the phase $\theta \propto \frac{2\pi}{\lambda}$ is a real function of the spacetime's position, while $C^{\mu}$ is in general a complex four-vector expressing the amplitude and polarization of the electromagnetic wave. 
Given this ansatz, we note that with $\lambda$ decreasing to zero, and $L$ and $\mathcal{R}$ fixed, the phase $\theta$ will get larger and larger, but $C^{\mu}$ will not vary very much. Therefore, we can express the dependence on $\lambda$ in Eq.~\eqref{eq:em_ansaz} by introducing the parameter $\epsilon=\lambda / d $, where $d={\rm min}(L, \, \mathcal{R})$, and expanding the solution in its powers 
\begin{equation}
A^{\mu}=(a^{\mu}+\epsilon b^{\mu}+\epsilon^2 c^{\mu}+\cdots) e^{i \frac{\theta}{\epsilon}} \, . \label{eq:em_ansaz_expansion}
\end{equation}
Note that $a^{\mu} e^{i \frac{\theta}{\epsilon}}$ is the leading order term and it constitutes the geometric optics approximation of our solution. The other higher-order terms (\setwd{h.o.t.}{acr:hot}) are all contained in $(\epsilon b^{\mu},\,  \epsilon^2 c^{\nu},\, \cdots)$, representing the post-geometric optics corrections (see e.g. \cite{ PhysRev.166.1263, PhysRev.166.1272, Harte:2018wni} for approaches beyond geometric optics).
Applying the ansatz in Eq.~\eqref{eq:em_ansaz_expansion}, the wave equation in geometric approximation gives 
\begin{align}
\ell_{\mu} a^{\mu}&=0  \label{eq:geom_phase}\\
a^\mu \ell_{\nu} \ell^{\nu} & =0 \label{eq:geom1}\\
- b^\mu \ell_{\nu} \ell^{\nu}  + i \left( a^{\mu} \nabla_{\nu}\ell^{\nu} \right. & \left. + 2 \ell_{\nu} \nabla^{\nu} a^{\mu} \right) =0 \, .\label{eq:geom2}
\end{align}
The first relation, Eq.~\eqref{eq:geom_phase} is the orthogonality relation between the vector amplitude $a^{\mu}$ and $\ell^{\mu}=\nabla^{\mu} \theta$, the vector normal to the surfaces of constant phase.
The relation Eq.~\eqref{eq:geom1} expresses the fact that in the geometric optics approximation one can consider the electromagnetic signals as travelling along null-like geodesic, whose tangent vector $\ell^{\mu}$ satisfies the geodesic equation\footnote{The geodesic equation~\eqref{eq:geodesicEQ} is related to the condition Eq.~\eqref{eq:geom1} as $0=\nabla_{\mu}(\ell^{\nu}\ell_{\nu})=(\nabla_{\mu}\ell^{\nu})\, \ell_{\nu}+\ell^{\nu}(\nabla_{\mu}\ell_{\nu})= \ell_{\nu}(\nabla_{\mu}\nabla^{\nu}\theta)+\ell^{\nu}(\nabla_{\mu}\nabla_{\nu}\theta)= \ell_{\nu} (\nabla^{\nu}\nabla_{\mu}\theta)+\ell^{\nu}(\nabla_{\nu}\nabla_{\mu}\theta)= \ell_{\nu} (\nabla^{\nu}\ell_{\mu})+\ell^{\nu}(\nabla_{\nu}\ell_{\mu})=2 \ell^{\nu}\nabla_{\nu}\ell_{\mu}$
, where we have used $\nabla_{\mu}(\nabla_{\nu}\theta)=\nabla_{\mu}(\partial_{\nu}\theta)=\partial_{\mu}\partial_{\nu}\theta-\Gamma^{\sigma}_{\mu \nu}\partial_{\sigma}\theta=\partial_{\nu}\partial_{\mu}\theta-\Gamma^{\sigma}_{\nu \mu}\partial_{\sigma}\theta=\nabla_{\nu}(\nabla_{\mu}\theta)$.} 
\begin{equation}
\ell^{\nu}\nabla_{\nu}\ell^{\mu}=0 \, . \label{eq:geodesicEQ}
\end{equation}
A generic geodesic can be represented as a parametric curve $\gamma(\lambda)$, where the parameter $\lambda$ spans the geodesic such that to a small variation of the parameter $d \lambda$ correspond a small displacement along the geodesic itself: 
$$d x^{\mu}= \ell^{\mu} \, d \lambda\, .$$
Therefore, the geodesic equation~\eqref{eq:geodesicEQ} can be expressed as the covariant derivative with respect to $\lambda$ as
\begin{equation}
\dfrac{D}{D \lambda}\ell^{\mu}=\dfrac{d^2 x^{\mu}}{d \lambda^2}+ \Gamma^{\mu}_{\sigma \rho} \dfrac{d x^{\sigma}}{d \lambda} \dfrac{d x^{\rho}}{d \lambda}=0 \, . \label{eq:geodesicEQ_par}
\end{equation}
It is worth noticing that Eq.~\eqref{eq:geodesicEQ_par} is satisfied if $\lambda$ is an affine parameter of the geodesic $\gamma(\lambda)$. However, the parametrisation of the geodesic is not unique, i.e. it is always possible to choose a different parametrisation $\tau$, such that $\gamma(\lambda) \to \gamma(\tau)$. In general the transformation introduces a new term in Eq.~\eqref{eq:geodesicEQ_par} proportional to the tangent vector, such that $\dfrac{D \ell^{\mu}}{D \tau}\propto \ell^{\mu}$: in this case $\tau$ is said a non-affine parameter. It is easy to show that the transformation 
\begin{equation}
\lambda \to A \cdot \lambda + B , \label{eq:affineparameter}
\end{equation}
with $A,\, B={\rm const} $, is the only possible transformation that leaves the Eq.~\eqref{eq:geodesicEQ_par} satisfied, namely it transforms an affine parameter into a new affine parameter (see problem 5 of Sec. 3 in~\cite{wald2010general}). 
Moreover, a different parametrisation of the geodesic $\gamma$ changes the value of the squared norm of the tangent vector. In general for time-like and space-like geodesics we prefer to use an affine parametrisation of the geodesic such that its tangent vector $k^{\mu}$ is normalised as $k^\mu k_\mu = \sigma$, with $\sigma=-1$ for time-like geodesics and $\sigma=1$ for space-like geodesics.
In the particular case of null geodesics we do not have a preferred, normalised parametrisation, since Eq.~\eqref{eq:geom1} holds, so we may always reparametrise $\gamma$ by an affine transformation, Eq.~\eqref{eq:affineparameter}. Then the null tangent vector $\ell^\mu$ transforms according to
 \bea
  \ell^\mu \to \frac{1}{A}\,\ell^\mu \label{eq:laffine}\, .
 \eea

The relation in Eq.~\eqref{eq:geom2} is the propagation equation for the vector amplitude: it is convenient to express $a^{\mu}$ as $a^{\mu}=a p^{\mu}$, where  $a=\sqrt{|a^{\mu}a_{\mu}|}$ is the scalar amplitude and $p^{\mu}=a^{\mu}/a$ is the polarization vector. Thus, Eq.~\eqref{eq:geom2} becomes
\begin{equation}
p^{\mu} \left(a  \nabla_{\nu}\ell^{\nu} + 2 \ell_{\nu} \nabla^{\nu} a \right) + 2 a \ell_{\nu} \nabla^{\nu} p^{\mu}=0\, . \label{eq:polartransp1}
\end{equation}
The term in parenthesis has an important physical meaning and it represents the flux conservation in geometric optics approximation. This can be easily proved by using the continuity equation $\nabla^{\mu}T_{\mu \nu}=0$ for the electromagnetic stress-energy tensor in the geometric optics approximation $T_{\mu \nu}= a^2 e^{2 i \theta} \ell_{\mu} \ell_{\nu}+ h.o.t.$. After some straightforward calculations, one get that at the leading order
\begin{equation}
0=\nabla^{\mu}T_{\mu \nu}=2 a \nabla^{\mu} a \,  \ell_{\mu} + a^2\nabla^{\mu}\ell_{\mu}=\nabla^{\mu}(a^2 \ell_{\mu})\, . \label{eq:photcons}
\end{equation}
The vector $a^2 \ell^{\mu}$ is the photon flux density and the volume integral $(8 \pi \hbar )^{-1}\int a^2 \ell^0 \sqrt{|-g|} d^3 x$ gives the number of photons (or geodesics) in the volume of integration on any $x^0 = {\rm const}$ hypersurface.
Implementing Eq.~\eqref{eq:photcons} in Eq.~\eqref{eq:polartransp1} we obtain the propagation equation for the polarization vector
\begin{equation}
\ell^{\nu} \nabla_{\nu} p^{\mu}=0\, , \label{eq:polartransp}
\end{equation}
or in other words, the polarization vector is parallel transported along the null geodesic.

To conclude, the geometric optics approximation can be summarised as follows:
\begin{itemize}
\item when an electromagnetic signal satisfies the conditions $\lambda \ll L\, \text{ and }\, \lambda \ll \mathcal{R}$ we can look for solutions to Eq.~\eqref{eq:curvER} of the form $A^{\mu}= a p^{\mu} e^{i \theta} + h.o.t.$;
\item in this approximation we can consider the photons as travelling along light rays (null geodesics), Eq.~\eqref{eq:geom1}, with the tangent vector $\ell_{\mu}=\nabla_{\mu} \theta$ being the normal to the surfaces of constant phase $\theta$;
\item the amplitude $a$ is governed by the evolution equation $$\ell^{\mu}\nabla_{\mu} a = -\frac{1}{2} a (\nabla_{\mu}\ell^{\mu})\, ,$$
which leads to the conservation of the photon number Eq.~\eqref{eq:photcons};
\item the polarization vector $p^{\mu}$ is perpendicular to the light rays, Eq.~\eqref{eq:geom_phase} and it is parallel transported along them Eq.~\eqref{eq:polartransp}.
\end{itemize}

\subsection{Geometric description of light beams}

The geometric optics approximation provides a description of light propagation in terms of rays. In this view, the image of an astronomical object is effectively presented as the cross section of a light beam, namely the bunch of rays emitted by the object and focused at the observer. However, the curved spacetime between the emitting object and the observer can bend the geodesics and hence produce deformations in the cross section of the beam. The result is that the apparent position, size, shape, and luminosity of the emitter will appear modified to the observer. 
These effects are described using the geodesic deviation equation, which is the equation describing the tendency of nearby geodesics to converge or diverge from each other due to the curvature of the spacetime. 
In this section we will introduce the geodesic deviation equation from a geometric prospective, without any restriction to its application, and we postpone to the next section, Sec.~\ref{sec:BGO}, the implementation of the concepts introduced here to describe a typical situation in observational astronomy.

As we just mentioned, the equation of geodesic deviation expresses the changes of the distance between a point $x^{\mu}(\lambda)$ on one geodesic $\gamma(\lambda)$ to a point $\tilde{x}^{\mu}(\lambda)$ on a nearby geodesic $\tilde{\gamma}(\lambda)$ at the same value\footnote{Here one has to assume that the two geodesics are labelled by the same affine parameter $\lambda$. However, since there is no unique way of relating the affine parameter on one geodesic to the affine parameter on another, we have a degeneracy in the separation vector definition. This will be clarified later.} of $\lambda$. The major assumption is that the two geodesics are close enough, so as we can define a deviation vector $\xi^{\mu}(\lambda)=\tilde{x}^{\mu}(\lambda) - x^{\mu}(\lambda)$ expressing the difference between the two points and such that the geodesic deviation equation is derived by expanding the geodesic equation for $\tilde{\gamma}(\lambda)$ up to linear order in $\xi^{\mu}$. Therefore, the geodesic deviation equation gives only the linear corrections in the deviation vector. Several authors have extended the geodesic deviation equation beyond the linear order, see e.g. \cite{Bazanski1, Puetzfeld:2015uxi, Vines:2014oba}, but the first-order is enough for the purposes of this work.

\begin{figure}[ht]
    \centering
    \includegraphics[width=0.9\linewidth]{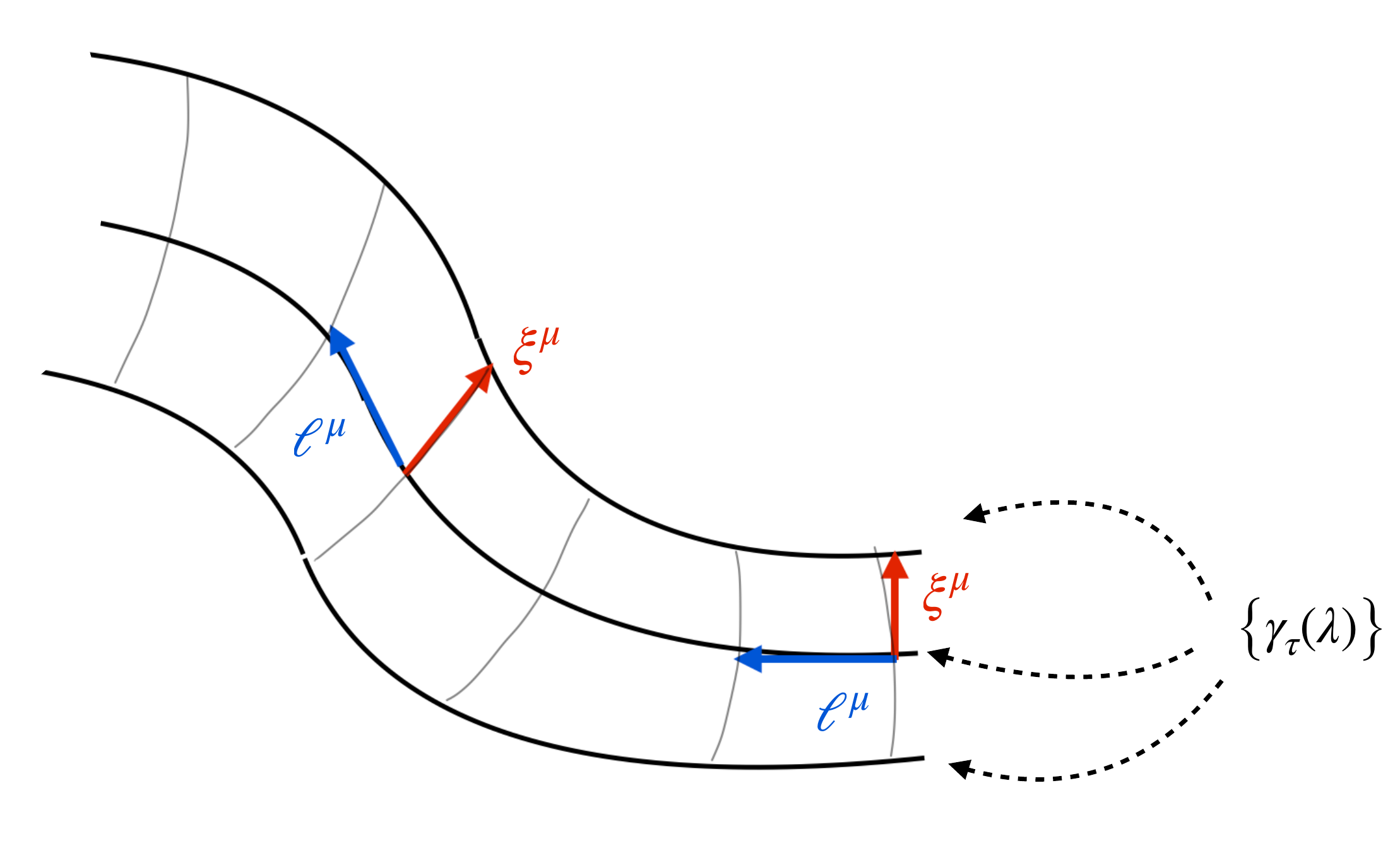}
    \caption{The tangent vector to the geodesics $\ell^{\mu}$ and the deviation vector $\xi^{\mu}$, representing the infinitesimal displacement from a nearby geodesic, characterise the geodesics of the family $\{\gamma_{\rm \tau}(\lambda)\}$.}\label{fig:congruence}
\end{figure}

This way of deriving the geodesic deviation equation requires the introduction of several details which make the derivation difficult to follow. Instead, we decided to present the geodesic deviation equation as derived in~\cite{wald2010general}: on the one hand this derivation has the advantage of having a clear physical interpretation, but on the other hand it hides the perturbative nature of the geodesic deviation equation. Here, we will try to solve this question clarifying where the linearisation in the deviation vector takes place.
Let us start by representing a light beam as a smooth one-parameter family of geodesics $\{\gamma_{\rm \tau}(\lambda)\}$, i.e. for each ${\rm \tau} \in \mathds{R}$ correspond a null geodesic $\gamma_{\tau}(\lambda)$ of the beam parametrised by the affine parameter $\lambda$, and such that the map $(\lambda, \tau) \to \{\gamma_{\rm \tau}(\lambda)\}$ is smooth. 
For our purposes we have considered light geodesics, but the derivation we will present is completely general and it does not depends on the nature of the geodesics.
Defining $\Sigma$ as the two-dimensional submanifold spanned by the geodesics of the family $\{\gamma_{\rm \tau}(\lambda)\}$, one can introduce the coordinate base $(\ell^{\mu}, \xi^{\mu})$: the vector $\ell^{\mu}=\frac{d x^{\mu}}{d \lambda}$ is tangent to the family of geodesics and it satisfies Eq.~\eqref{eq:geodesicEQ}). The vector  $\xi^{\mu}=\frac{ d x^{\mu}}{d \tau}$ is the deviation vector and it represents the infinitesimal displacement between two nearby geodesics, see Fig~\ref{fig:congruence}. 
To evaluate the change of the vector field $\xi^{\mu}$, along the flow defined by the tangent vector to the geodesics $\ell^{\mu}$, let us calculate the Lie derivative of $\xi^{\mu}$ with respect to $\ell^{\mu}$ on $\Sigma$
\begin{equation}
\mathcal{L}_{\ell} \xi^{\mu} = \ell^{\nu}\nabla_{\nu} \xi^{\mu}-\xi^{\nu}\nabla_{\nu} \ell^{\mu}=\ell^{\nu}\partial_{\nu} \xi^{\mu}-\xi^{\nu}\partial_{\nu} \ell^{\mu}\, ,\label{eq:lie_gde}
\end{equation}
where last equality follows from the symmetry of the Christoffel symbols $\Gamma^{\lambda}_{\mu \nu}$.
Now, since $\ell^\mu=\frac{\partial x^{\mu}}{\partial \lambda}$ and $\xi^{\mu}=\frac{\partial x^{\mu}}{\partial \tau}$, it is easy to see that the Lie derivative vanishes, implying the commutation of the two vector fields
\begin{equation}
\ell^{\nu}\nabla_{\nu} \xi^{\mu}=\xi^{\nu}\nabla_{\nu} \ell^{\mu}\, . \label{eq:commutation_vector}
\end{equation}

Let us remark that the perturbative nature of the geodesic deviation equation is already assumed when we use the Lie derivative. In fact, from the very definition of the Lie derivative we have
\begin{equation}
- \mathcal{L}_{\ell} \xi^{\mu}=\mathcal{L}_{\xi} \ell^{\mu}=\lim_{\Delta \tau \to 0} \dfrac{1}{\Delta \tau} \left[ \dfrac{\partial x^{\mu}_{0}}{\partial x^{\nu}_{\tau}}\ell^{\nu}(x_{\tau})-\ell^{\mu}(x_{0})\right]\, ,\label{eq:pro1}
\end{equation}
where $\xi^{\mu}$ points from $x^{\mu}_0 \in \gamma_0$ to a nearby point $x^{\mu}_{\tau} \in \gamma_{\tau}$.
We can now see that Eq.~\eqref{eq:commutation_vector}, which is the starting point of this derivation, expresses the vanishing of the linear order expansion of the flow of $\ell^{\mu}$ along the vector field $\xi^{\mu}$.

We can also go back from the right hand side of Eq.~\eqref{eq:pro1} to the Lie derivative in Eq.~\eqref{eq:lie_gde} by expanding the first term as an infinitesimal coordinate transform $x^{\mu}_{\tau}$ to $x^{\mu}_0=x^{\mu}_{\tau}- \xi^{\mu} \Delta \tau + \calO(\Delta \tau^2)$ acting on $\ell^\mu$:
\begin{equation}
 \dfrac{\partial x^{\mu}_{0}}{\partial x^{\nu}_{\tau}}\ell^{\nu}(x_{\tau})= \ell^{\mu}(x_{\tau})-\ell^{\nu}(x_{\tau}) \partial_{\nu} \xi^{\mu}(x_0) \Delta \tau + \calO(\Delta \tau^2)\, .\label{eq:pro2}
\end{equation}
Expressing $\ell^{\mu}(x_{\tau})=\ell^{\mu}(x_0)+\xi^{\nu}(x_0)\partial_{\nu}\ell^{\mu}(x_0) \, \Delta \tau + \mathcal{O}(\Delta \tau^2)$ in Eq.~\eqref{eq:pro2}, we indeed obtain the previous expression of the Lie derivative
\begin{equation}
\mathcal{L}_{\xi} \ell^{\mu}= \dfrac{1}{\Delta \tau} \left[\ell^{\mu}(x_{0})+\xi^{\nu}\partial_{\nu}\ell^{\mu}\Delta \tau-\ell^{\nu}\partial_{\nu}\xi^{\mu}\Delta \tau+\calO(\Delta \tau^2)-\ell^{\mu}(x_{0})\right]=\xi^{\nu}\partial_{\nu}\ell^{\mu}-\ell^{\nu}\partial_{\nu}\xi^{\mu}\, .
\end{equation}


Let us move back on the derivation. Following the interpretation that $\xi^{\mu}$ represents the displacement between nearby geodesics, the left hand side of Eq.~\eqref{eq:commutation_vector} defines the relative velocity between geodesics. Similarly, the relative acceleration between the geodesics of the family is $w^{\mu}=\ell^{\rho}\nabla_{\rho} (\ell^{\nu}\nabla_{\nu} \xi^{\mu})$. From Eq.~\eqref{eq:commutation_vector} then we have
\begin{align}
w^{\mu}&=\ell^{\rho}\nabla_{\rho}(\xi^{\nu}\nabla_{\nu} \ell^{\mu}) \nonumber \\
 & = (\ell^{\rho}\nabla_{\rho}\xi^{\nu})(\nabla_{\nu} \ell^{\mu}) + \ell^{\rho}\xi^{\nu}\nabla_{\rho}\nabla_{\nu} \ell^{\mu} \nonumber \\
& = (\xi^{\rho}\nabla_{\rho}\ell^{\nu})(\nabla_{\nu} \ell^{\mu}) + \ell^{\rho}\xi^{\nu}(\nabla_{\nu}\nabla_{\rho} \ell^{\mu}+R\UD{\mu}{\sigma \rho \nu }\ell^{\sigma}) \nonumber \\
& = (\xi^{\rho}\nabla_{\rho}\ell^{\nu})(\nabla_{\nu} \ell^{\mu}) + \xi^{\nu}\nabla_{\nu}(\ell^{\rho}\nabla_{\rho} \ell^{\mu})-(\xi^{\nu}\nabla_{\nu}\ell^{\rho})(\nabla_{\rho} \ell^{\mu})+R\UD{\mu}{\sigma \rho \nu }\ell^{\sigma}\ell^{\rho}\xi^{\nu} \nonumber \\
& = R\UD{\mu}{\sigma \rho \nu }\ell^{\sigma}\ell^{\rho}\xi^{\nu}\, .
 \label{eq:acc_gde}
\end{align}
The result is the geodesic deviation equation (\setwd{GDE}{acr:GDE}), 
\begin{equation}
\ell^{\rho}\nabla_{\rho} (\ell^{\nu}\nabla_{\nu} \xi^{\mu})=R\UD{\mu}{\sigma \rho \nu }\ell^{\sigma}\ell^{\rho}\xi^{\nu}\, ,\label{eq:GDE} 
\end{equation}
which relates the relative acceleration between infinitesimally close geodesics with the spacetime curvature, \cite{wald2010general}. The GDE for timelike geodesics plays an important role in the foundation of General Relativity, since it can be used to characterize the spacetime curvature as the relative motion of free falling bodies. This was covered by many authors, see e.g. \cite{pirani, szekeres, Bazanski1, Bazanski2, Aleksandrov1979, CiufoliniDemianski, Vines:2014oba, Puetzfeld:2015uxi, Flanagan:2018yzh}). 
For analysis on the GDE for null geodesics see \cite{Bartelmann_2010, Clarkson:2016zzi, Clarkson:2016ccm, Korzynski:2018, Uzun:2018yes, Grasso:2018mei}.

\subsection{Properties of the GDE}
Let us discuss now two general properties of the  GDE solutions which hold irrespectively of the underlying geometry, as shown in \cite{Vines:2014oba, Korzynski:2018, Grasso:2018mei}.
The first property is derived by multiplying the GDE, Eq.~\eqref{eq:GDE}, by $\ell_{\mu}$ to obtain
\begin{equation}
\nabla_{\ell} \nabla_{\ell} \left(\ell_{\mu} \xi^\mu \right)=0\, ,
\label{eq:gde_prop_in}
\end{equation}
with $\nabla_{\ell}=\ell^{\sigma} \nabla_{\sigma}$, and we use the fact that $\ell^{\mu}$ is a solution of Eq.~\eqref{eq:geodesicEQ} and $\ell_{\mu} R\UD{\mu}{\alpha \beta \nu}\ell^{\alpha}\ell^{\beta}=0$ from the symmetry of the Riemann tensor. Then we have 
\bean
\xi^\mu \, \ell_\mu = C + D\,\lambda,
\eean
with $C,D = \const$. In this way we have defined 2 constants of motion for the GDE, namely the quantities 
\begin{align}
D = (\nabla_{\ell}\xi^\mu)\, \ell_\mu \label{eq:gde_prop1B}\\
C = \xi^\mu \, \ell_\mu - D\,\lambda\, , \label{eq:gde_prop1A}
\end{align}
are conserved along the geodesics.

\begin{figure}[ht]
    \centering
    \includegraphics[width=0.9\linewidth]{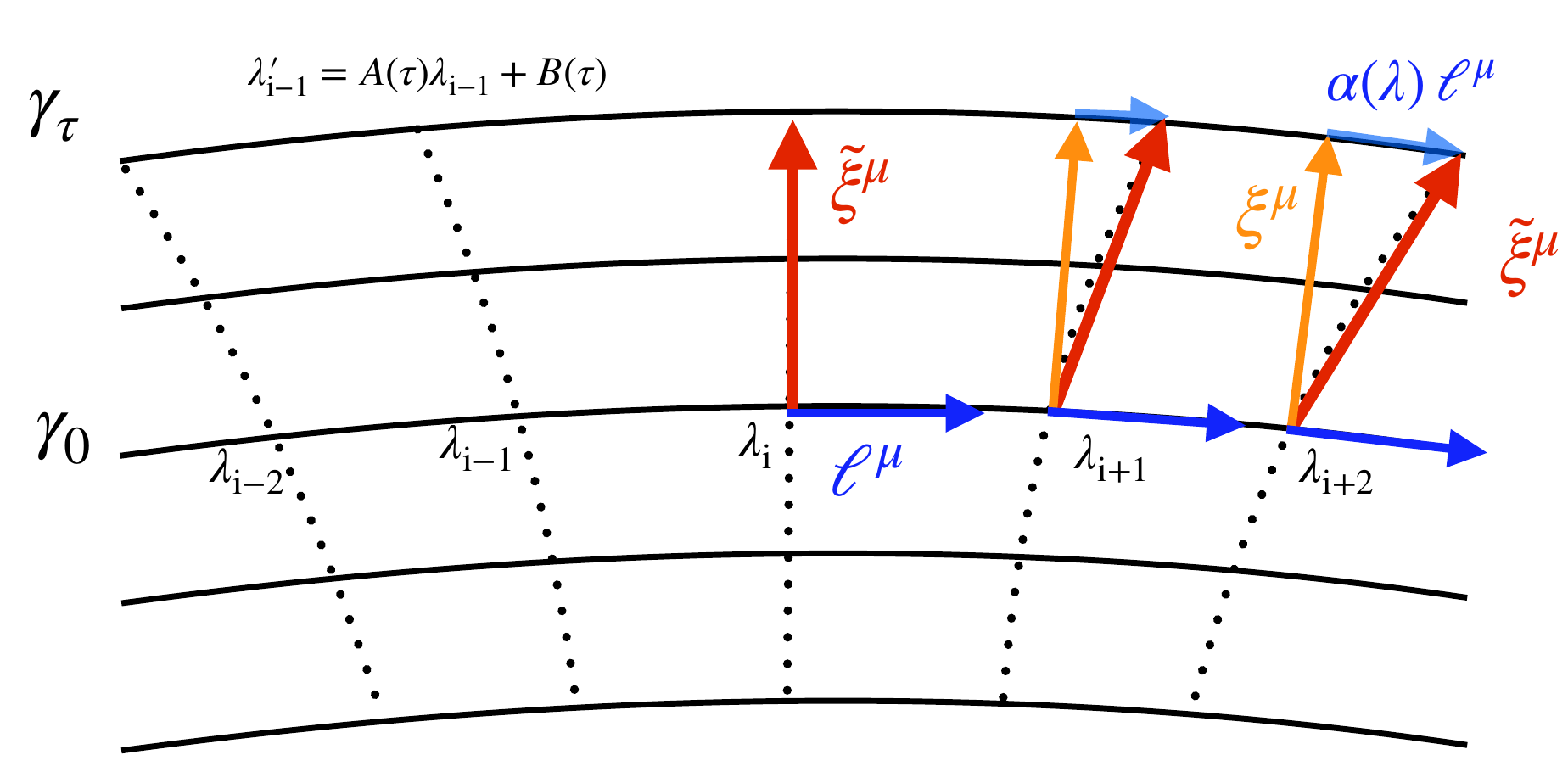}
    \caption{The two deviation vectors $\xi^{\mu}$ and $\tilde{\xi}^{\mu}=\xi^{\mu}+\alpha(\lambda)\ell^{\mu}$ identify the same geodesic $\gamma_{\tau}$. This invariance of the GDE is related to the gauge freedom one has in choosing the affine parameter. Indeed, the different deviation vector $\tilde{\xi}^{\mu}$ can be obtained by introducing the different affine parametrisation $\lambda'$ of the geodesic $\gamma_{\tau}$.
}\label{fig:prop2}
\end{figure}

The second property states that if $\xi^\mu(\lambda)$ is a solution of the GDE, then also $\tilde{\xi}^\mu =  \xi^\mu + \alpha(\lambda)\,\ell^\mu$ is  a solution. The form of the proportionality function $\alpha(\lambda)$ is easily obtained by inserting $\tilde{\xi}^\mu$ into Eq.~\eqref{eq:GDE} to find that $\alpha(\lambda)= (E + F\,\lambda)$, with $E,F = \const$. In other words, we have that
\bea
\tilde{\xi}^\mu =  \xi^\mu + (E + F\,\lambda)\,\ell^\mu \label{eq:gde_prop2}
\eea
is a solution of Eq.~\eqref{eq:GDE}.
This ``gauge freedom'' of adding terms proportional to $\ell^\mu$ to the solution of the GDE is a direct consequence of the freedom we have in choosing the affine parametrisation of a geodesic. In fact, as previously discussed, if $\lambda$ is the affine parameter of the geodesic $\gamma_0 \in \{\gamma_{\tau}\}$, then $A(\tau)\, \lambda + B(\tau)$ is the only possible form for the affine parameter of any of the other geodesics of the family $\{\gamma_{\tau}\}$.
Geometrically, $\tilde{\xi}^\mu$ corresponds to the same congruence of geodesics as $\xi^\mu$, but with a change of parametrisation of the geodesics around $\gamma_0$, see Fig.~\ref{fig:prop2}. 

Assuming that the character of the geodesics is conserved, i.e. the geodesics of the family are all of the same type, we can draw some conclusions regarding $C$ and $D$. Let us start by using Eq.~\eqref{eq:gde_prop1A} to write Eq.~\eqref{eq:gde_prop2} as
\begin{equation}
\tilde{C} + \lambda \tilde{D}=\tilde{\xi}^{\mu}\ell_{\mu}=C + \lambda D+(E+F \lambda)\ell^{\mu}\ell_{\mu}\, ,
\end{equation}
where $C + \lambda D=\xi^{\mu}\ell_{\mu}$, and the equality is satisfied for
\begin{align}
\tilde{C} &= C + E\, \ell^{\mu}\ell_{\mu} \\
\tilde{D} &= D + F\, \ell^{\mu}\ell_{\mu} \, .
\end{align}
We distinguish two cases: 
\begin{itemize} 
\item for a time-like or space-like family we have $\ell^\mu \ell_\mu=\epsilon \neq 0$, implying that it is always possible to choose a different reparametrisation of the geodesics around $\gamma_0$ such that the vectors $\xi^\mu$ and $\ell^{\mu}$ stay perpendicular along $\gamma_0$, namely
\begin{align}
\tilde{C} &= C + \dfrac{-C}{\ell^{\mu}\ell_{\mu}}\, \ell^{\mu}\ell_{\mu}=0 \label{eq:gdeRel1}\\
\tilde{D} &= D + \dfrac{-D}{\ell^{\mu}\ell_{\mu}}\, \ell^{\mu}\ell_{\mu}=0 \label{eq:gdeRel2}\, .
\end{align}
\item for a null family one has $\ell^\mu \ell_\mu=0$ for all geodesics, implying that for any choice of $E$ and $F$ we have $\tilde{C} = C$ and $\tilde{D} = D$.
\end{itemize} 
We conclude that for a congruence of null geodesics there is no\footnote{Note that we are not saying that $\xi^\mu \ell_{\mu}\neq 0$ for null geodesics. This is possible and it represent a specific choice of the initial conditions for the null geodesics of the family. What we meant is that if $\xi^\mu \ell_{\mu}\neq 0$, then it is not possible to chose a proper affine reparametrisation to transform $\xi^\mu$ into a vector orthogonal to $\ell^\mu$.} affine reparametrisation of the geodesics around $\gamma_0$ which makes the deviation vector $\xi^\mu$ perpendicular to  $\ell^{\mu}$, \cite{Korzynski:2018, Grasso:2018mei}. In this sense the null families represent a special class of families of geodesics.

Let us proceed in our analysis by noting that Eq.~\eqref{eq:gde_prop1B} can be expressed as $D=\frac{1}{2}\xi^{\nu}\nabla_{\nu}(\ell^\mu \ell_\mu)$, which follows multiplying by $\ell_{\mu}$ the relation in Eq.~\eqref{eq:commutation_vector}, and using the equality\footnote{It is derived as follows: $\nabla_{\nu}( \ell^\mu \ell_\mu)=(\nabla_{\nu} \ell^\mu)\ell_\mu+(\nabla_{\nu} \ell_\mu)\ell^\mu=(\nabla_{\nu} \ell^\mu)\ell_\mu+[\nabla_{\nu}(g_{\mu \sigma} \ell^\sigma)]g^{\mu \rho} \ell_\rho$. Using $\nabla_{\nu}g_{\mu \sigma}=0$, we have
$\nabla_{\nu}( \ell^\mu \ell_\mu)=(\nabla_{\nu} \ell^\mu)\ell_\mu+(\nabla_{\nu} \ell^\sigma)\delta\DU{ \sigma}{ \rho} \ell_\rho=2(\nabla_{\nu} \ell^\mu)\ell_\mu=2(\nabla_{\nu} \ell_\mu)\ell^\mu$.} $\xi^{\nu}(\nabla_{\nu} \ell^\mu)\ell_\mu=\frac{1}{2}\xi^{\nu}\nabla_{\nu}(\ell^\mu \ell_\mu)$. Again we have different conclusions depending on the character of the family: 
\begin{itemize}
\item for a time-like or space-like family we have a preferred normalisation of the tangent vector $\ell^\mu \ell_\mu=\epsilon$, which in general may differ among the geodesics of the family. In other words, the value of $\ell^\mu \ell_\mu=\epsilon$ changes along $\xi^{\mu}$, implying that $D=\xi^{\nu}\nabla_{\nu}(\ell^\mu \ell_\mu)\neq 0$. To impose $D=0$ we need to perform a reparametrisation of the other geodesics of the family such that the normalisation of the tangent vector remains constant: this is precisely what we demanded in Eq.~\eqref{eq:gdeRel2}.
\item for a null family one has $\ell^\mu \ell_\mu=0$ for all geodesics, implying $D=0$ and $C = \xi^\mu \, \ell_\mu={\rm const}$ irrespectively of the parametrisation. 
\end{itemize} 
The condition $D=0$ for null geodesics leads to the \emph{flat lightcone approximation} (\setwd{FLA}{acr:FLA}) for the time of arrival of the electromagnetic signals\footnote{It will be clarified later when we introduce the semi-null frame.}, which is a direct consequence of the linearity of the GDE in $\xi^{\mu}$, \cite{Grasso:2018mei}. 
Indeed, if $\gamma_0$ is a null geodesic with tangent vector $\ell^{\mu}$ and $\gamma$ is a nearly displaced geodesic of the same family, with tangent vector $k^{\mu}=\ell^{\mu}+\nabla_{\ell} \xi^{\mu}$, the condition for $\gamma$ to remain null reads 
\begin{equation}
g_{\mu \nu}(\ell^{\mu}+ \nabla_{\ell} \xi^{\mu})(\ell^{\nu}+ \nabla_{\ell} \xi^{\nu})= 2 \ell_{\mu}\nabla_{\ell} \xi^{\mu}+\nabla_{\ell} \xi_{\mu}\nabla_{\ell} \xi^{\mu}=0\, , \label{eq:nulldisplaced}
\end{equation} 
where we have already removed the term $\ell_{\mu}\ell^{\mu}=0$. Since we are considering small displacements, the term $\nabla_{\ell} \xi_{\mu}\nabla_{\ell} \xi^{\mu}$ is quadratic in $\xi^{\mu}$ and it can be neglected. The null condition for $\gamma$ reduces to $\ell_{\mu}\nabla_{\ell} \xi^{\mu}=0$, that from Eq.~\eqref{eq:gde_prop1B} reads $D=0$.


\section{The bilocal geodesic operators}
\label{sec:BGO}



In the geometric optics regime, the geodesic equation~\eqref{eq:geodesicEQ} and the GDE~\eqref{eq:GDE} are the two master equations governing light propagation in the presence of curvature.
In the following we apply these concepts to describe a typical situation in observational astronomy.

Consider the physical system consisting of an observer $\calO$ and a source $\mathcal{S}$ far apart and moving freely along their timelike worldlines. 
We denote the regions of spacetime where the observer and the source are moving $N_{\mathcal{O}}$ and $N_{\mathcal{S}}$, and we suppose that $N_{\mathcal{O}}$ and $N_{\mathcal{S}}$ are causally connected, meaning that any signal emitted by $\mathcal{S}$ is received by $\mathcal{O}$ at any later time.
If  $L$ is typical size of $N_\calO$ and $N_\calE$, it must be much smaller than the characteristic curvature scale of the spacetime $\mathcal{R}$, i.e. $L \ll \mathcal{R}$.
In this case, we may 
 effectively treat both $N_\calO$ and $N_\calE$ as flat, and use special relativity to describe the effects on light propagation in these regions. From a geometric prospective, the local flatness of the two regions allows us to identify points in $N_\calO$ and $N_\calE$ with points in the corresponding tangent spaces $T_{x_\calO} \calM $ and $ T_{x_\calS} \calM$, up to quadratic terms in $x^\mu$. 

\bfi
\includegraphics[width=0.9\textwidth]{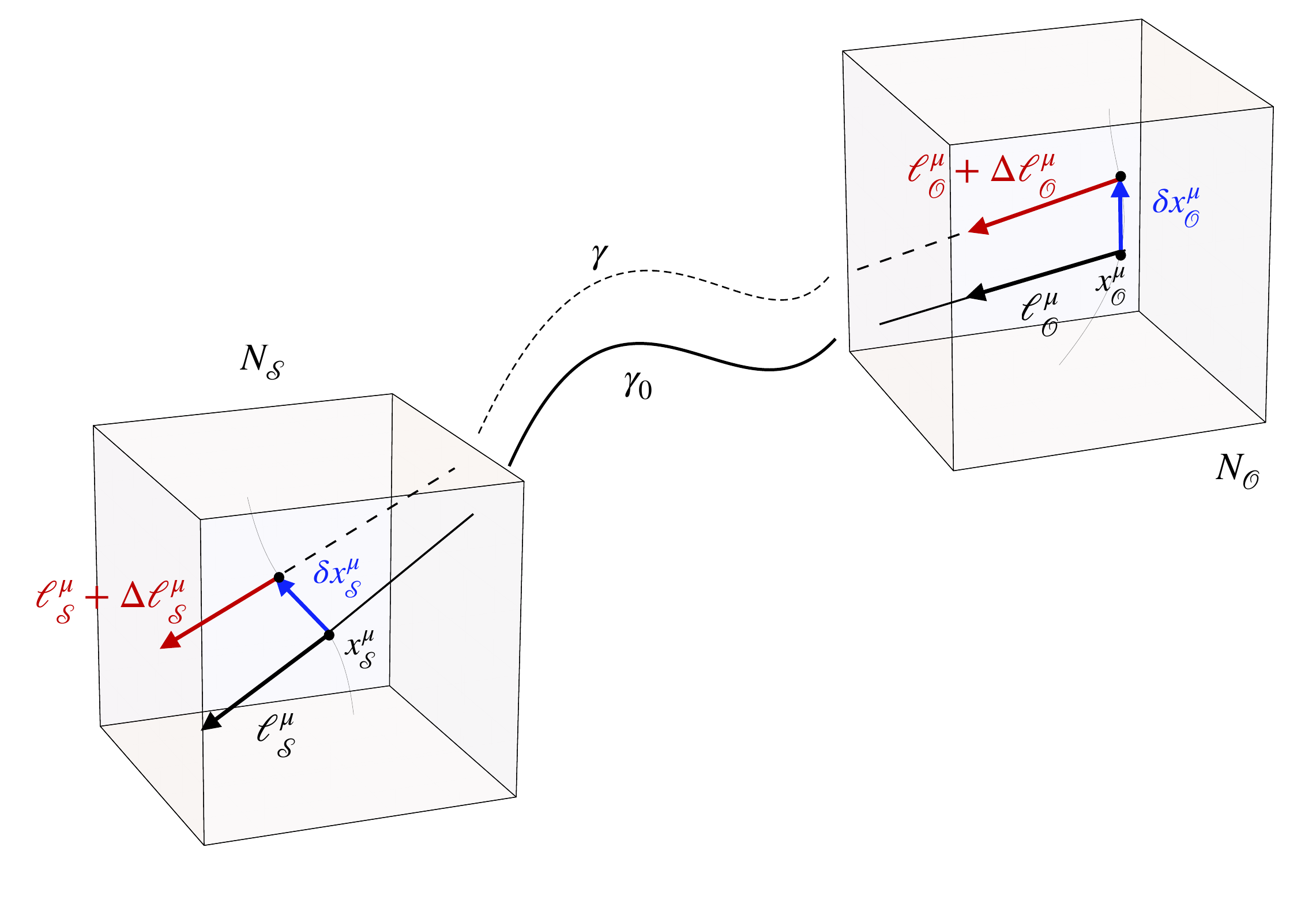}
\caption{The observer $\calO$ and the source $\calS$ are free to move along their worldlines in the two small regions $N_\calO$ and $N_\calE$. The point $x^{\mu}_{\calS}$ on the $\calS$'s worldline is connected to $x^{\mu}_{\calO}$ on the $\calO$'s worldline by the null geodesic $\gamma_0$. $\gamma_0$ can be identified by the initial position and initial tangent vector at the observer $(x^{\mu}_{\calO}, \ell^{\mu}_{\calO})$. Since the observer and source are free to move, they will be connected at a later time by another geodesic $\gamma$ identified by the displacement vectors $(\delta x_\calO^\mu, \Delta \ell_\calO^\mu)$.}
\label{fig:geometricsetup}
\efi

Since $N_\calO$ and $N_\calE$ are causally connected, it is possible to find a \emph{fiducial null geodesic} $\gamma_0$  going from $\calE$ to $\calO$, i.e. such that $\gamma(\lambda_{\mathcal{S}})=x^{\mu}_{\mathcal{S}}$ and $\gamma(\lambda_{\mathcal{O}})=x^{\mu}_{\mathcal{O}}$ are the source's and observer's positions, respectively\footnote{Note that the two values $\lambda_{\mathcal{S}}$ and $\lambda_{\mathcal{O}}$, for which $\gamma(\lambda_{\mathcal{S}})=x^{\mu}_{\mathcal{S}}$ and $\gamma(\lambda_{\mathcal{O}})=x^{\mu}_{\mathcal{O}}$, may change if we consider a different affine parameter for $\gamma_0$. However, in geometric optics we are interested only in the question whether or not a null geodesic passes through a given event and what null direction it follows at that moment.}. The vector $\ell_\calO^\mu$ is the tangent vector to $\gamma_0$ at $\calO$ and $\ell_\calE^\mu$ is the corresponding tangent vector at $\calE$.
To simplify the notation we will denote $T_{x_\calO} \calM$ by $T_\calO \calM$ and $T_{x_\calS} \calM$ by $T_\calS \calM$. 
Now, we generalise the definition of light beam by considering all null geodesics connecting points from $N_\calO$ with $N_\calE$, which are contained in a four-dimensional tube around $\gamma_0$. We consider that this tube is sufficiently narrow such that we can use the first-order geodesic deviation equation, Eq.~\eqref{eq:GDE}, for describing the deviation between these geodesics.
Within this assumptions, the geodesics are uniquely specified by giving their initial points and initial tangent vectors in $N_\calO$ (or in $N_\calE$).
Alternatively, we can characterize the geodesics by their deviation from the fiducial null geodesic $\gamma_0$: we use this second method of  identification, see Figure \ref{fig:geometricsetup}, defining the \emph{initial displacement vector} $\delta x(\lambda_\calO) \in T_\calO \calM$ as the displacement between two nearby geodesics of the family $\gamma$ ad $\gamma_0$ at $N_{\calO}$
\bea
\delta x^{\mu}(\lambda_\calO)\equiv \delta x_\calO^\mu =y^{\mu}_\calO - x^{\mu}_\calO\, , \label{eq:deltaxOdef}
\eea
where $y^{\mu}_\calO=\gamma(\lambda_\calO)$ and $x^{\mu}_\calO= \gamma_0(\lambda_\calO)$.
Note that the displacement $\delta x^{\mu}_\calO$ can be in all spatial and temporal directions.
The variation of $\delta x^{\mu}_\calO$ along the fiducial geodesic gives the \emph{initial direction deviation vector} $\Delta \ell_\calO^\mu$, defined as 
 \bea
\Delta \ell^\mu_\calO \equiv \left.\nabla_{\ell}\delta x^{\mu}(\lambda)\right|_\calO= \delta \ell_\calO^\mu + \Gamma\UD{\mu}{\nu\sigma}(x_\calO)\, \ell_\calO^\nu \,\delta x_\calO^\sigma\, , \label{eq:DeltalOdef}
\eea
where $\Delta \ell^\mu_\calO\equiv \Delta \ell^\mu(\lambda_\calO)$, $\delta \ell_\calO^\mu\equiv\delta \ell^\mu(\lambda_\calO)= \left.\frac{d \delta x^\mu}{d \lambda}\right|_\calO$, and $\Gamma\UD{\mu}{\nu\sigma}(x_\calO)$ are the Christoffel symbols at $\calO$.  
The  pair $(\delta x_\calO^\mu, \Delta \ell_\calO^\mu)$ labels all the geodesics in the vicinity of $\gamma_0$ and will be referred to as \emph{the displacement vectors}.
Let us notice that the choice of setting the initial displacement and direction deviation at $N_{\calO}$ is arbitrary. In the rest of the chapter we adopt this choice but in principle we could choose to parametrise the geodesics of the family by starting from $N_{\calE}$ giving $(\delta x_\calE^\mu, \Delta \ell_\calE^\mu)$. This second method is described  in \cite{Grasso:2021iwq} and it is one of the original results of this thesis.

Since the geodesics are expected to be confined within the narrow tube all along $\gamma_0$, the initial displacement vectors must be small. In this case, their propagation from $\calO$ to $\calE$ is described by the GDE
\bea
\nabla_{\ell} \nabla_{\ell} \delta x^{\mu} - R\UD{\mu}{\alpha\beta\nu}\,\ell^\alpha\,\ell^\beta\,\delta x^\nu = 0\, , \label{eq:GDE_delta}
\eea
with the initial data
\bea
\left. \delta x^\mu(\lambda)\right|_\calO &=& \delta x_\calO^\mu  \label{eq:GDEID1}\\
\left. \nabla_{\ell} \delta x^\mu(\lambda)\right|_\calO &=& \Delta \ell_\calO^\mu. \label{eq:GDEID2}
\eea
The solution gives the displacements at the other end for $\lambda=\lambda_\calE$: $\delta x^\mu(\lambda_\calE)=\delta x_\calE^\mu$ and $\left. \nabla_{\ell} \delta x^\mu(\lambda)\right|_\calS=\Delta \ell_\calE^\mu $. The combination $R\UD{\mu}{\alpha\beta\nu}\,\ell^\alpha\,\ell^\beta$ is the \emph{optical tidal matrix} expressing the spacetime curvature along the line of sight $\gamma_0$.
The condition for applicability of the GDE excludes the possibility of multiple imaging for light rays contained within the tube. 

Due to the linearity of the GDE, the solutions at $\lambda_{\calS}$ are given as linear combination of the initial conditions $(\delta x^{\mu}_{\calO}, \, \Delta \ell^{\mu}_{\calO})$ 
\bea
\delta x_\calE^\mu = { W_{XX} }\UD{\mu}{\nu}\,\delta x_\calO^\nu + { W_{XL} }\UD{\mu}{\nu}\,\Delta \ell_\calO^\nu \label{eq:positiondeviation1} \\
\Delta \ell_\calE^\mu = { W_{LX} }\UD{\mu}{\nu}\,\delta x_\calO^\nu + { W_{LL} }\UD{\mu}{\nu}\,\Delta \ell_\calO^\nu \label{eq:directiondeviation1},
\eea
with $\WXX$, $\WXL$, $\WLX$, $\WLL$ being bilocal operators  (also known as 2-point tesors \cite{SyngeBook} or bitensors \cite{Poisson2011, Vines:2014oba}), acting from $T_\calO \calM$ to $T_\calE \calM$. 
We refer to the four operators $W_{XX}$, $W_{XL}$, $W_{LL}$ and $W_{LX}$ as the \emph{bilocal geodesic operators} (\setwd{BGO}{acr:BGO}), \cite{Grasso:2018mei}. In the context of timelike geodesics the first two are the \emph{Jacobi propagators} $K$ and $H$ introduced in \cite{DeWittBrehme, Dixon2, Vines:2014oba}. Moreover, the BGO can also be related to the Synge's worldfunction \cite{SyngeBook, Dixon2, Vines:2014oba}. Recently the BGO defined along a timelike geodesic have  been used as a tool to study of the gravitational waves memory effect \cite{Flanagan:2018yzh}. Here, we will focus exclusively on the application of the BGO to describe null geodesics as presented in \cite{Grasso:2018mei}. The notation we introduced in Eqs.~\eqref{eq:positiondeviation1}-\eqref{eq:directiondeviation1} highlights their relation with the resolvent operator, or the Wro\'nski matrix \cite{Fleury:2014gha} for the GDE, 
\begin{equation}
\mathcal{W}=\begin{pmatrix}
\WXX{}\UD{\mu}{\nu} && \WXL{}\UD{\mu}{\sigma}\\
\WLX{}\UD{\rho}{\nu} && \WLL{}\UD{\rho}{\sigma}
\end{pmatrix}\, .
\end{equation}
$\mathcal{W} = \mathcal{W}(\calE, \calO)$ is the linear mapping between vector sums of two copies of the tangent space, i.e.
\beq
\mathcal{W}: T_\calO \calM \oplus T_\calO \calM \to T_\calE \calM \oplus T_\calE \calM\, ,
\eeq
defined by the relation 
\begin{equation}
\begin{pmatrix}
\delta x^{\mu}_{\calS}\\
\Delta \ell^{\rho}_{\calS}
\end{pmatrix}= \begin{pmatrix}
\WXX{}\UD{\mu}{\nu} && \WXL{}\UD{\mu}{\sigma}\\
\WLX{}\UD{\rho}{\nu} && \WLL{}\UD{\rho}{\sigma}
\end{pmatrix}\begin{pmatrix}
\delta x^{\nu}_{\calO}\\
\Delta \ell^{\sigma}_{\calO}
\end{pmatrix}=\mathcal{W}\begin{pmatrix}
\delta x^{\nu}_{\calO}\\
\Delta \ell^{\sigma}_{\calO}
\end{pmatrix}\, .
\label{eq:deviations_compact}
\end{equation}
It is also a symplectic mapping, as noted by Uzun \cite{Uzun:2018yes}, since in GR the ordinary differential equations (\setwd{ODE}{acr:ODE}) for null geodesics can be expressed as a Hamiltonian system, both in general and in the first-order perturbation theory \cite{Fleury:2014gha}. In contrast to other approaches, here we evaluate displacements in all four dimensions, including time. The Wro\'nski matrix formalism extends to the fully four-dimensional GDE without any problems, preserving its properties (as shown in \cite{Julius-spherical}).

It follows easily from the geodesic deviation equation (\ref{eq:GDE_delta}) and from Eqs.~(\ref{eq:positiondeviation1})-(\ref{eq:directiondeviation1}) that the BGO can be expressed as solutions to the GDE along $\gamma_0$ and with initial data at $\lambda_\calO$. 
Let us start by using the fact $\nabla_{\ell} \delta x^{\mu}(\lambda)=\Delta \ell^{\mu}(\lambda)$ to express the GDE~\eqref{eq:GDE_delta} as a system of two first-order ODE as
\begin{equation}
\left\{ \begin{matrix}
\nabla_{\ell} \delta x^{\mu}= \Delta \ell^{\mu}\\
\\
\nabla_{\ell} \Delta \ell^{\rho}= R\UD{\rho}{\alpha\beta\nu}\,\ell^\alpha\,\ell^\beta \delta x^{\nu}
\end{matrix}\right.\, , 
\end{equation}
with initial conditions
\begin{align}
\delta x^{\mu}(\lambda_{\mathcal{O}})&=\delta x^{\mu}_{\mathcal{O}}\\
\Delta \ell^{\rho}(\lambda_{\mathcal{O}})&= \Delta \ell^{\rho}_{\mathcal{O}}\, .
\end{align}
In a more compact form the system becomes
\begin{equation}
\nabla_{\ell} \begin{pmatrix}
\delta x^{\mu}\\
\Delta \ell^{\rho}
\end{pmatrix}= \begin{pmatrix}
0 && \delta\UD{\mu}{\sigma}\\
R\UDDD{\rho}{\ell}{\ell}{\nu} && 0
\end{pmatrix}\begin{pmatrix}
\delta x^{\nu}\\
\Delta \ell^{\sigma}
\end{pmatrix}\, ,
\label{eq:GDE_deviations}
\end{equation}
where we defined $R\UD{\rho}{\alpha\beta\nu}\,\ell^\alpha\,\ell^\beta=R\UD{\rho}{\ell \ell \nu}$.
Making use of Eq.~\eqref{eq:deviations_compact}, one obtains a matrix ODE for $\mathcal{W}$
\begin{equation}
\nabla_{\ell} \mathcal{W}= \begin{pmatrix}
0 && \mathbb{1} \\
R_{\ell \ell} && 0
\end{pmatrix} \cdot \mathcal{W}\, ,
\label{eq:GDE_W}
\end{equation}
where the dot indicates usual matrix product operation. The initial conditions for the BGO are
\begin{equation}
\mathcal{W}= \mathbb{1}_{8 \times 8}\, .
\end{equation}

Let us clarify that the action of the covariant derivative is intended as acting on bitensors \cite{Poisson2011}, i.e. 
\begin{equation}
\nabla_{\ell} \mathcal{W}= \begin{pmatrix}
\partial_{\ell} \WXX{}\UD{\mu}{\nu}+ \Gamma^{\mu}_{\alpha \beta}\WXX{}\UD{\alpha}{\nu}\ell^{\beta}  && \partial_{\ell} \WXL{}\UD{\mu}{\sigma}+ \Gamma^{\mu}_{\alpha \beta}\WXL{}\UD{\alpha}{\sigma}\ell^{\beta}  \\
\partial_{\ell} \WLX{}\UD{\rho}{\nu} + \Gamma^{\rho}_{\alpha \beta}\WLX{}\UD{\alpha}{\nu}\ell^{\beta}  && \partial_{\ell} \WLL{}\UD{\rho}{\sigma}+ \Gamma^{\rho}_{\alpha \beta}\WLL{}\UD{\alpha}{\sigma}\ell^{\beta}
\end{pmatrix}\, .
\end{equation}

The relations in Eq.~\eqref{eq:GDE_W} are the evolution equations for the BGO along $\gamma_0$. They show that the BGO can be expressed as non-local functionals of the Riemann tensor along the line of sight.  
Even though the GDE and the matrix equations~\eqref{eq:GDE_W} are linear, the BGO are \emph{nonlinear} functionals of the curvature tensor along $\gamma_0$. 
This can be easily checked as follows: let us take two solutions of Eq.~\eqref{eq:GDE_W}, $\mathcal{W}^{(1)}$ and $\mathcal{W}^{(2)}$, each ones corresponding to two different optical tidal tensor functions,
$R^{(1)}_{\ell \ell}$ and $R^{(2)}_{\ell \ell}$
\begin{align}
\nabla_{\ell} \mathcal{W}^{(1)}=& \begin{pmatrix}
0 && \mathbb{1} \\
R^{(1)}_{\ell \ell} && 0
\end{pmatrix} \mathcal{W}^{(1)} \\
\nabla_{\ell} \mathcal{W}^{(2)}=& \begin{pmatrix}
0 && \mathbb{1} \\
R^{(2)}_{\ell \ell} && 0
\end{pmatrix} \mathcal{W}^{(2)}\, .
\end{align}
Then let us see if the linear combination of these solutions $a \mathcal{W}^{(1)}+b \mathcal{W}^{(2)}$, with $a, \, b\, \in \mathbb{R}$, is also a solution of Eq.~\eqref{eq:GDE_W} for the same linear combination of the optical tidal tensor functions, i.e.
\begin{equation}
\nabla_{\ell} (a \mathcal{W}^{(1)}+ b \mathcal{W}^{(2)}) = \begin{pmatrix}
0 && \mathbb{1} \\
a R^{(1)}_{\ell \ell}+ b R^{(2)}_{\ell \ell} && 0
\end{pmatrix} (a \mathcal{W}^{(1)} + b \mathcal{W}^{(2)})\, .
\end{equation}
Rearranging the terms we finally get
\begin{align}
\nonumber a\left[\nabla_{\ell} \mathcal{W}^{(1)}- \begin{pmatrix}
0 && \mathbb{1} \\
a R^{(1)}_{\ell \ell} && 0
\end{pmatrix} \mathcal{W}^{(1)}\right]+ & b \left[\nabla_{\ell} \mathcal{W}^{(2)}- \begin{pmatrix}
0 && \mathbb{1} \\
b R^{(1)}_{\ell \ell} && 0
\end{pmatrix} \mathcal{W}^{(2)}\right] =\\
& b \begin{pmatrix}
0 && 0 \\
a R^{(1)}_{\ell \ell} && 0
\end{pmatrix}  \mathcal{W}^{(2)} +a \begin{pmatrix}
0 && 0 \\
 b R^{(2)}_{\ell \ell} && 0
\end{pmatrix}  \mathcal{W}^{(1)}\, , \label{eq:gde_nonlin}
\end{align}
from which it is easy to see that a linear combination of the solutions of~\eqref{eq:GDE_W} \emph{does not} satisfy the same equations proving the nonlinearity of the BGO with respect to the optical tidal tensor\footnote{In the case $a=b=1$, the left hand side of Eq~\eqref{eq:gde_nonlin} vanishes while the right hand side is in general non zero.}.
The nonlinearity of the BGO with respect to the curvature reflects the fact that, although they describe small deviations from the fiducial geodesics, the BGO captures all nonlinear effects of light bending combined along $\gamma_0$.

\subsection{Algebraic properties of the BGO.} 
From its very definition, the $\mathcal{W}$ matrix satisfies the following properties
\begin{align}
\mathcal{W}(\mathcal{O}, \mathcal{S})&=\mathcal{W}^{-1}(\mathcal{S}, \mathcal{O}) \label{eq:Winversion}\\
\mathcal{W}(\mathcal{S}, \mathcal{O})&=\mathcal{W}(\mathcal{S}, p_{\lambda}) \, \mathcal{W}(p_{\lambda}, \mathcal{O})\, ,
\label{eq:Wcomposition}
\end{align}
with $p_{\lambda}$ an arbitrary point on the fiducial geodesic $\gamma_0$, \cite{Grasso:2018mei}. We also have that $\mathcal{W}$ is a symplectic mapping, \cite{Uzun:2018yes}.

Moreover, the properties of the GDE \eqref{eq:gde_prop1A}-\eqref{eq:gde_prop2} can be immediately translated to corresponding properties of the BGO, which hold irrespective of the spacetime geometry or whether $\gamma_0$ is null or not, \cite{Grasso:2018mei}. 
From the first property, Eqs.~\eqref{eq:gde_prop1A} and~\eqref{eq:gde_prop1B}, we have that for any initial data $\delta x_\calO^\mu$ and $\Delta \ell_\calO^\mu$ the values of $C$ and $D$ need to remain equal in $\calO$ and $\calE$. This means that
\begin{align}
\ell_{\calO\,\mu}\,\Delta \ell_\calO^\mu &= \ell_{\calE\,\mu}\,\Delta \ell_\calE^\mu \label{eq:Dl_conserv}\\
\ell_{\calO\,\mu}\,\delta x_\calO^\mu - \lambda_\calO\,\ell_{\calO\,\mu}\,\Delta \ell_\calO^\mu &= \ell_{\calE\,\mu}\,\delta x_\calE^\mu - \lambda_\calE\,\ell_{\calE\,\mu}\,\Delta \ell_\calE^\mu\, . \label{eq:dx_conserv}
\end{align}
We make use of Eqs.(\ref{eq:positiondeviation1})-(\ref{eq:directiondeviation1}) in order to express $\delta x_\calE^\mu$ and $\Delta \ell_\calE^\mu$ by $\delta x_\calO^\mu$ and $\Delta \ell_\calO^\mu$.
The resulting equations are equivalent to the following 4 relations:
\bea
\ell_{\calE\,\mu}\,{ W_{XX} }\UD{\mu}{\nu} 
 &=& \ell_{\calO\,\nu} \label{eq:Wprop5} \\
\ell_{\calE\,\mu}\,{ W_{XL} }\UD{\mu}{\nu} &=& (\lambda_\calE - \lambda_\calO)\,\ell_{\calO\,\nu} \label{eq:Wprop7} \\
\ell_{\calE\,\mu}\,{ W_{LX} }\UD{\mu}{\nu} &=& 0 \label{eq:Wprop6} \\
\ell_{\calE\,\mu}\,{ W_{LL} }\UD{\mu}{\nu} &=& \ell_{\calO\,\mu} \,.  \label{eq:Wprop8}
\eea 
The ``inverted'' relations are obtained by considering the solution (\ref{eq:gde_prop2}) at $\calO$ and $\calE$: we have $\delta \tilde{x}_\calO^\mu = \delta x_\calO^\mu + (E + \lambda_\calO\, F)\,\ell_\calO^\mu$, 
$\Delta \tilde{\ell}_\calO^\mu =  \Delta \ell_\calO^\mu + F\,\ell_\calO^\mu$ and $\delta \tilde{x}_\calE^\mu =  \delta x_\calE^\mu + (E + \lambda_\calE\,F)\,\ell_\calE^\mu$,  $\Delta \tilde{\ell}_\calE^\mu = \Delta \ell_\calE^\mu + F\,\ell_\calE^\mu$. We substitute these equations to (\ref{eq:positiondeviation1})-(\ref{eq:directiondeviation1}) and 
assuming the resulting relations must hold for all $E$ and $F$ we get
\bea
{ W_{XX} }\UD{\mu}{\nu}\,\ell_\calO^\nu &=& \ell_\calE^\mu \label{eq:Wprop1}\\
{ W_{XL} }\UD{\mu}{\nu}\,\ell_\calO^\nu &=& (\lambda_\calE - \lambda_\calO)\,\ell_\calE^\mu \label{eq:Wprop3} \\
{ W_{LX} }\UD{\mu}{\nu}\,\ell_\calO^\nu &=& 0 \label{eq:Wprop2} \\
{ W_{LL} }\UD{\mu}{\nu}\,\ell_\calO^\nu &=& \ell_\calE^\mu \label{eq:Wprop4}\, .
\eea

Finally, let us note that two of the BGO undergo rescaling under the affine reparametrisations of the fiducial null geodesic $\gamma_0$. Namely, under the transformation Eq.~\eqref{eq:affineparameter} we have the following rescaling for the BGO\footnote{The derivation of these relations uses the fact that under the affine reparametrisation the tangent to the geodesic transforms as Eq.~\eqref{eq:laffine}. Then one can use the GDE, Eq.~\eqref{eq:GDE_W}, together with Eqs.~\eqref{eq:Wprop1}-\eqref{eq:Wprop4} to obtain the relations Eqs.~\eqref{eq:Wresc1}-\eqref{eq:Wresc4}.}
\begin{align}
 W_{XX} &\to \tilde{W}_{XX} = W_{XX} \label{eq:Wresc1}\\
 W_{XL} &\to  \tilde{W}_{XL} = A\cdot W_{XL} \label{eq:Wresc2}\\ 
 W_{LX} &\to  \tilde{W}_{LX} = \frac{1}{A} \cdot W_{LX} \label{eq:Wresc3}\\
 W_{LL} &\to  \tilde{W}_{LL} = W_{LL}\label{eq:Wresc4}\, .
\end{align}

\subsection{BGO for light propagation and the quotient space}
The considerations we have made so far are independent of the character of the geodesics we are considering, so they are valid for BGO describing the properties of all types of geodesics in the vicinity of $\gamma_0$. 
Since our goal is the application of the BGO to describe light propagation, from now on we will consider only families of null geodesics.
Let us start by noting that the BGO distinguish differently parametrised geodesics sharing the same path. However, from the point of view of geometric optics, an affine reparametrisation of null geodesics is just a gauge freedom, as we have already discussed in Eq.\eqref{eq:gde_prop2}. The change in the affine parameter of the geodesics around $\gamma_0$ transforms solutions of the GDE $\xi^{\mu}$ as $\tilde{\xi}^{\mu} = \xi^{\mu}+\alpha(\lambda) \ell^{\mu}$, with $\xi^{\mu}$ and $\tilde{\xi}^{\mu}$ pointing at the same displaced geodesic. Here, we want to isolate this gauge freedom in order to consider only geodesics having different path. In other words, we want to identify the initial data for which the position and directional deviations only differ by a multiple of  $\ell_\calO^\mu$
\bea
\left(\begin{array}{l}
\delta x_\calO^\mu\\
\Delta \ell_\calO^\mu
\end{array} \right) \sim \left(\begin{array}{l}
\delta x_\calO^\mu +  C_1 \ell_\calO^\mu \\
\Delta \ell_\calO^\mu +  C_2 \ell_\calO^\mu
\end{array} \right)\, , \label{eq:identification1}
\eea
for some non-vanishing constants $C_1$ and $C_2$. Substituting Eq.~\eqref{eq:identification1} in Eqs.~\eqref{eq:positiondeviation1}-\eqref{eq:directiondeviation1}, and making use of Eqs.~\eqref{eq:Wprop1}-\eqref{eq:Wprop4}, we obtain that adding this type of terms in $N_\calO$ leads to a similar change in the final data
\bea
\left(\begin{array}{l}
\delta x_\calE^\mu\\
\Delta \ell_\calE^\mu
\end{array} \right) \sim \left(\begin{array}{l}
\delta x_\calE^\mu +  D_1\,\ell_\calE^\mu \\
\Delta \ell_\calE^\mu +  D_2\,\ell_\calE^\mu,
\end{array} \right)\, , \label{eq:equiv3}
\eea
with constants $D_1$ and $D_2$ related to $C_1$ and $C_2$.
The relation $Y^{\mu} \sim X^{\mu}$ in Eqs.~\eqref{eq:identification1}-\eqref{eq:equiv3} defines equivalence classes $[X]$ in both $T_\calO \calM$ and $T_\calS \calM$. To remove this gauge freedom, one can consider the BGO as maps between vectors $[X]$ in the quotient spaces $\calQ_\calO = T_\calO \calM / \ell_\calO$ and $\calQ_\calE = T_\calE \calM / \ell_\calE$, see Fig.~\ref{fig:quotientspaces}.
The equivalence relation $Y^{\mu} \sim X^{\mu}$ effectively suppresses one dimension (the one along the tangent vector) in the tangent spaces in $N_{\calS}$ and $N_{\calO}$, leaving only three non-trivial directions in $\mathcal{Q}$. 

Let us consider the subspaces $\ell_{\calO}^\perp \subset T_\calO \calM$ and $\ell_{\calS}^\perp \subset T_\calS \calM$ consisting of all vectors orthogonal to $\ell_{\calO}^\mu$ and $\ell_{\calS}^\mu$, respectively. In analogy to $\calQ$, it is possible to define the two-dimensional \emph{perpendicular spaces} $\calP_\calO = \ell_\calO^\perp / \ell_\calO$ and $\calP_\calE = \ell_\calE^\perp / \ell_\calE$ as the subspaces orthogonal to $\ell_\calO^\mu$ and $\ell_\calE^\mu$, respectively \cite{Korzynski:2018, Grasso:2018mei}. 
The physical meaning of the quotient spaces $\calQ$ and $\calP$ will be clear once we introduce a reference frame.
\bfi
\centering
\includegraphics[width=\linewidth]{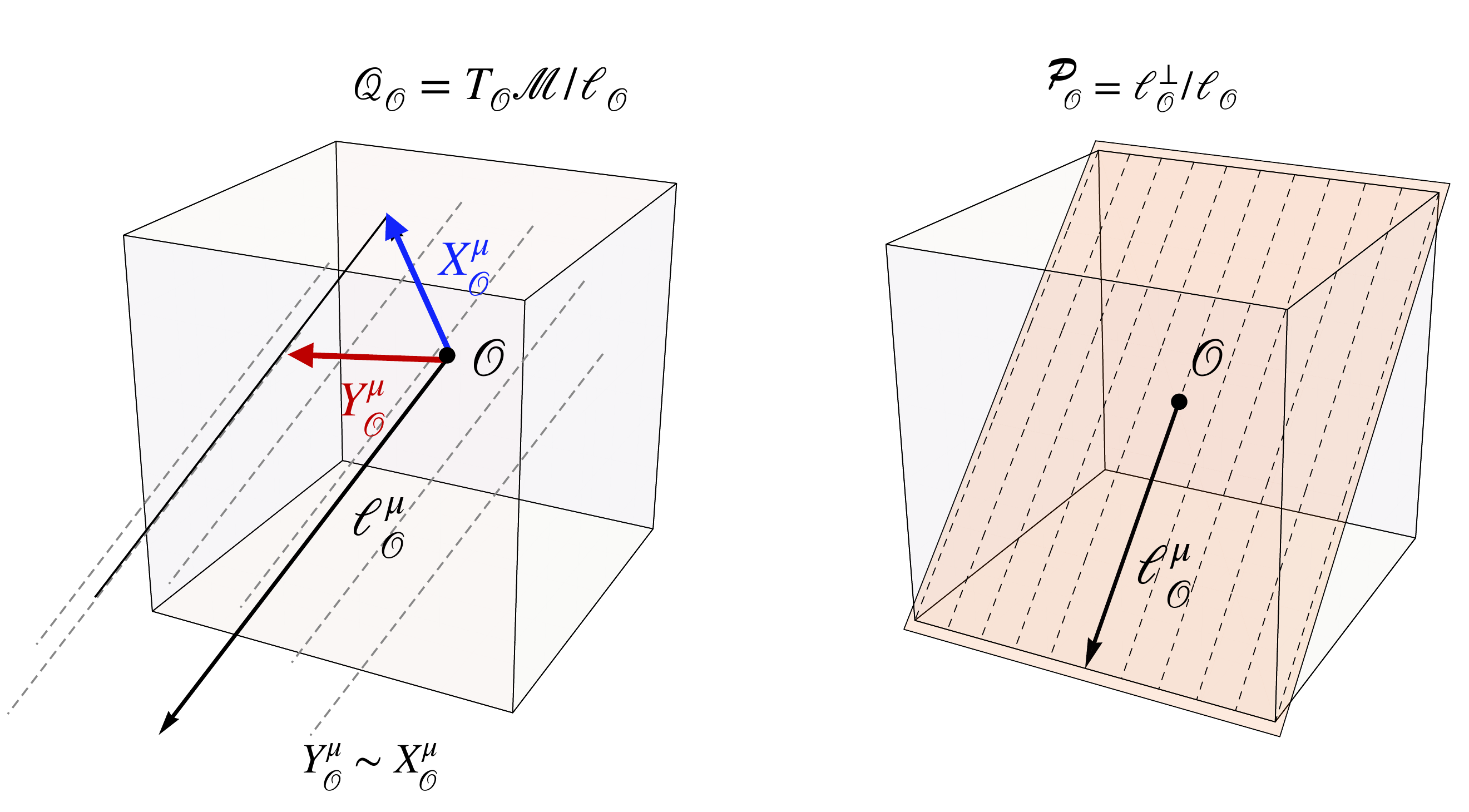}
\caption{Geometry of the quotient spaces $\calQ_\calO$ and $\calP_\calO$. Elements $[X] \in \calQ_\calO$ correspond to the vectors $X^{\mu}$ and $Y^{\mu}$ in $T_\calO \calM$ identified by the relation $Y^{\mu} \sim X^{\mu}$, i.e. such that $Y^{\mu}=X^{\mu}+c\ell^{\mu}_{\calO}$. Geometrically is the space containing all geodesics in $N_\calO$ parallel to $\gamma_0$. Elements $\bm{X} \in \calP_\calO$ corresponds to the vectors $X^{\mu}$ and $Y^{\mu}$ in $T_\calO \calM$, which are perpendicular to $\ell_\calO$ (i.e. $X^{\mu}\ell_\calO{}_{\mu}=0$ and $Y^{\mu}\ell_\calO{}_{\mu}=0$) and identified by $Y^{\mu} \sim X^{\mu}$. Geometrically, $\calP_\calO$ is the space containing the geodesics parallel to $\gamma_0$ and lying on the null hypersurface orthogonal to $\ell_\calO$.}
\label{fig:quotientspaces}
\efi


\subsection{The semi-null frame and the observer's sky}

Up to this point, the GDE formulation in terms of bilocal operators we just presented is completely covariant, namely it was derived without invoking explicitly a coordinate system or a frame of reference. However, in order to relate the BGO to actual observable quantities, we need to introduce a reference frame. 
In the context of relativistic geometric optics, it is customary to introduce a frame that relates to the results of observations at $\calO$. The standard approach is to use the Sachs orthonormal frame, consisting of the observer four-velocity $u_\calO^\mu$, the direction vector $r^\mu$, being the direction from which the observer sees the light coming, and two perpendicular, spatial vectors $\phi_{\bm{A}}^\mu$ spanning the so called Sachs screen space \cite{perlick-lrr, sachs, ehlers-jordan-sachs}. However, here we use a different frame called the \emph{semi-null frame} (\setwd{SNF}{acr:SNF}) \cite{Grasso:2018mei}, in which the two quotient spaces $\calQ$ and $\calP$ have a simple physical interpretation. The SNF consists of $u_\calO^\mu$, the same two perpendicular, spatial vectors $\phi_{\bm{A}}^\mu$ and the null vector $\ell^\mu$ instead of $r^\mu$. It is not orthonormal and we can check that the products of the basis vectors read
\begin{align}
\nonumber \ell^\mu\,\ell_\mu &= 0 \\
\nonumber \phi_{\bm{A}}^\mu\,\ell_\mu &= 0 \\
\nonumber u_\calO^\mu\,\phi_{\bm{A}\,\mu} &= 0 \\
 \phi_{\bm{A}}^\mu\,\phi_{\bm{B}\,\mu} &= \delta_{\bm{A}\bm{B}} \label{eq:SNF_relations}\\
\nonumber u_\calO^\mu\,u_{\calO\,\mu} &= -1 \\
\nonumber \ell^\mu\,u_{\calO\,\mu} &= Q\, , 
\end{align}
where we have introduced the constant\footnote{Note that the sign of the constant $Q$ depends on the temporal orientation of $\ell^{\mu}$. The standard convention in cosmology is to consider the tangent vector past-oriented, from the observer $\calO$ to the source $\calS$, which cause to have $Q>0$. However, we will later consider the case when $\ell^{\mu}$ is future-oriented and in that case $Q<0$.} $Q$ for the product of $\ell^\mu$ and $u_\calO^\mu$.
We denote the frame indices by boldface letters: capital Latin indices $\bm{A}$, $\bm{B}, \ldots$,  running over the spatial components $\bm{1}$ and $\bm{2}$,
lower case Latin indices $\bm{i}$, $\bm{j}, \ldots$, running over $\bm{1}$, $\bm{2}$ and $\bm{3}$, and the boldface Greek indices $\bm{\mu}$, $\bm{\nu}, \ldots$,  running over all 4 dimensions from $\bm{0}$ to $\bm{3}$.
The associated coframe $\psi^{\bm{\alpha}}_{\mu}$ is composed of the tetrad of vectors $\psi^{\bm{\alpha}}_{\mu}=(\frac{\ell_{\mu}}{Q}, \phi^{\bm{A}}_{\mu}, \frac{u_{\mu}}{Q}+\frac{\ell_{\mu}}{Q^2})$. In the SNF the displacement vector $\delta x^{\mu}$ has components $\delta x^{\bm{\mu}}=\delta x^{\nu}\hat{\psi}^{\bm{\mu}}_{\nu}=(\frac{\delta x^{\nu}\ell_{\nu}}{Q}, \delta x^{\nu}\hat{\phi}^{\bm{A}}_{\nu}, \frac{\delta x^{\nu}\hat{u}_{\calO\, \nu}}{Q}+\frac{\delta x^{\nu}\ell_{\nu}}{Q^2})$, where hatted vectors are parallel transported along the fiducial geodesic $\gamma_0$.

In the SNF the presence of the quotient spaces have a natural explanation: the first three components of the GDE (\ref{eq:GDE_delta}) in the SNF\footnote{The optical tidal matrix in the SNF components is defined as $R\UD{\bm{\mu}}{\ell \ell \bm{\nu}}=\hat{\psi}^{\bm{\mu}}_{\rho}R\UD{\rho}{\ell \ell \sigma}\hat{\phi}^{\sigma}_{\bm{\nu}}$ and from the symmetries of the Riemann tensor follows that $R\UD{\bm{0}}{\ell \ell \bm{\nu}}=R\UD{\bm{\mu}}{\ell \ell \bm{3}}=0$.} 
\begin{align}
\frac{d^2 \delta x^{\bm{0}}}{ d \lambda^2} = & 0 \label{eq:gdeSNF0}\\
\frac{d^2 \delta x^{\bm{A}}}{ d \lambda^2} = & R\UD{\bm{A}}{\ell \ell \bm{0}}\delta x^{\bm{0}}+R\UD{\bm{A}}{\ell \ell \bm{B}}\delta x^{\bm{B}} \label{eq:gdeSNFA}\, ,
\end{align}
decouple from the fourth one, $\frac{d^2 \delta x^{\bm{3}}}{ d \lambda^2} = R\UD{\bm{3}}{\ell \ell \bm{0}}\delta x^{\bm{0}}+R\UD{\bm{3}}{\ell \ell \bm{B}}\delta x^{\bm{B}}$. 
Moreover, from $\frac{d^2 \delta x^{\bm{0}}}{ d \lambda^2} = 0$ we have that $\delta x^{\bm{0}}=\frac{\delta x^{\mu}\ell_{\mu}}{Q}= \const $, according to Eq.~\eqref{eq:gde_prop1A} for null geodesics in the flat lightcones approximation, in which we have  $\frac{d \delta x^{\bm{0}}}{ d \lambda} \propto \ell_{\mu}\Delta \ell^{\mu}=0$. 

Geometrically, the condition $\ell_{\calO\,\mu}\,\delta x_{\calO}^\mu = \ell_{\calE\,\mu}\,\delta x_{\calE}^\mu $ defines foliations of $N_\calO$ and $N_\calE$ by families of null hypersurfaces, see Fig.~\ref{fig:foliations}, implying that the observers located on a leaf in $N_\calO$ can only perceive the events lying on the corresponding leaf in $N_\calE$, as explained in \cite{Grasso:2018mei}.
\bfi
\centering
\includegraphics[width=.9\textwidth]{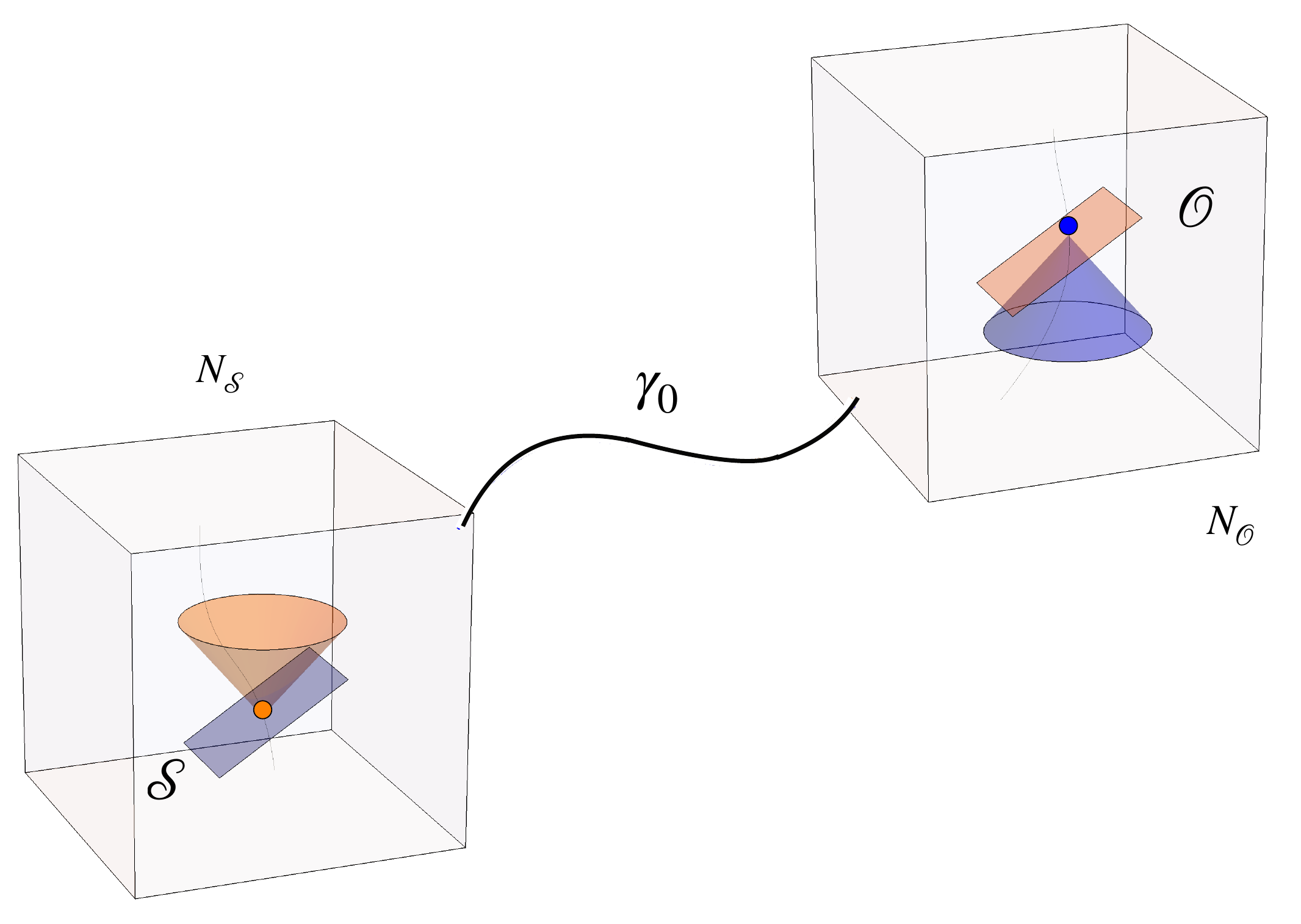}
\caption{The past light cone (blue) in $N_{\calO}$ degenerates to the flat null hypersurface (blue plane) in $N_{\calS}$. Similarly, the future light cone (orange) in $N_{\calS}$ degenerates to the flat null hypersurface (orange plane) in $N_{\calO}$. $\calO$ can observe only those events on the corresponding hypersurface in $N_{\calS}$.}
\label{fig:foliations}
\efi
The interpretation of the two null foliations is straightforward: at two ends of $\gamma_0$ the foliations $\delta x^{\mu}\ell_{\mu}= \const$  are the degenerate families of light cones centred at the opposite ends of $\gamma_0$. In other words, the past lightcone of the point $p$ in $N_\calO$ degenerate to a flat hypersurface in $N_\calE$ due to the large distance between the two regions and their small size. Similarly, the future light cone of any point on that null hypersurface will degenerate to the null hypersurface containing $p$ in $N_\calO$. These clarifies the name used for the condition $\ell_{\mu}\Delta \ell^{\mu}=0$. 
In the special case $\delta x^{\bm{0}}=0$, Eqs.~\eqref{eq:gdeSNFA} decouple from Eq.~\eqref{eq:gdeSNF0}, and their solutions $(\delta x^{\bm{A}}, \Delta \ell^{\bm{A}})$ form a subspace of solutions in $\calP$.

Vectors expressed in the SNF have a very simple representation in the quotient spaces $\calQ_\calO = T_\calO \calM / \ell_\calO$ and $\calQ_\calS = T_\calS \calM / \ell_\calS$. In fact, for vectors $[X] \in \calQ$ we can ``forget'' about the fourth component $X^{\bm 3}$, namely $(X^{\bm 0}, X^{\bm 1}, X^{\bm 2}, X^{\bm 3}) \to (X^{\bm 0}, X^{\bm 1}, X^{\bm 2})$. Similarly, vectors in any perpendicular subspace $\calP$ along $\gamma_0$ have additionally vanishing first component, i.e. $(0,X^{\bm 1}, X^{\bm 2})$. 

The introduction of a frame is mandatory to perform measurements like the positions of celestial objects: the observer $\calO$ sees the source $\calS$ in the direction corresponding to the line of sight $\gamma_0$ as
\begin{equation}
r^\mu_0 = \frac{1}{\ell_\sigma\,u_\calO^\sigma}\,\ell^\mu + u_\calO^\mu\, . \label{eq:dir}
\end{equation}
The direction $r^\mu_0$ serves as a reference point on the observer screen, indeed in the SNF gives $r^{\bm{A}}_0=(0,0)$. Similarly, for the geodesic $\gamma$ emitted by a another source $\mathcal{S}'$ in $N_{\calS}$, the observer sees the light from the direction 
\begin{equation}
r^\mu=\frac{1}{k_\sigma\,u_\calO^\sigma}\,k^\mu + u_\calO^\mu\, ,
\end{equation}
where $k^{\mu}$ is the tangent vector of the geodesic. 
The position of $\mathcal{S}'$ is identified by the observer $\calO$ measuring the angle between $r^\mu_0$ and $r^\mu$, that for a source which lies close\footnote{The approximation is valid for $\delta\theta^{\bm A} \ll 1\, {\rm rad}$. The relation for larger angles requires the use of the standard trigonometric formulae.} to $r_0^\mu$ is simply
\bea
 \delta \theta^{\bm A} \approx r^{\bm A}\, . \label{eq:positionapprox}
\eea
Therefore, all objects in the observer's view can be conceived as being projected on an ideal sphere, representing the observer's sky, where the apparent positions of objects in the sky are determined from the transversal components of $r^\mu$ in the semi-null frame of the observer, denoted as $r^{\bm{A}}$.
The expression Eq.~\eqref{eq:dir} defines an observer-dependent mapping from the set of null tangent vectors ${\cal N}_\calO = \left\{ \ell \in T_\calO \calM | \ell^\mu\,\ell_\mu = 0, \ell^0 < 0 \right\}$ to the observer's sky of directions $\textrm{Dir}(u_\calO) = \left\{ r\in T_\calO \calM | r^\mu\,r_\mu = 1, u_\calO^\mu\,r_\mu = 0 \right\}$, i.e. the set of normalised, purely spatial vectors for the observer \cite{perlick, low, Korzynski:2018, Grasso:2018mei}.

The introduction of the observer's sky provides an observer-dependent method to identify the apparent position of sources. Now we will see how observations made by observers with different four-velocities and at different points in $N_\calO$ may be compared. This is not a simple task in a generic spacetime because the position on the sky is a vector in the observer-dependent space of directions. Let us split the task in two simpler problems:
\begin{enumerate}
  \item how do we compare position vectors at different points, 
  \item how do we compare directions on the sky measured by
observers boosted with respect to each other.
\end{enumerate}
The first issue is overcome by the assumed flatness of $N_\calO$, which allows to use the parallel propagation of the frame for $\calO$ to identify $T_\calO \calM$ as the tangent space at all points. 
Thus, we can introduce a parallel propagated SNF $\phi_{\bm{\alpha}}^\mu=(u_\calO^\mu, \phi_{\bm{A}}^\mu, \ell_\calO^\mu)$ from $\calO$ throughout the whole region $N_\calO$ to compare vector or tensor defined at different points.  
From now on all equations are expressed in this type of parallel transported SNF at $N_\calO$ (and a similar one at $N_\calE$). The tangent vector of $\gamma_0$ is $\ell^{\mu}_{\calO}$, while for the other null geodesics the tangent vector is simply $k_\calO^\mu = \ell^\mu_\calO + \Delta \ell_\calO^\mu$. Then, the direction vector $r^{\bm{\mu}}$ in the SNF of the observer $u_\calO^\mu$ become 
\begin{align}
 r^{\bm{A}} =r^{\mu}\phi_{\mu}^{\bm{A}}&=\left(\dfrac{\ell^\mu_\calO + \Delta \ell_\calO^\mu}{u_{\calO\,\sigma}\,\left(\ell_\calO^\sigma + \Delta \ell_\calO^\sigma\right)}+u^{\mu}_{\calO}\right)\phi_{\mu}^{\bm{A}}= \frac{\Delta \ell^{\bm{A}}_\calO}{u_{\calO\,\sigma}\,\left(\ell_\calO^\sigma + \Delta \ell_\calO^\sigma\right)} \nonumber \\
  & = \frac{\Delta \ell^{\bm{A}}_\calO}{u_{\calO\,\sigma}\,\ell_\calO^\sigma\left(1+\frac{u_{\calO\,\sigma}\,\Delta \ell_\calO^\sigma}{u_{\calO\,\sigma}\,\ell_\calO^\sigma}\right)}= \frac{\Delta \ell^{\bm{A}}_\calO}{u_{\calO\,\sigma}\,\ell_\calO^\sigma}\,\left(1 - \frac{\Delta \ell_\calO^{\bm{0}}}{u_{\calO\,\sigma}\,\ell_\calO^\sigma} + \Delta \ell_\calO^{\bm{3}}\right)^{-1}\, , \label{eq:position1}
\end{align}
where in the last equality we have used Eq.~\eqref{eq:SNF_relations} to express
\begin{equation}
u_{\calO\,\sigma}\,\Delta \ell_\calO^\sigma= u_{\calO\,\sigma}\,(\phi^{\sigma}_{\bm{\nu}}\Delta \ell_\calO^{\bm{\nu}})=- \Delta \ell_\calO^{\bm{0}}+u_{\calO\,\sigma}\,\ell_{\calO}^{\sigma}\Delta \ell_\calO^{\bm{3}}\, .
\end{equation}
Let us remark that we are in the regime of the first-order GDE, so we only need to consider linear terms in the displacement (and its derivatives). This is indeed the condition of the FLA, i.e. $\ell_{\calO\, \mu}\Delta\ell^{\mu}_{\calO}=0$, that we have considered in Eq.~\eqref{eq:nulldisplaced}. If we express this condition in the SNF we have $0=\ell_{\calO\, \mu}(\phi^{\mu}_{\bm{\nu}}\Delta\ell^{\bm{\nu}}_{\calO})=\ell_{\calO\, \mu}(u^{\mu}_{\calO}\Delta\ell^{\bm{0}}_{\calO})$, or in other words the FLA gives that $\Delta\ell^{\bm{0}}_{\calO}$ is an higher-order correction.
Now, we can expand Eq.~\eqref{eq:position1} in terms of the components of the direction deviation vector $\Delta \ell$, to obtain
\begin{equation}
r^{\bm{A}} = \frac{\Delta \ell^{\bm{A}}_\calO}{u_{\calO\,\sigma}\,\ell_\calO^\sigma}\,\left(1 - \Delta \ell_\calO^{\bm{3}}\right)+ h.o.t.=\frac{\Delta \ell^{\bm{A}}_\calO}{u_{\calO\,\sigma}\,\ell_\calO^\sigma} + h.o.t.\, \label{eq:direction_rA}
\end{equation}
From Eq.~\eqref{eq:positiondeviation1}, the direction deviation vector $\Delta \ell^{\bm{A}}_\calO$ is related to the BGO as
\begin{equation}
\Delta \ell^{\bm{A}}_\calO=(\WXL^{-1}){}\UD{\bm{A}}{\bm{\nu}}\,\left[ \delta x^{\bm{\nu}}_{\calS}-\WXX{}\UD{\bm{\nu}}{\bm{\sigma}}\delta x^{\bm{\sigma}}_{\calO} \right]\, .\label{eq:Dl_inBGO_with_Ps}
\end{equation}
Should be emphasised that the vector $\left[ \delta x^{\mu}_{\calS}-\WXX{}\UD{\mu}{\sigma}\delta x^{\sigma}_{\calO} \right]$ is an element of the perpendicular space $\ell^{\perp}_{\calS}$, in fact from Eq.~\eqref{eq:dx_conserv} for null geodesics in the FLA and Eq.~\eqref{eq:Wprop5} follow that \cite{Grasso:2018mei}
\begin{equation}
\ell_{\calS}{}_{\mu} \delta x^{\mu}_{\calS}-\ell_{\calS}{}_{\mu}\WXX{}\UD{\mu}{\sigma}\delta x^{\sigma}_{\calO}=\ell_{\calS}{}_{\mu} \delta x^{\mu}_{\calS}-\ell_{\calO}{}_{\sigma}\delta x^{\sigma}_{\calO}=0\, .
\end{equation}
Therefore, although $\delta x^{\mu}_{\calS}$ and $\WXX{}\UD{\mu}{\sigma}\delta x^{\sigma}_{\calO}$ are not necessarily orthogonal to $\ell^{\mu}_{\calS}$, the combination $\left[ \delta x_{\calS}-\WXX(\delta x_{\calO}) \right]^{\mu}$ certainly is\footnote{Since $\Delta \ell_\calO \in \calP_{\calO}$, this implies also that $\WXL{}\UD{\bm{A}}{\bm{B}} : \calP_{\calO} \to \calP_{\calS}$ is the operator mapping direction deviations in $\calP_{\calO}$ to images in $\calP_{\calS}$.}, and it can be pulled back to the quotient space $\calP_{\calS}$ to finally obtain
\begin{equation}
r^{\bm{A}} = \frac{(\WXL^{-1}){}\UD{\bm{A}}{\bm{B}}}{u_{\calO\,\sigma}\,\ell_\calO^\sigma}\,\left[ \delta x_{\calS}-\WXX(\delta x_{\calO}) \right]^{\bm{B}}\, .\label{eq:rA_wBGO}
\end{equation}

The second question is how we compare directions registered by another observer $\calO'$ with a different four-velocity $U^\mu_{\calO}$. As for $\calO$, we introduce a SNF $f_{\bm{\alpha}}^\mu=(U_{\calO}^\mu, f_{\bm{A}}^\mu, \ell_\calO^\mu)$ adapted to the observer $\calO'$ that is used to define directions on $U_{\calO}^\mu$'s sky. The relation between directions measured by the two observers is contained in the expression of the spatial vectors $f_{\bm{A}}^\mu$ in components of $\phi_{\bm{\alpha}}^\mu$
\begin{equation}
f_{\bm{A}}^\mu=\omega\UD{\bm{B}}{\bm{A}}\phi_{\bm{B}}^\mu+C_{\bm{A}}\,\ell_\calO^\mu\, ,\label{eq:boostRel}
\end{equation}
where the coefficients $\omega\UD{\bm{B}}{\bm{A}}$ and $C_{\bm{A}}$ are found using the relations for the SNF. In particular, from $f_{\bm{A}}^\mu (f_{\bm{B}})_\mu=\delta_{\bm{A}\,\bm{B}}$ follows that $\omega\UD{\bm{B}}{\bm{A}}\omega\UD{\bm{D}}{\bm{C}}\delta_{\bm{B}\,\bm{D}}=\delta_{\bm{A}\,\bm{C}}$. In the case that $f_{\bm{A}}^\mu$ and $\phi_{\bm{A}}^\mu$ have the same orientation, i.e. ${\rm det}\left(\omega\UD{\bm{B}}{\bm{A}}\right)>0$, then the matrix coefficient $\bm{\omega}\in SO(2)$. In conclusion, Eq.~\eqref{eq:boostRel} tell us that the two screen vectors are related by a rotation around the direction vector $r^{\mu}$ and possibly by a component along $\ell^{\mu}_{\calO}$.
We can always choose the spatial vectors $f_{\bm{A}}^\mu$ aligned along $\phi_{\bm{A}}^\mu$, such that $f_{\bm{A}}^\mu = \phi_{\bm{A}}^\mu + C_{\bm{A}}\,\ell_\calO^\mu$, with appropriate $C_{\bm{1}}$ and $C_{\bm{2}}$ \cite{Korzynski:2018}. The two pairs of vectors $f_{\bm{A}}^\mu $ and $ \phi_{\bm{A}}^\mu$ belong to the same equivalence classes in $\calP_{\calO}$. This way both $u_\calO^\mu$ and $U^\mu$ may use the fiducial null vector $\ell_\calO^\mu$ to provide the reference direction on their skies and the screen vectors $\phi_{\bm{A}}^\mu$ and $f_{\bm{A}}^\mu$ as the two perpendicular vectors on the celestial sphere. Now, the two spatial components of the direction vector $r^{\bm{A}}$ can be used to compare the registered directions on the sky between the two observers. 
In general, to any SNF $(u^\mu,\phi_{\bm{ A}}^\mu, \ell_\calO^\mu)$ we can take $\bm{ u} = [u]$, $\bm{\phi}_{\bm{ A}} = [\phi_{\bm{ A}}]$ to obtain a frame $(\bm{ u},\bm{\phi}_{\bm{ A}})$ in $\calQ_\calO$ and a frame $(\bm{\phi}_{\bm{ A}})$ in $\calP_\calO$. By parallel propagating the SNF and repeating this procedure we obtain similar parallel propagated frames $(\hat u^\mu,\hat \phi_{\bm{ A}}^\mu,\hat \ell_\calO^\mu)$, $(\hat {\bm{ u}},\hat {\bm{\phi}}_{\bm{ A}})$ and $(\hat{\bm{ \phi}}_{\bm{ A}})$, in $T_p \calM$, $\calQ_p$ and $\calP_p$, respectively.


\section{Observables with the BGO: momentary observables and drift effects}
\label{sec:observables}
In this section we apply the machinery of the BGO to compute multiple observables within the same framework. We show the derivation of the angular diameter distance, the parallax distance, the position drift, and the redshift drift as functionals of the BGO, following the results in \cite{perlick-lrr, Korzynski:2018, Grasso:2018mei, Korzynski:2019oal}. We also recall the definition of the redshift and  the luminosity distance, as these are fundamental quantities in cosmology.

\subsection{The redshift}
The redshift is a dimensionless quantity which measures the relative difference in the light wavelength between the emission and the observation points. In our system, the photons travelling along $\gamma_0$ from the source $\calS$ to the observer $\calO$ will experience the redshift $z$
\begin{equation}
1+z = \dfrac{\left(\ell_{\sigma } u^{\sigma}\right)|_{\cal S}}{\left(\ell_{\sigma } u^{\sigma}\right)|_{\cal O}}\, ,
\label{eq:redshift_def}
\end{equation}
where $\ell^{\sigma}$ is the tangent to $\gamma_0$, and $u^{\sigma}_{\mathcal{O}}$ and $u^{\sigma}_{\mathcal{S}}$ are the observer and source four-velocities.
In practical applications we distinguish the following sources of the redshift: 
\begin{itemize}
\item when $\calS$ and $\calO$ are moving with respect to each other, we have \emph{the relativistic Doppler effect},
\item when $\calS$ is immersed in a different gravitational potential than $\calO$, i.e. when one end of $\gamma_0$ is in a region of the spacetime more curved then the other end, we have \emph{the gravitational redshift},
\item when the spacetime between $\calS$ and $\calO$ is expanding, we have  \emph{the cosmological redshift}.
\end{itemize}
In FLRW spacetimes, the cosmological redshift is directly related to the scale factor $1+z=a_0/a$ and can be used as an independent variable to express other quantities, such as distance measurements.

\subsection{The angular diameter distance}
In astronomy there are several method to measure the distance of faraway objects, each based on different techniques. The \emph{angular diameter distance} (or area distance\footnote{Actually, the angular diameter distance and the area distance have different definitions, as pointed out in \cite{perlick-lrr}. However, here we will consider the two as synonyms to the definition in Eq.~\eqref{eq:D_ang_def}.}) is a measure of distance based on the idea that the farther away an object is, the smaller it appears to be, and it is defined as
\begin{equation}
D_{ang} = \sqrt{\left|\dfrac{A_{\calS}}{ \Omega_{\calO}}\right|}\, ,
\label{eq:D_ang_def}
\end{equation}
with $A_{\calS}$ being the area of the cross-section $C$ of the emitting body measured in its own frame $\phi_{\calS}{}^{\mu}_{\bm{\nu}}$, and $ \Omega_{\calO}$ is the solid angle\footnote{Note that $A_{\calS}$ and $\Omega_{\calO}$ are signed quantities depended on the orientation. To remove this dependence we have introduced an absolute value in Eq.~\eqref{eq:D_ang_def}.} occupied by the image $I$ in the observer's celestial sphere, see Fig.~\ref{fig:magnificationmatrix}. 
\bfi
\centering
\includegraphics[width=0.9\textwidth]{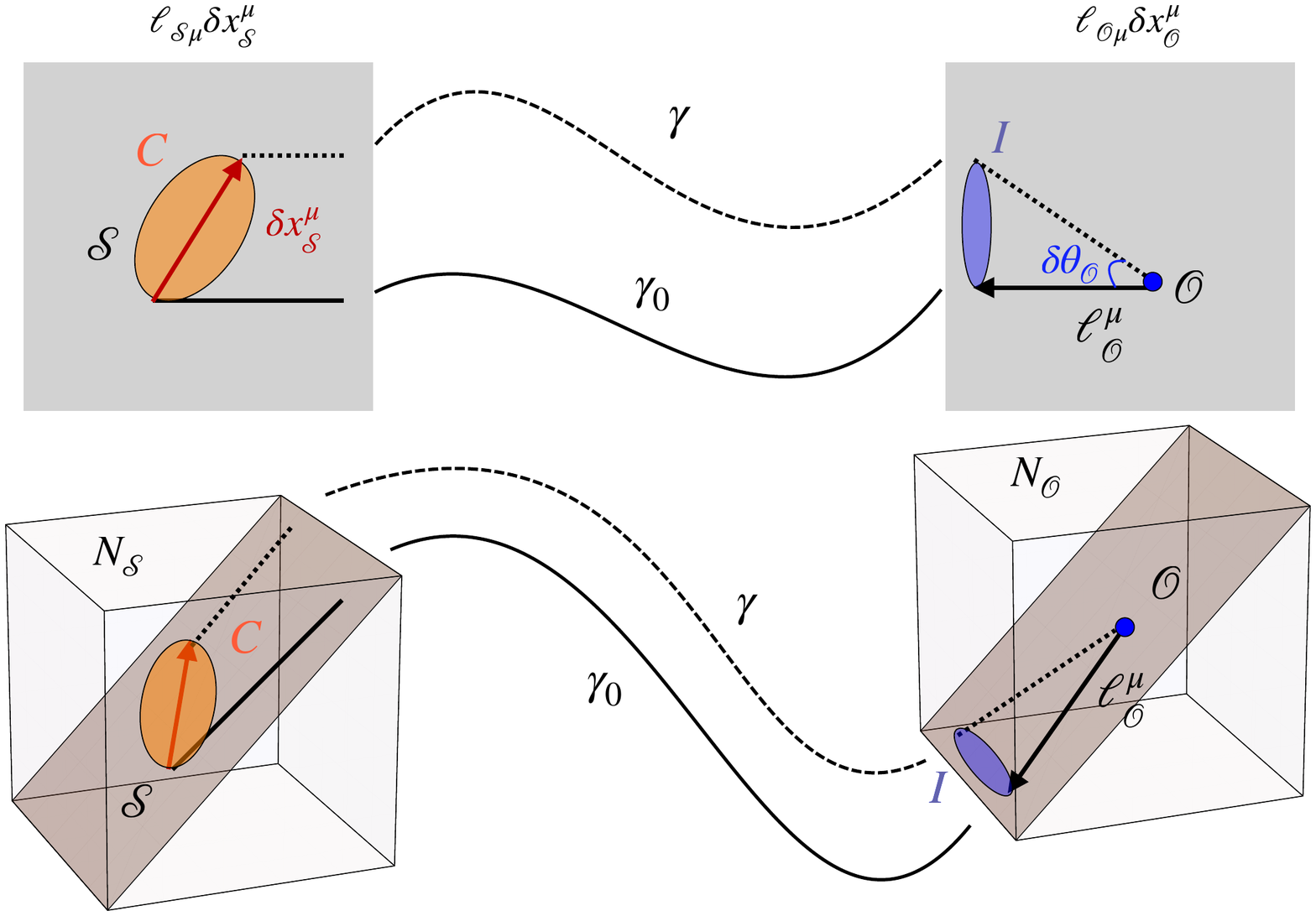}
\caption{On the null hypersurface $\ell_{\calS}{}_{\mu}\delta x^{\mu}_{\calS}=0$ the cross section $C$ of the source $\calS$ has area $A_{\calS}$. On the corresponding null hypersurface $\ell_{\calO}{}_{\mu} \delta x^{\mu}_{\calO}=0$, the observer $\calO$ measures the solid angle $\Omega_{\calO}$ occupied by the image $I$ of the source.}
\label{fig:magnificationmatrix}
\efi
Now, the area and the solid angle can be expressed in terms of the deviation vector $\delta x^{\mu}$ as
\begin{align}
A_{\calS}= & \int_{C} \delta x^{\bm{1}}_{\calS} \wedge \delta x^{\bm{2}}_{\calS} \nonumber \\
\Omega_{\calO} =& \int_{I} \delta \theta^{\bm{1}}_{\calO} \wedge \delta \theta^{\bm{2}}_{\calO}=\frac{1}{(u_{\calO\,\sigma}\,\ell_\calO^\sigma)^2}\left[{\rm det}\left(\WXL {}\UD{\bm A}{\bm B}\right)\right]^{-1}\,\int_{C} \delta x^{\bm{1}}_{\calS} \wedge \delta x^{\bm{2}}_{\calS}\, , \label{eq:area-angle_Dang}
\end{align}
where we have used Eq.~\eqref{eq:rA_wBGO} with $\delta x^{\mu}_{\calO}=0$ to express small angles $\delta \theta^{\bm{A}}\sim r^{\bm{A}}$.
Inserting Eqs.~\eqref{eq:area-angle_Dang} in the definition Eq.~\eqref{eq:D_ang_def} we obtain the angular diameter distance in terms of BGO as
\begin{equation}
D_{ang} = \left(\ell_{\sigma} u^{\sigma}\right)|_{\cal O} \left| \det \left(\WXL {}\UD{\bm A}{\bm B}\right) \right|^{\frac{1}{2}}\, .
\label{eq:D_ang_BGO}
\end{equation}

By direct comparison with the standard definition of the angular diameter distance, see e.g. \cite{perlick-lrr}, we have that the two-by-two submatrix of $\WXL$ is the well-known Jacobi operator $\mathcal{D}\UD{\bm A}{\bm B}$, namely the map between physical separations $\delta x^{\bm{A}}_{\calS}$ at the source position and direction deviations $ \Delta \ell^{\bm{A}}_{\calO}$ at the observer position.

\subsection{The luminosity distance}
Similar to the angular diameter distance, the \emph{luminosity distance} is based on the idea that the farther away an object is located, the fainter appears its light. Indeed, if we consider an isotropic light emission, the energy flux of the light $F$ decreases with distance from the source $D$ according to the flux-luminosity relation $F= L / (4 \pi D^2)$. The luminosity distance is then defined as the ratio between the luminosity $L$ of the source and the flux measured at the observer
\begin{equation}
D_{\rm lum}=\sqrt{\dfrac{L}{4 \pi F}}\, . \label{eq:D_lum_def}
\end{equation}
As we did for $D_{\rm ang}$, also in this case we can express eq.~\eqref{eq:D_lum_def} in terms of geometrical quantities by inverting the role of observer and emitter and calculating the flux of photons. However, in this case we prefer to use the well-known result obtained by Etherington, see \cite{etherington, etherington2}, that relates the luminosity distance $D_{\rm lum}$ to the angular diameter distance $D_{\rm ang}$ via the \emph{distance duality relation}
\begin{equation}
D_{\rm lum}=(1+z)^2 D_{\rm ang}\, . \label{eq:etherington_rel}
\end{equation}

\subsection{The parallax and the parallax distance}
Another distance estimator in astronomy is the parallax distance, called this way because it takes into account the apparent change in position of a source on the celestial sphere when viewed from at least two different viewpoints, known as the parallax effect. Unlike angular diameter distance or luminosity distance, it does not require knowledge of the source's properties. For this reason, parallax is an attractive method of measuring distance. However, studying the parallax of distant objects is complex and requires accurate astrometric measurements \cite{Ding:2009xs, gaiadr2-parallaxes, Hobbs:2019arz}.
On top of that, the parallax has a straightforward interpretation only in a flat space and in non-relativistic context: its generalization to general relativity is more cumbersome, generating confusion on its interpretation \cite{mccrea, weinberg-letter, kasai, rosquist, rasanen, Marcori:2018cwn}. A covariant treatment of cosmic parallax was proposed by R\"{a}s\"{a}nen in \cite{rasanen}, where the author distinguishes different definitions of parallax by the distance between observation points $\delta x^{\mu}_{\calO}$. 
In the following we will focus on two definitions of parallax: the classic parallax, that we use to define the (classical) parallax distance, and the position drift.

Let us examine the classic parallax of a source $\calS$ as it is seen by a number of observers in $N_{\calO}$. The observers are chosen such that they all perceive signals emitted exactly at the same moment by $\calS$. In other words, all observation points lie on the same null hypersurface $\delta x_\calO^\mu\,\ell_{\calO\,\mu} = \const$:
\bfi
\centering
\includegraphics[width=0.8\textwidth]{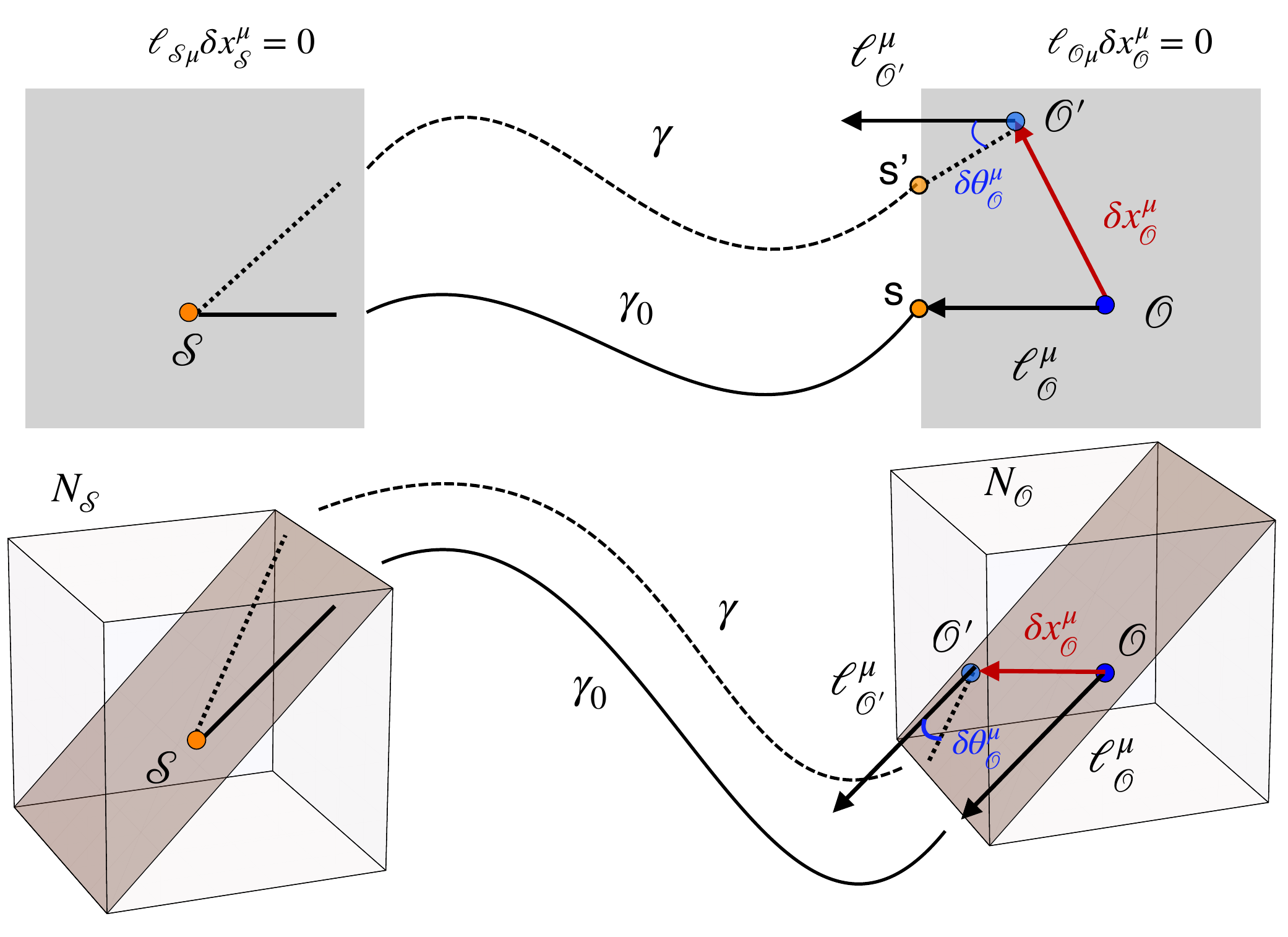}
\caption{The worldlines of two observers $\calO$ and $\calO'$ cross the same null hypersurface $\delta x_\calO^\mu\,\ell_{\calO\,\mu} = 0$ and observe the source $\calS$, which lies on the corresponding null hypersurface $\delta x_\calS^\mu\,\ell_{\calS\,\mu} = 0$ in $N_{\calS}$. $\calO$ and $\calO'$ are displaced by $\delta x^{\mu}_{\calO}$, and they perceive the source in the apparent positions $\sf s$ and $\sf s'$, respectively. The difference between $\sf s$ and $\sf s'$ gives the parallax angle $\delta \theta^{\mu}_{\calO}$. On the screen $\hat{\phi}^{\mu}_{\bm{A}}$, the displacement $\delta x^{\bm{A}}_{\calO}$ and the angular distance $\delta \theta^{\bm{A}}_{\calO}$ are related by the parallax matrix $\Pi\UD{\bm{A}}{\bm{B}}$.}
\label{fig:parallaxmatrix}
\efi
for simplicity we assume that the observers are comoving\footnote{This way we do not need to consider the aberration effects when comparing the results of their measurements.} and they all perform the measurement when their worldlines cross the null hypersurface $\delta x_\calO^\mu\,\ell_{\calO\,\mu} = 0$. Therefore, at the moment of observation we have $[\delta x_\calO] \in \calP_\calO$ for the equivalence class of their displacement vectors, see Fig.~\ref{fig:parallaxmatrix}. 

The difference in the apparent position of the source, as measured by the two observes $\calO$ and  $\calO'$, defines the classic parallax. Using Eq.~\eqref{eq:rA_wBGO} with $\delta x^{\mu}_{\calS}=0$ the classic parallax takes the form
\bea
\delta \theta_\calO^{\bm A} = - \dfrac{1}{\ell_{\calO\, \sigma} u^{\sigma}_{\calO}}\left(\WXL^{-1}\right){}\UD{\bm A}{\bm B} \, \WXX{}\UD{\bm B}{\bm C} \, \delta x_\calO^{\bm C}\, .\label{eq:parallax_dev}
\eea
\bfi
\centering
 \includegraphics[width=0.8\textwidth]{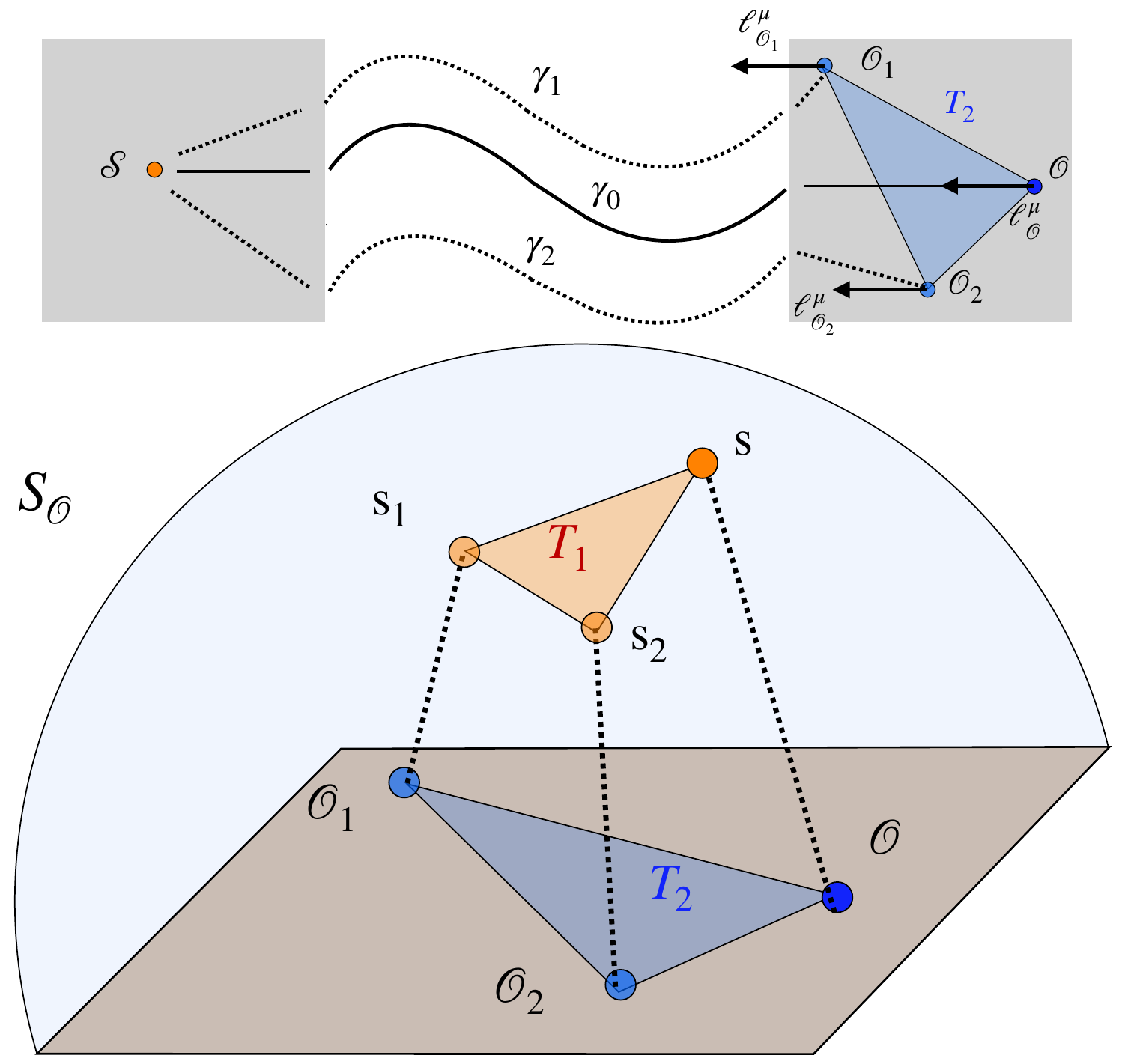}
\caption{The source $\calS$ is observed by $\calO$, $\calO_{\rm 1}$, and $\calO_{\rm 2}$ in $N_{\calO}$. The positions of the three observers form the blue triangle $T_{\rm 2}$ in the space perpendicular to $\ell_{\calO}$. Similarly, on the shared celestial sphere $S_{\calO}$, the apparent positions of the source $\sf s$, $\sf s_{\rm 1}$, and $\sf s_{\rm 2}$ form the orange triangle $T_{\rm 1}$. The parallax distance is defined as the ratio between the area of $T_{\rm 2}$ and the solid angle of $T_{\rm 1}$.}
\label{fig:Dpar}
\efi
The product
\begin{equation}
\Pi\UD{\bm A}{\bm B}= \dfrac{1}{\ell_{\calO\, \sigma} u^{\sigma}_{\calO}}\left(\WXL^{-1}\right){}\UD{\bm A}{\bm B}\, \WXX{}\UD{\bm B}{\bm C}\, ,\label{eq:parallax_matrix}
\end{equation}
defines the observer-dependent \emph{parallax matrix}, namely the map between perpendicular displacement on the observer's side $\delta x^{\bm{A}}_{\calO}$, and two-dimensional angles $\delta \theta^{\bm{A}}_{\calO}$ measuring the observed position on the sky in comparison with the position observed by $\calO'$ at $\calO$, see Fig.~\ref{fig:Dpar}.

In astronomy the parallax has been used to measure the distances to objects up to few kiloparsecs \cite{2001A&A...369..339P,  Riess:2014uga}. In the following we present the parallax distance formula in terms of BGO\footnote{As noted in  \cite{rasanen, Grasso:2018mei}, for curved spacetimes the trigonometric parallax angle depends on the direction of the baseline $\delta x^{\mu}_{\calO}$. To overcome this problem, we use the baseline-averaged definition of parallax distance as presented in \cite{Grasso:2018mei}.} by using the definition of the classic parallax discussed earlier. Let us consider an observer $\calO$ and two additional observers $\mathcal{O}_{1}$ and $\mathcal{O}_{2}$, comoving with $\calO$ and such that their displacement with respect to $\calO$ are on $\calP_\calO$. We also introduce a screen frame $[\hat{\phi}_{\bm{A}}] \in \calP_\calO$, which is parallel transported on $N_{\calO}$. The three observers define a triangle $T_2$ on the screen space perpendicular to $\ell_{\calO}$ with area $A_\calO= \int_{T_2} \delta x^{\bm{1}}_\calO \wedge \delta x^{\bm{2}}_\calO$. 
Now, the observers measure the apparent position of a source $\calS$ and, using the parallel transported frame, they combine their observations on a shared celestial sphere $S_\calO$, see Fig.~\ref{fig:Dpar}. The combined observation form a solid triangle $T_1$ on $S_\calO$. 
Denoting as $\Omega_\calO=\int_{T_1} \delta \theta^{\bm{1}}_\calO \wedge \delta \theta^{\bm{2}}_\calO$ the solid angle taken up by $T_1$, we define the parallax distance as
\begin{equation}
D_{par} = \sqrt{\left|\frac{A_\calO}{\Omega_\calO}\right|}\, . \label{eq:D_par_def}
\end{equation}
Using Eq.~\eqref{eq:parallax_dev} to express the solid angle, i.e. 
\begin{equation}
\Omega_\calO=\int_{T_1} \delta \theta^{\bm{1}}_\calO \wedge \delta \theta^{\bm{2}}_\calO= \dfrac{1}{(\ell_{\calO\, \sigma} u^{\sigma}_{\calO})^2}{\rm det}\left[\left(\WXL^{-1}\right){}\UD{\bm A}{\bm B} \WXX{}\UD{\bm B}{\bm C}\right] \,\int_{T_2} \delta x^{\bm{1}}_\calO \wedge \delta x^{\bm{2}}_\calO\, ,
\end{equation}
the expression for the parallax distance becomes
\begin{equation}
D_{par} =\left(\ell_{\sigma} u^{\sigma}\right)|_{\cal O} \frac{\left| \det \left(\WXL {}\UD{\bm A}{\bm B}\right) \right|^{\frac{1}{2}}}{\left| \det \left(\WXX {}\UD{\bm A}{\bm B}\right) \right|^{\frac{1}{2}}}\, .
\label{eq:D_par_BGO}
\end{equation}

\subsection{Position drift} 
Among the various definitions of parallax, there is also the case where a single observer measures the temporal changes in the position of the source in the sky. 
This momentary rate of change of the source's position in the observer's sky is the \emph{position drift} $\delta_{\calO} r^{\bm{A}}$ \cite{Korzynski:2018, Grasso:2018mei} and is one of the real-time measurements commonly referred to as \emph{drift effects}. In contrast to the classic parallax, the position drift depends on the four-velocities of both the observer and the emitter, involving also the observer's four-acceleration \cite{Korzynski:2018, Hellaby:2017soj, Marcori:2018cwn}.
A general formula for the position drift has already been presented in \cite{Korzynski:2018}, while a special case for spherically symmetric metrics was presented in \cite{Quercellini:2009ni, Quercellini:2010zr}. Here we present the general formula for the position drift in terms of BGO as derived in \cite{Grasso:2018mei}.

As was said many times, the position drift measures the temporal change of directions on the observer's sky as the observer moves along its worldline. This means that, contrary to the other observables considered so far, we are actually measuring changes in the direction vector as the observer crosses different null hypersurfaces $\delta x^{\sigma}\ell_{\sigma}=\const$. This implies that, having fixed a reference frame, we need to find a way to transport it along the observer's worldline in order to be able to measure the changes in the directions registered on the next null hypersurface. The choice we make is to use the Fermi-Walker transport, which reduces to the usual parallel transport if the $\calO$'s worldline is a geodesic. The Fermi-Walker transport of vectors in the observer's sphere $\textrm{Dir}(u_\calO)$ along the worldline defines our ``fixed directions on the sky'' \cite{Hellaby:2017soj, Korzynski:2018}. 
The Fermi-Walker derivative of the direction vector $r_0^{\mu}$ in Eq.~\eqref{eq:dir} is expressed as
\begin{equation}
\delta_{\calO} r^{\mu}= u_{\calO}^{\sigma}\nabla_{\sigma} r_0^{\mu} - u_{\calO}^{\mu} w_{\calO\, \sigma}r_0^{\sigma}+w_{\calO}^{\mu} u_{\calO\, \sigma}r_0^{\sigma}\, ,\label{eq:Fermi-Walker_der}
\end{equation}
where $u_{\calO}^{\mu}$ and $w_{\calO}^{\mu}$ are the observer's four-velocity and four-acceleration respectively. 
From Eq.~\eqref{eq:dir} follows that the last term in Eq.~\eqref{eq:Fermi-Walker_der} vanishes, since $u_{\calO\, \sigma}r^{\sigma}_0=0$, while the covariant derivative $u_{\calO}^{\sigma}\nabla_{\sigma} r_0^{\mu}$ is\footnote{The relation $\Delta \ell^{\mu}=\delta \tau_{\calO} u^{\sigma}_{\calO}\nabla_{\sigma}\ell^{\mu}_{\calO}$ follows from the definition $\Delta \ell^{\mu}_{\calO}=\ell^{\sigma}\nabla_{\sigma} \delta x^{\mu}_{\calO}$ by using Eqs.~\eqref{eq:commutation_vector} and expressing $\delta x^{\mu}_{\calO}=u^{\mu}_{\calO} \delta \tau_{\calO}$.}
\begin{equation}
u^{\sigma}_{\calO}\nabla_{\sigma} r^{\mu}_0=\dfrac{u^{\sigma}_{\calO}\nabla_{\sigma}\ell^{\mu}_{\calO}}{\left(\ell^{\rho}_{\calO}u_{\calO}{}_{\rho}\right)} +u^{\sigma}_{\calO}\nabla_{\sigma} u^{\mu}_{\calO}=\dfrac{\dfrac{\Delta \ell^{\mu}_{\calO}}{\delta \tau_{\calO}}}{\left(\ell^{\rho}_{\calO}u_{\calO}{}_{\rho}\right)} +w^{\mu}_{\calO}\, ,\label{eq:uDr}
\end{equation}
with $\delta \tau_{\calO}$ the observer's proper time. 
The pull-back to $\calP_{\calO}$ of Eq.~\eqref{eq:Fermi-Walker_der} defines the position drift $\delta_{\calO} r^{\bm{A}}$ measured with respect to inertially dragged fixed directions
\begin{equation}
\delta_{\calO} r^{\bm{A}}= \dfrac{\dfrac{\Delta \ell^{\bm{A}}_{\calO}}{\delta \tau_{\calO}}}{\left(\ell^{\rho}_{\calO}u_{\calO}{}_{\rho}\right)} +w^{\bm{A}}_{\calO}\, .\label{eq:position_drift_def}
\end{equation} 
The first term in Eq.~\eqref{eq:position_drift_def} can be obtained from Eq.~\eqref{eq:Dl_inBGO_with_Ps} as
\bea
\dfrac{\Delta \ell_{\calO}^{\bm A}}{\delta \tau_{\calO}} = \left(\WXL^{-1}\right){}\UD{\bm A}{\bm B}\,\left[\frac{1}{1+z}\,u_\calE - \WXX(u_\calO) \right]^{\bm{B}}\, , \label{eq:rA_in_tau}
\eea
where we expressed the two displacements as
\begin{align}
\delta x^{\mu}_{\calO} = & u^{\mu}_{\calO} \delta \tau_{\calO}\\
\delta x^{\mu}_{\calS} = & u^{\mu}_{\calS} \delta \tau_{\calS} \, , \label{eq:dx_udt}
\end{align}
and we used the relation\footnote{The relation is obtained from the condition $\delta x^{\mu}_{\calO}\ell_{\calO\, \mu}=\delta x^{\mu}_{\calS}\ell_{\calS\, \mu}$, by using Eq.~\eqref{eq:dx_udt} and the definition of redshift Eq.~\eqref{eq:redshift_def}.}
\begin{equation}
\delta \tau_{\calS}=\dfrac{\delta \tau_{\calO}}{1+z}\, , \label{eq:tE-tO_formula}
\end{equation}
between the proper time as measured at the observer $\delta \tau_{\calO}$ and the proper time as measured at the source $\delta \tau_{\calS}$, \cite{perlick, kermack_mccrea_whittaker_1934, Korzynski:2018}.
Combining (\ref{eq:position_drift_def}) and (\ref{eq:rA_in_tau}) yields
\bea
\delta_\calO r^{\bm A} = \dfrac{1}{\ell_{\calO\, \sigma} u^{\sigma}_{\calO}} \left(\WXL^{-1}\right){}\UD{\bm A}{\bm B}\,\left[\frac{1}{1+z}\,u_\calE - \WXX(u_\calO) \right]^{\bm{B}} + w_\calO^{\bm A}\, . \label{eq:positionDRIFT_BGO}
\eea
Note that the last term is the perpendicular component of the observer's four-acceleration. It corresponds to the special relativistic effect of the position drift due to the drift of the aberration \cite{rasanen,Korzynski:2018,Marcori:2018cwn}.  
For a longer discussion of the position drift formula and its physical and astrophysical consequences see \cite{Korzynski:2018, Grasso:2018mei}.

\subsection{The redshift drift formula}
The last observable under consideration is the redshift drift, a real time observable expressing the temporal changing of the redshift, due to cosmic expansion and proper motion of the source and the observer. The formulation of the redshift drift was firstly proposed by Sandage in 1962 \cite{sandage}, and later applied by A. Loeb \cite{loeb} as a tracer of the expansion of the Friedmann-Lema\^{i}tre-Robertson-Walker Universe.  Here, we will present the derivation of the general formula of the redshift drift in terms of BGO, \cite{Julius}.

Let us consider two consecutive measurements of the redshift $z$ as taken by the observer $\mathcal{O}$ at the two instants $\tau_{\mathcal{O}}$ and $\tau_{\mathcal{O}} + \delta \tau_{\mathcal{O}}$. In the time lapse $\delta \tau_{\mathcal{O}}$ the observer and the source are shifted along their worldlines by:
\begin{equation}
\begin{array}{l}
\delta x^{\mu}_{\mathcal{O}}=u^{\mu}_{\mathcal{O}} \delta \tau_{\mathcal{O}} \\
\delta x^{\mu}_{\mathcal{S}}=u^{\mu}_{\mathcal{S}} \delta \tau_{\mathcal{S}}=\dfrac{1}{1+z}u^{\mu}_{\mathcal{S}} \delta \tau_{\mathcal{O}}\, ,
\end{array}
\end{equation}
where we have used the relation between $\delta \tau_{\mathcal{S}}$ and $\delta \tau_{\mathcal{O}}$ in Eq.~\eqref{eq:tE-tO_formula}. 
Now, the redshift drift is obtained varying with respect to the observer proper time the definition of the redshift. For our convenience, let us take the logarithm of the redshift in Eq.~\eqref{eq:redshift_def}
\begin{equation}
\log(1+z)=\log(\left.\ell^{\mu}u_{\mu}\right|_{\mathcal{S}})-\log(\left.\ell^{\mu}u_{\mu}\right|_{\mathcal{O}})\, ,
\end{equation}
and do its variation
\begin{equation}
\delta \log(1+z)=\dfrac{(\Delta \ell^{\mu}_{\mathcal{S}}u_{\mathcal{S}}{}_{\mu}+\ell^{\mu}_{\mathcal{S}}\Delta u_{\mathcal{S}}{}_{\mu})}{\ell^{\mu}_{\mathcal{S}}u_{\mathcal{S}}{}_{\mu}}-\dfrac{(\Delta \ell^{\mu}_{\mathcal{O}}u_{\mathcal{O}}{}_{\mu} + \ell^{\mu}_{\mathcal{O}}\Delta u_{\mathcal{O}}{}_{\mu})}{\ell^{\mu}_{\mathcal{O}}u_{\mathcal{O}}{}_{\mu}}\, .
\label{eq:red_drift_intermediate1}
\end{equation}
The two terms
\begin{equation}
\begin{array}{l}
\Delta u^{\mu}_{\mathcal{O}}=w^{\mu}_{\mathcal{O}} \delta \tau_{\mathcal{O}}\, , \\
\Delta u^{\mu}_{\mathcal{S}}=w^{\mu}_{\mathcal{S}} \delta \tau_{\mathcal{S}}=\dfrac{1}{1+z}w^{\mu}_{\mathcal{S}} \delta \tau_{\mathcal{O}}
\end{array}\, ,
\end{equation}
define the four-acceleration of the observer $w_{\calO}$ and the emitter $w_{\calS}$.

The variation in Eq.~\eqref{eq:red_drift_intermediate1} can be reshuffled as
\begin{equation}
\begin{array}{l l}
\delta \log(1+z)= & \\
\left(\dfrac{1}{1+z}\dfrac{\ell^{\mu}_{\mathcal{S}}w_{\mathcal{S}}{}_{\mu}}{\ell^{\mu}_{\mathcal{S}}u_{\mathcal{S}}{}_{\mu}}-\dfrac{\ell^{\mu}_{\mathcal{O}}w_{\mathcal{O}}{}_{\mu}}{\ell^{\mu}_{\mathcal{O}}u_{\mathcal{O}}{}_{\mu}}\right)\delta \tau_{\mathcal{O}}+\left(\dfrac{\Delta \ell^{\mu}_{\mathcal{S}}u_{\mathcal{S}}{}_{\mu}}{\ell^{\mu}_{\mathcal{S}}u_{\mathcal{S}}{}_{\mu}}-\dfrac{\Delta \ell^{\mu}_{\mathcal{O}}u_{\mathcal{O}}{}_{\mu}}{\ell^{\mu}_{\mathcal{O}}u_{\mathcal{O}}{}_{\mu}}\right)
\end{array}\, .
\label{eq:red_drift_intermediate2}
\end{equation}
The first term is a special relativistic term representing the Doppler effect along the line of sight caused by the four-acceleration of the observer and the emitter 
\begin{equation}
\Xi_{\rm Doppler}=\left[\dfrac{1}{1+z}\dfrac{\left(\ell^{\mu}w_{ \mu}\right)|_{\mathcal{S}}}{\left(\ell^{ \mu} u_{ \mu}\right)|_{\mathcal{S}}}-\dfrac{\left(\ell^{ \mu}w_{ \mu}\right)|_{\mathcal{O}}}{\left(\ell^{ \mu} u_{ \mu}\right)|_{\mathcal{O}}}\right]
\label{eq:doppler}\, .
\end{equation}
The second term contains the effects of the spacetime curvature on the redshift drift and it can be expressed in terms of the BGO. Let us start by writing the second term in the matrix form:
\begin{equation}
\dfrac{\Delta \ell^{\mu}_{\mathcal{S}}u_{\mathcal{S}}{}_{\mu}}{\ell^{\mu}_{\mathcal{S}}u_{\mathcal{S}}{}_{\mu}}-\dfrac{\Delta \ell^{\mu}_{\mathcal{O}}u_{\mathcal{O}}{}_{\mu}}{\ell^{\mu}_{\mathcal{O}}u_{\mathcal{O}}{}_{\mu}}=-\left(\dfrac{u_{\mathcal{O}}{}_{\nu}}{\ell^{\mu}_{\mathcal{O}}u_{\mathcal{O}}{}_{\mu}}\ \  \dfrac{u_{\mathcal{S}}{}_{\mu}}{\ell^{\mu}_{\mathcal{S}}u_{\mathcal{S}}{}_{\mu}}\right) \cdot\left(\begin{matrix}
\Delta \ell^{\nu}_{\mathcal{O}}\\
-\Delta \ell^{\mu}_{\mathcal{S}}
\end{matrix} \right)\, .
\label{eq:second_term1}
\end{equation}
The vector $\left( \Delta \ell^{\nu}_{\mathcal{O}}\ \ -\Delta \ell^{\mu}_{\mathcal{S}} \right)^{\rm T}$ is expressed in terms of the BGO using  Eqs.~\eqref{eq:positiondeviation1}-\eqref{eq:directiondeviation1}
\begin{equation}
\left\{\begin{array}{l}
\Delta \ell^{\nu}_{\mathcal{O}}= \WXL^{-1}{}\UD{\nu}{\rho} \delta x^{\rho}_{\mathcal{S}}- \WXL^{-1}{}\UD{\nu}{\rho}\WXX{}\UD{\rho}{\sigma} \delta x^{\sigma}_{\mathcal{O}}\\
-\Delta \ell^{\mu}_{\mathcal{S}}= -\WLX{}\UD{\mu}{\rho} \delta x^{\rho}_{\mathcal{O}} - \WLL{}\UD{\mu}{\nu} \WXL^{-1}{}\UD{\nu}{\rho} \delta x^{\rho}_{\mathcal{S}} + \WLL{}\UD{\mu}{\nu}\WXL^{-1}{}\UD{\nu}{\rho}\WXX{}\UD{\rho}{\sigma} \delta x^{\sigma}_{\mathcal{O}}
\end{array}\right. \, ,
\end{equation}
from which we finally get
\begin{equation}
\begin{pmatrix}
\Delta \ell^{\nu}_{\mathcal{O}}\\
-\Delta \ell^{\mu}_{\mathcal{S}}
\end{pmatrix}=\begin{pmatrix}
- \WXL^{-1}{}\UD{\nu}{\rho}\WXX{}\UD{\rho}{\sigma} & \WXL^{-1}{}\UD{\nu}{\rho} \\
\WLL{}\UD{\mu}{\nu}\WXL^{-1}{}\UD{\nu}{\rho}\WXX{}\UD{\rho}{\sigma}-\WLX{}\UD{\mu}{\sigma} & - \WLL{}\UD{\mu}{\nu} \WXL^{-1}{}\UD{\nu}{\rho}
\end{pmatrix} \begin{pmatrix}
\delta x^{\sigma}_{\mathcal{O}}\\
\delta x^{\rho}_{\mathcal{S}}
\end{pmatrix}
\label{eq:U_matrix}
\end{equation}

Denoting
\begin{equation}
U=\begin{pmatrix}
- \WXL^{-1}{}\UD{\nu}{\rho}\WXX{}\UD{ \rho}{ \sigma} & \WXL^{-1}{}\UD{ \nu}{ \rho} \\
\WLL{}\UD{ \mu}{ \nu}\WXL^{-1}{}\UD{ \nu}{ \rho}\WXX{}\UD{ \rho}{ \sigma}-\WLX{}\UD{ \mu}{ \sigma} & - \WLL{}\UD{ \mu}{ \nu} \WXL^{-1}{}\UD{ \nu}{ \rho}
\end{pmatrix} \, 
\end{equation}
as the large $8 \times 8$ block matrix containing the BGO, Eq.~\eqref{eq:U_matrix} becomes
\begin{equation}
\begin{pmatrix}
\Delta \ell^{\nu}_{\mathcal{O}}\\
-\Delta \ell^{\mu}_{\mathcal{S}}
\end{pmatrix}=U \begin{pmatrix}
\delta x^{\sigma}_{\mathcal{O}}\\
\delta x^{\rho}_{\mathcal{S}}
\end{pmatrix}=U\begin{pmatrix}
 u^{\sigma}_{\mathcal{O}} \delta \tau_{\mathcal{O}}\\
\dfrac{\delta \tau_{\mathcal{O}}}{1+z} u^{\rho}_{\mathcal{S}}
\end{pmatrix}\, ,
\label{eq:U_matrix_2}
\end{equation}
and it can then inserted in Eq.~\eqref{eq:second_term1} that finally becomes
\begin{equation}
\begin{array}{l c}
-(\dfrac{u_{\mathcal{O}}{}_{\nu}}{\ell^{\mu}_{\mathcal{O}}u_{\mathcal{O}}{}_{\mu}}\ \  \dfrac{u_{\mathcal{S}}{}_{\mu}}{\ell^{\mu}_{\mathcal{S}}u_{\mathcal{S}}{}_{\mu}}) \cdot\left(\begin{matrix}
\Delta \ell^{\nu}_{\mathcal{O}}\\
-\Delta \ell^{\mu}_{\mathcal{S}}
\end{matrix} \right)= & \\
-\dfrac{\delta \tau_{\mathcal{O}}}{\ell^{\mu}_{\mathcal{O}}u_{\mathcal{O}}{}_{\mu}}(u_{\mathcal{O}}{}_{\nu}\ \  \dfrac{u_{\mathcal{S}}{}_{\mu}}{1+z}) \cdot U \cdot \begin{pmatrix}
 u^{\sigma}_{\mathcal{O}}\\
\dfrac{u^{\rho}_{\mathcal{S}}}{1+z} 
\end{pmatrix} \, ,  & \\
\end{array}
\end{equation}
where we invite the reader to notice that this derivation was made considering $U$ with upper-down indices distribution.

Finally, Eq.~\eqref{eq:red_drift_intermediate2} in terms of the new defined quantities gives the expression of the redshift drift $\frac{\delta \log(1+z)}{\delta \tau_{\mathcal{O}}} \equiv \zeta$ in terms of the BGO
\begin{equation}
\zeta= \Xi_{\rm Doppler}- (u_{\mathcal{O}}{}_{\nu}\ \  \dfrac{u_{\mathcal{S}}{}_{\mu}}{1+z}) \cdot U \cdot \begin{pmatrix}
 u^{\sigma}_{\mathcal{O}}\\
\dfrac{u^{\rho}_{\mathcal{S}}}{1+z} 
\end{pmatrix}\, .
\label{eq:z_DRIFT_BGO}
\end{equation}
The expression Eq.~\eqref{eq:z_DRIFT_BGO} is completely general in the sense that it was derived from general considerations and without referring to a specific model. Of course, the specific expression of the BGO is dictated by the particular form of the spacetime in which they are calculated, but once the BGO are computed, they can be used to calculate the redshift drift with the formula above. 


In conclusion, the BGO are fundamental objects describing multiple effects on light propagation in the regime of geometric optics. Let us notice that, although the BGO formalism is independent of the frame used, the observables depend on the emitter and observer kinematics, as shown by the explicit dependence of $u^{\mu}_{\mathcal{O}}$, $u^{\mu}_{\mathcal{S}}$, $w^{\mu}_{\mathcal{O}}$ and $w^{\mu}_{\mathcal{S}}$ in Eqs.~\eqref{eq:D_ang_BGO}-\eqref{eq:z_DRIFT_BGO}. Indeed, it is possible to apply the Lorentz transformations to change reference frame, but this modifies the observables introducing special relativistic effects such as the Doppler effect or aberration.
In this sense, the BGO provide a unified framework for computing all optical observables, such us those in Eqs.~\eqref{eq:D_ang_BGO},
 \eqref{eq:D_par_BGO}, \eqref{eq:positionDRIFT_BGO}, and \eqref{eq:z_DRIFT_BGO}. 
Moreover, while there are already analogous formulas for $D_{\rm ang}$ and $D_{\rm par}$, where instead of the BGO we have the magnification and the parallax matrix (see \cite{Korzynski:2019oal} for the comparison), there was no general formula for the position drift and the redshift drift: Eqs.~\eqref{eq:positionDRIFT_BGO} and \eqref{eq:z_DRIFT_BGO} look the same for each spacetime model considered. 

%% file: template/paper_bigonlight.tex

\chapter{Paper I: ``{\tt BiGONLight}: light propagation with bilocal operators in Numerical Relativity''} 
\label{chap:bigonlight}

The chapter presents {\tt BiGONLight}, a {\tt Wolfram} package designed to implement the BGO framework in $3 + 1$ form to compute optical observables from numerically generated spacetimes. The package is completely general: it can simulate light propagation in geometric optics approximation in any spacetime, with no assumptions on the gauge or coordinate system used. 
It was specifically designed to be compatible with the full-GR codes in numerical relativity based on the ADM formalism. Nevertheless, it also takes advantage of Mathematica's symbolic algebra manipulation to compute the BGO from the analytical expression of the metric tensor.
The output of the package are the BGO, which are then combined with the observer $\calO$ and source $\calS$ four-velocities ($u^{\mu}_{\calO}$, $u^{\mu}_{\calS}$) and four-accelerations ($w^{\mu}_{\calO}$, $w^{\mu}_{\calS}$) to compute observables.

This work is conceptually divided into two parts. In the first part, {\tt BiGONLight} is presented together with the theoretical formulation of the BGO framework in  $3+1$ form. In the second part, the code is tested by calculating observables in three well-known cosmological models: the $\Lambda$CDM and the Szekeres (analytical) spacetimes, and a dust Universe obtained from numerical simulation.\\

\textit{\textbf{Author's contribution}}\\
\newline
{\tt BiGONLight} is my original contribution in \cite{Grasso:2021iwq} and my main achievement in this thesis. On the $13^{th}$ of July 2021, I released the stable version of the package (v1.0), which is publicly available on the GitHub repository {\color{blue}{https://github.com/MicGrasso/bigonlight}} under the GPL-3.0 license.
The {\tt BiGONLight} package is a collection of {\tt Wolfram} functions that can be used in a {\tt Mathematica} notebook to compute observables.
The procedure to compute observables with {\tt BiGONLight} can be summarised as follows:
\begin{enumerate}[(i)]
	\item the user provides the metric $g_{\mu \nu}$, and the observer $\calO$ and source $\calS$ kinematics as input. They can be given already in $3+1$ decomposition or as four-dimensional quantities. In the last instance, the user can use the functions {\tt ADM[]} and {\tt Vsplit[]} to perform the $3+1$ decomposition of $g_{\mu \nu}$ and of the vectors ($u^{\mu}$, $w^{\mu}$), respectively;
	\item set the initial conditions ($x^{\mu}_{\calO}$, $\ell^{\mu}_{\calO}$) for the photon's geodesic, provided as $3+1$ components or using {\tt Vsplit[]} and {\tt InitialConditions[]};
	\item obtain the expression of the geodesic equation in $3+1$ with the functions {\tt GeodesicEquations[]} and {\tt EnergyEquations[]}. The functions implement the $3+1$ geodesic equations obtained by Vincent et. al. in \cite{Vincent:2012kn};
	\item solve the geodesic equations numerically with {\tt SolveGeodesic[]} and {\tt SolveEnergy[]};
	\item set the initial conditions for the SNF, directly in ADM components or using {\tt SNF[]}. Then the function {\tt PTransportedFrame[]} computes the parallel transported SNF along the geodesic by solving the $3+1$ parallel transport equations Eq.~($31$) in \cite{Grasso:2021iwq};
	\item compute the optical tidal matrix projected into the SNF with the {\tt OpticalTidalMatrix[]} function, as expressed in Eq.~($39$) in \cite{Grasso:2021iwq};
	\item compute the expressions of the evolution equations for the BGO with {\tt BGOequations[]}, as in Eq.~($44$) in \cite{Grasso:2021iwq}, and solve them with {\tt SolveBGO[]} for obtaining $\mathcal{W}(\calS, \calO)$.
\end{enumerate} 
Step (vii) is the starting point for obtaining the observables by combining the BGO with the observer and source four-velocities and four-accelerations as discussed in Sec.~\ref{sec:observables}.

Other of my original contributions in \cite{Grasso:2021iwq} are the expressions of the parallel transport equation and the optical tidal matrix in terms of ADM quantities (Eq.~($31$) and Eq.~($39$) in \cite{Grasso:2021iwq}), and the explicit transformation relations from forward to backward integrated BGO (Eqs.~($49$)-($52$) in \cite{Grasso:2021iwq}). I also obtained the specific form of the optical tidal matrix and the BGO in the SNF, and I used these expressions to simplify the calculations of their components.

I have also performed the tests computing the redshift, the angular diameter distance, the parallax distance, and the redshift drift for two analytical spacetimes corresponding to the $\Lambda$CDM and the (axially-symmetric) Szekeres models (presented in \cite{Meures:2011ke}), and for a numerically evolved dust Universe (EdS). The tests were designed jointly by E. Villa and me. I have performed the numerical calculations of the observables with {\tt BiGONLight} and the comparisons with their analytical expressions in the $\Lambda$CDM and EdS models (Fig.~$2$ and Figs.~$4$-$5$ in \cite{Grasso:2021iwq}). The numerical evolution of the dust Universe was done by me with the {\tt Einstein Toolkit} and the {\tt FLRWSolver}, \cite{loffler2012einstein, macpherson2017}.

In the axially-symmetric Szekeres model we were considering, there are no analytical expressions for the observables: the redshift and the angular diameter distance can be obtained numerically as shown in \cite{Meures:2011gp} and the results compared with those obtained with {\tt BiGONLight} (Fig.~$3$ in \cite{Grasso:2021iwq}). Conversely, there was no known way to calculate the redshift drift. To this end, I derived an ODE whose solution gives the redshift drift for a geodesic along the axis of symmetry of the model (Eq.~($96$) in \cite{Grasso:2021iwq}). Finally, I have compared the redshift drift obtained from this result with the one obtained with the package. 
The results were discussed jointly by E. Villa and me, and published in \cite{Grasso:2021iwq}.

\includepdf[pages=-]{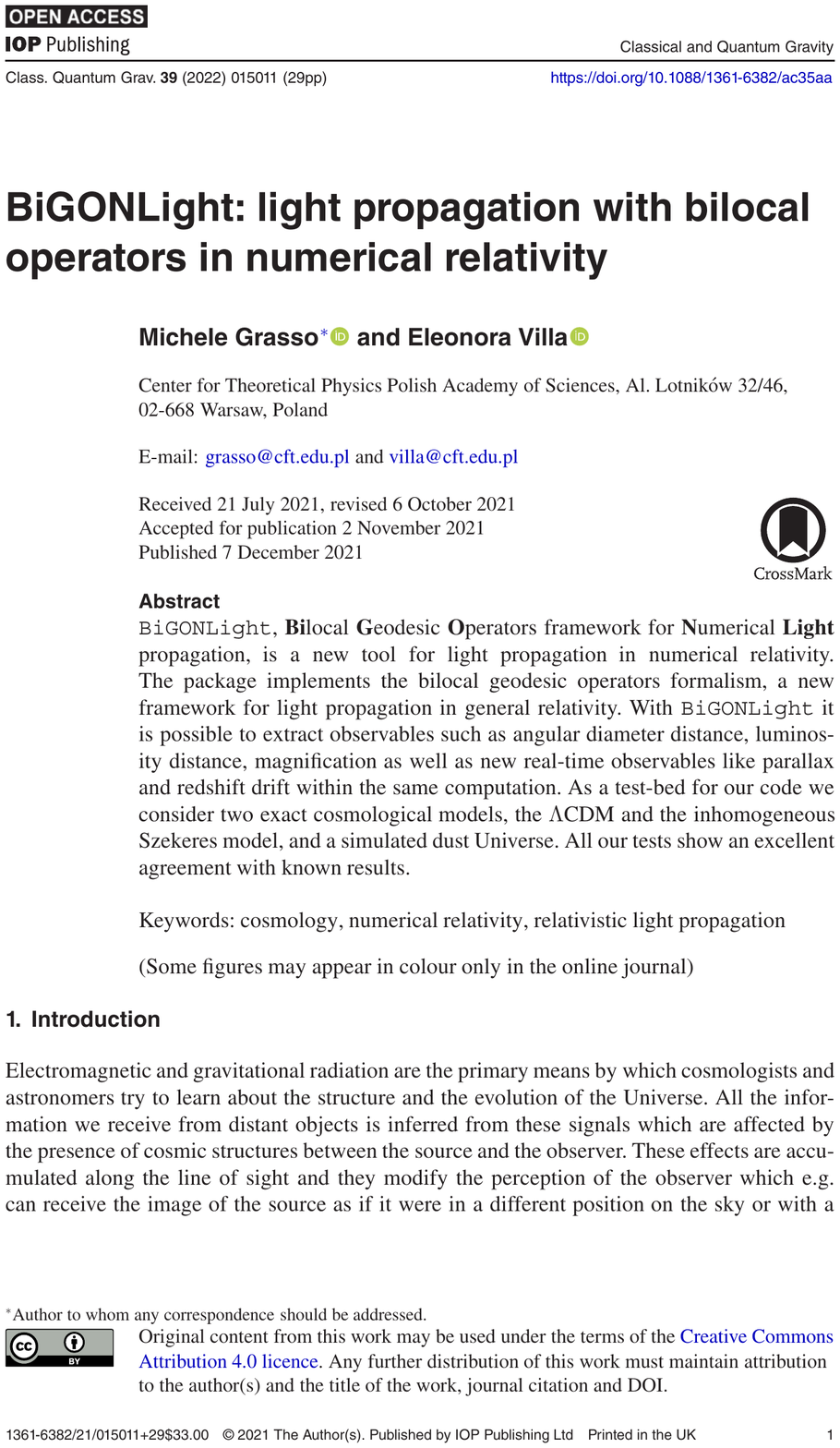}

%% file: template/paper_nonlinear.tex

\chapter{Paper II: ``Isolating nonlinearities of light propagation in inhomogeneous cosmologies''} 
\label{chap:nonlinearities} 

This chapter presents the application of {\tt BiGONLight} to study the nonlinear contributions to light propagation in an inhomogeneous cosmological model. The ultimate goal of this work is to isolate and quantify how different sources of inhomogeneities contribute to nonlinearities in cosmological observables. To this end, instead of using a realistic model of the Universe, it is preferable to have a toy model, whose properties can be tuned easily by changing its parameters. 
The \emph{wall Universe}, in which the matter is condensed in a sequence of plane-symmetric perturbations around a homogeneous distribution (FLRW background), is the toy model employed in this investigation. The analytical expression of its metric in PN approximation is given in \cite{Villa:2011vt} and is used in this analysis at three different approximations: linear PT, Newtonian, and PN.
The forms of the metric at Newtonian and PN approximations are provided as input metric in {\tt BiGONLight} to compute numerically the observables at Newtonian and PN order, respectively. The observables at linear PT are obtained analytically from the expressions of the BGO at linear order.
The nonlinear contributions are determined as relative differences between observables computed within these three different approximations.

The study answers the following four questions: 
\begin{enumerate}
	\item what are the Newtonian and PN corrections to the linear PT observables?
	\item what is the impact of the size of inhomogeneities?
	\item how much do the free parameters of the model affect the comparison?
	\item how important are the nonlinear PN corrections?
\end{enumerate}
Although many other authors have already examined the first question  (see e.g. \cite{Dyer:1974, Barausse:2005nf, Bonvin:2006, Meures:2011gp, Umeh:2012pn, Macpherson:2021gbh}), the other three questions penetrate deeply into the origin of nonlinearities, whose contributions are precisely evaluated by {\tt BiGONLight}.\\

\textit{\textbf{Author's contribution}}\\
\newline
The work described in this Chapter was performed in collaboration with E. Villa, M. Korzy\'nski, and  S. Matarrese.
The idea of this analysis was proposed by E. Villa and later discussed with the other authors. The $\Lambda$CDM extension of the metric in \cite{Villa:2011vt} (originally formulated for an EdS background) have been obtained jointly by me and E. Villa (Eq.~$2$ in \cite{Grasso:2021zra}). I have performed the simulations for light propagation and the computations of the observables in the three approximations. 
E. Villa and I did the preliminary analysis on the comparisons of the observables, and the results were discussed with all the other authors in order to draw conclusions. 
The comparison of the metric in Eq.~($7$) in \cite{Grasso:2021zra} with the Szekeres metric in \cite{Meures:2011ke} was done jointly by E. Villa and me. We also derived the analytical expressions of the linear observables with the BGO, and we did the match with the known formulas in the literature (i.e. without BGO). The draft was written by E. Villa and me and jointly published by all authors.

\includepdf[pages=-]{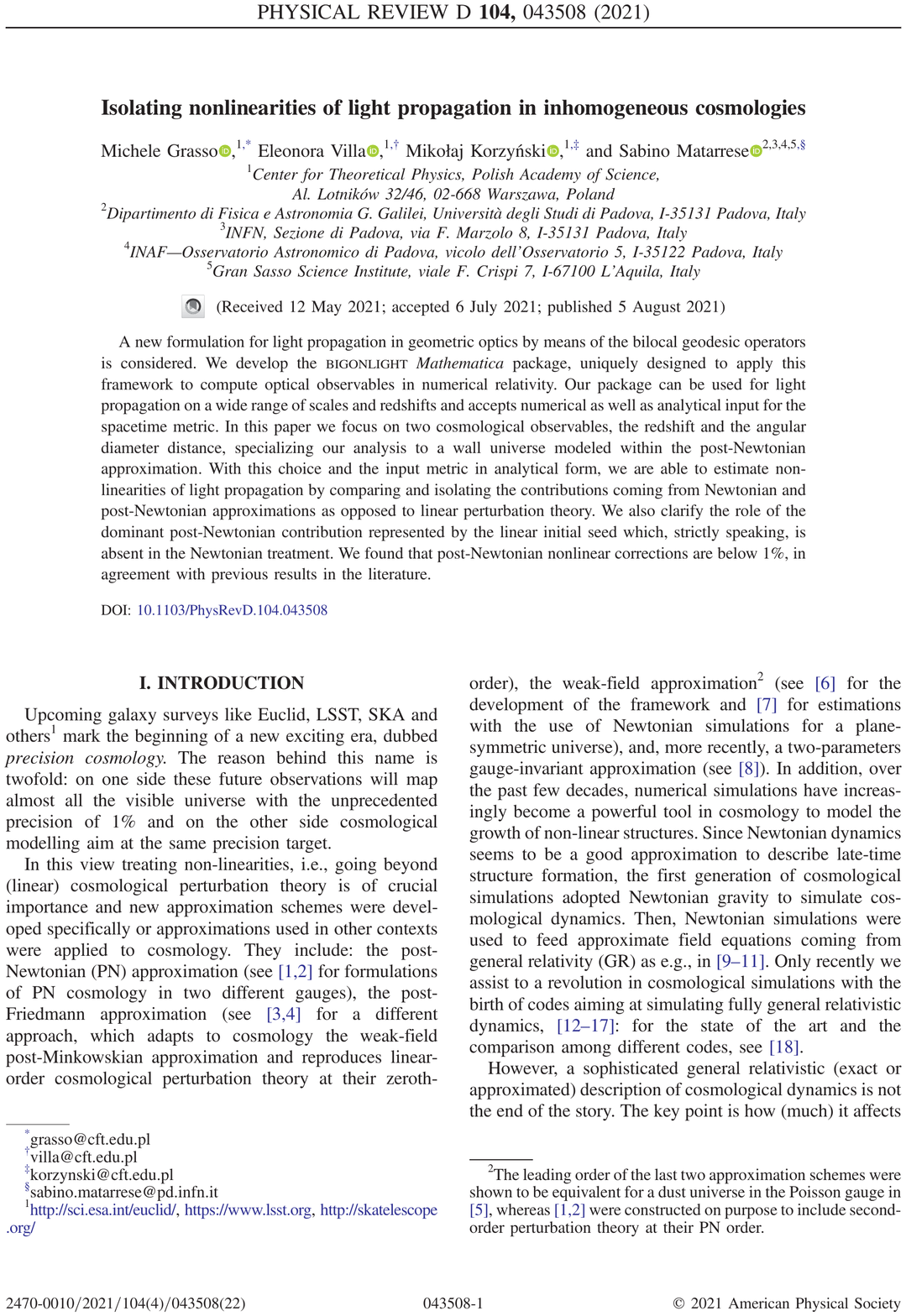}


%% file: template/conclusion.tex
\chapter{Summary}
\label{chap:conclusion}

This dissertation deals with the computation of optical observables in cosmological simulations using the new {\tt Wolfram} package {\tt BiGONLIght}.
Numerical simulations have become an increasingly important instrument in modern cosmology for reconstructing the Universe's large-scale structure. To test the validity of cosmological theories, it is essential to correctly simulate the interaction of light with these structures to determine the origin of nonlinear relativistic effects measured in observations. In the past, this has been done with various methods for gravitational lensing observations. With the possibility of more precise measurements on the one hand and the observation of new quantities on the other, a unique approach to light propagation is needed to keep pace with this revolution in observational cosmology.
The main feature of the package I have created is that it allows the direct implementation of the BGO formalism for computing multiple observables in a single calculation. This is possible because the BGO provide a unified framework for describing all possible optical effects caused by gravity on light propagation. Once computed along a geodesic, the BGO can be used to compute observables such as magnification, shear, and angular diameter distance, as well as new real-time observables such as parallax, redshift drift, and position drift resulting from temporal variations in the positions of sources and observers.

In the first paper, we introduce {\tt BiGONLIght} and show how it is applied to compute multiple observables in numerical relativity. In order to be compatible with most of the full-GR codes employed in numerical relativity, the package is designed to implement the BGO framework in the $3+1$ form. To this end, I express the parallel transport equation, the optical tidal matrix $R\UD{\bm{\mu}}{\ell \ell \bm{\nu}}$, and the evolution equation for the BGO in terms of the ADM quantities (Eqs.~($31$),~($39$), and~($44$) in \cite{Grasso:2021iwq}). These results, together with the transformations from forward to backward integrated BGO (see Eqs.~($49$)-($52$) in \cite{Grasso:2021iwq}), are my main theoretical contribution to this paper.
Together with the $3+1$ geodesic equation (presented in \cite{Vincent:2012kn}), these formulas are encoded in {\tt BiGONLIght} as {\tt Mathematica} functions.
These functions take as input the ADM quantities (namely the spatial metric $\gamma_{i j}$, the extrinsic curvature $K_{i j}$, the lapse $\alpha$ and the shift $\beta^i$) and the $3+1$ components of the velocities and accelerations  of the observer $\calO$ and source $\calS$ to obtain the ODEs for computing the geodesics, to perform the parallel transport of a tetrad of vectors, and to compute the BGO. 
The user can provide the input as interpolated data from a numerical simulation or as analytical expressions of the components of the metric and the four-vectors (velocities and accelerations of $\calO$ and $\calS$). For this second case, I have included functions in the package that perform the $3+1$ splitting of the four-dimensional metric tensor and the four-vectors to obtain the ADM quantities. This hybrid design makes {\tt BiGONLIght} highly adaptable to study different types of problems in both numerical simulations and analytical (perturbation and/or exact) approaches. 

The solutions of the ODEs for the geodesics, parallel transport, and GDE for BGO are found numerically using {\tt BiGONLIght} functions that solve these ODEs within a certain numerical precision. The user sets the precision via the precision control options implemented in {\tt Mathematica}. 
The final output of {\tt BiGONLight} are the BGO $\mathcal{W}(\calS, \, \calO)$ computed along the geodesics from the observer $\calO$ to the source $\calS$. These can be used to obtain observables as described in Sec.~\ref{sec:observables}. Furthermore, the ready-to-use transformations from forward-integrated BGO $\mathcal{W}(\calO, \, \calS)$ to backward-integrated BGO $\mathcal{W}(\calS, \, \calO)$ make {\tt BiGONLight} potentially adaptable to perform light propagation on-the-fly with a simulation of spacetime, that is forward in time by construction. The package is currently designed to perform light propagation in post-processing rather than on-the-fly, with the advantage of processing inputs from a variety of numerical codes for cosmological dynamics.
The procedure for computing observables with {\tt BiGONLIght} is described in \cite{Grasso:2021iwq} and implemented in a set of example notebooks publicly available on the GitHub repository {\color{blue}{https://github.com/MicGrasso/bigonlight}} under the GPL-3.0 license. 
The release of {\tt BiGONLIght} is the main result of my research and the most important achievement of this thesis.

We test {\tt BiGONLIght} by computing observables in three different cosmological models: the two homogeneous $\Lambda$CDM and EdS models and the inhomogeneous Szekeres model. The spacetime metrics corresponding to the $\Lambda$CDM and the Szekeres model are provided analytically, while the metric for the EdS model is obtained from the numerical evolution of a homogeneous dust Universe performed with the {\tt Einstein Toolkit} and the {\tt FLRWSolver}.
In the homogeneous $\Lambda$CDM model, I have calculated the redshift, the angular diameter distance, the parallax distance, and the redshift drift using {\tt BiGONLIght} and I have compared these results with those obtained using analytical expressions from the literature. This test shows that {\tt BiGONLIght} can accurately reproduce the analytical results with a relative difference of the order of $10^{-22} \div 10^{-31}$.
In the inhomogeneous Szekeres model, the metric presented in \cite{Meures:2011ke, Meures:2011gp} is provided as an analytical input to {\tt BiGONLIght} to compute the redshift and the angular diameter distance. I compare these results with those obtained by numerically solving the equations derived in \cite{Meures:2011gp}. I also test the calculation of the redshift drift using {\tt BiGONLIght}. In this case, the comparison is made with the formula I derived for the specific structure of light propagation in the Szekeres model under consideration (Eq.~($96$) in \cite{Grasso:2021iwq}). 
These other code tests in the Szekeres metric also show a good agreement between the observables obtained with {\tt BiGONLIght} and those obtained with other methods, with a relative difference of the order of $\sim 10^{-22}$.
The last group of code tests is performed considering the homogeneous EdS model evolved numerically with the {\tt Einstein Toolkit} and {\tt FLRWSolver}. The data resulting from this simulation are passed as numerical input to {\tt BiGONLIght} and used to calculate the observables. These tests also differ from the other two groups, because in this case I have performed light propagation forward in time, namely from the source $\calS$ to the observer $\calO$, and obtained the forward integrated BGO $\mathcal{W}(\calO, \calS)$. Then I use the transformations in Eqs.~($49$)-($52$) in the paper \cite{Grasso:2021iwq} to obtain $\mathcal{W}(\calS, \calO)$ and compute observables. The tests on the observables (redshift, angular diameter distance, parallax distance, and redshift drift) show that the precision of the numerical simulation of the spacetime, which in our case is $\sim 10^{-10}$, determines the accuracy of the observables computed with {\tt BiGONLIght}. The very same results are obtained if $\mathcal{W}(\calS, \calO)$ are directly computed, as was done for the tests in $\Lambda$CDM and Szekeres models.

In the second paper, we show how {\tt BiGONLight} can be used to isolate and quantify various nonlinear contributions to light propagation. The analysis is performed in a toy model of the Universe, where the inhomogeneities in the density fluctuations form a sequence of plane-symmetric perturbations around a homogeneous $\Lambda$CDM background. The nonlinearities of light propagation are quantified by considering the relative difference of observables, defined as $\Delta O({\rm b, a})=(O^{\rm b}-O^{\rm a})/O^{\rm a}$, where ${\rm a,\, b}={\rm Lin, \, N,\, PN}$ denotes the following three different approximations: linear observables $O^{\rm Lin}$, obtained using standard first-order perturbation theory, Newtonian observables $O^{\rm N}$, obtained using the Newtonian approximation of the plane-parallel metric as the analytical input to {\tt BiGONLight}, and post-Newtonian observables $O^{\rm PN}$, obtained using the post-Newtonian approximation of the plane-parallel metric as the analytical input to {\tt BiGONLight}. The expressions of the plane-parallel metric in Newtonian and PN approximations are given in \cite{Villa:2011vt}, and we extended them providing the corresponding metrics with a $\Lambda$CDM background. These metrics fit our analysis well, as the terms from all three approximations are easily identifiable and can be used directly as input in the package to compute observables. For our analysis, we only consider the redshift $z$ and the angular diameter distance $D_{\rm ang}$ computed in the three different methods. After a preliminary analysis, we decided to fix the setting for light propagation as discussed in Sec.~IV in \cite{Grasso:2021zra}. 

The variations $\Delta O({\rm Lin, N})$ and $\Delta O({\rm PN, N})$ are calculated by varying the free parameters of the model. The gravitational potential $\phi_0(q^{\rm 1})$ is a free function that provides the spatial profile of the matter distribution. We consider a sinusoidal distribution $\phi_0(q^{\rm 1})=\mathcal{I} \sin(\frac{2 \pi}{k}q^{\rm 1})$, where $\mathcal{I}$ and $k$ denote the amplitude and scale of the inhomogeneities, respectively. We vary $(k, \,\mathcal{I})$ as described in Sec.~V and according to the list of values in Table~$1$ in \cite{Grasso:2021zra}. Another free parameter of the model is $a_{\rm nl}$, which is related to the primordial non-Gaussianity parameter $f_{\rm nl}$. It expresses the deviations from a Gaussian distribution of the primordial fluctuations and an estimate of its value is $a_{\rm nl}=0.46 \pm 3.06$, see \cite{planck2019anl}. In the PN metric \cite{Villa:2011vt}, a specific value of $a_{\rm nl}$ can modulate the effects of some of the PN terms. We compute $O^{\rm PN}$ for the four values of $a_{\rm nl}=0.46,\, 3.52,\,-2.6,\,1 $, corresponding to the reference value, the two extremes of the confidence interval, and for perfect Gaussian distribution. The other (cosmological) parameters are set using the values measured by Planck satellite \cite{planck2018param}.
We isolate the various sources of nonlinear corrections in the observables by analysing the dependence of $\Delta O$ on these freely specifiable quantities.

My original contributions to this paper are all simulations in the three approximations with {\tt BiGONLight} and calculations of $\Delta O$, for the various choices of the free parameters $(\mathcal{I}, \, k, \, a_{\rm nl})$.
These results have led to the following findings:
\begin{enumerate}[(i)]
\item We quantify the nonlinear corrections in the observables from Newtonian and PN approximations by computing $\Delta O({\rm Lin, N})$ and $\Delta O({\rm PN, N})$. In general, our results are consistent with similar results in the literature, as $\Delta z({\rm Lin, N})$ and $\Delta D_{\rm ang}({\rm Lin, N})$ are well below $1\%$. However, we note a different behaviour in the two observables. For the redshift, the Newtonian corrections contribute most to the nonlinearities, with $\Delta z({\rm Lin, N})\sim 10^{2}\Delta z({\rm PN, N})$. On the other hand, for the angular diameter distance we find that $\Delta D_{\rm ang}({\rm Lin, N})\sim \Delta D_{\rm ang}({\rm PN, N})$, i.e. $D_{\rm ang}^{\rm Lin}$ and $D_{\rm ang}^{\rm PN}$ have similar corrections with respect to $D_{\rm ang}^{\rm N}$.
\item We estimate the effects of the scale of perturbations $k$ by computing $\Delta O$ for different values of $k$. The observables at linear, Newtonian, and PN order are computed for $k=500\, {\rm Mpc}, \,300\, {\rm Mpc}, \,100\, {\rm Mpc}, \,50\, {\rm Mpc}, \,300\, {\rm Mpc}$. We find that the amplitude of $\Delta D_{\rm ang}$ decreases monotonically with the scale for both linear-Newtonian and PN-Newtonian comparisons. On the other hand, the amplitude for $\Delta z$ increases for $500 \, {\rm Mpc} \leq k \leq 100 \, {\rm Mpc}$ and decreases  for $100 \, {\rm Mpc} < k \leq 30  \, {\rm Mpc}$, with a maximum amplitude of $\Delta z$ for $k=100 \, {\rm Mpc}$.
\item We constrain the dependence on primordial non-Gaussianity in $\Delta O ({\rm PN, N})$ and \\
$\Delta O^{\rm PN}(a_{\rm nl_1}, a_{\rm nl_2})$ with $O^{\rm PN}$ computed with the four different values of $a_{\rm nl}$. In both cases, we found that the primordial non-Gaussianity parameter has negligible effects in our comparison, i.e. $\Delta O^{\rm PN}(a_{\rm nl_1}, a_{\rm nl_2}) \ll \Delta O ({\rm PN, N})$.
\item Finally, we examine the relative difference $\Delta O({\rm a, \Lambda CDM })$ of the observables in the three approximations ${\rm a} = {\rm Lin, \, N, \, PN}$ with respect to the observable for the $\Lambda$CDM background. This comparison for the angular diameter distance shows that the leading contribution to the PN corrections is the linear PN term $ - \frac{5}{3 c^2} \phi_0$, known as ``initial seeds''.
\end{enumerate}
The results in (iv) motivate the findings in (i) and (iii). In (i), the result $\Delta D_{\rm ang}({\rm Lin, N})\sim \Delta D_{\rm ang}({\rm PN, N})$ is related to the fact that the initial seed is only present in the linear metric and the PN metric (as the leading correction), but is absent in the Newtonian metric. On the other hand, in (iii), the dependence on $a_{\rm nl}$ is negligible since the primordial non-Gaussian parameter can only trigger the nonlinear PN terms, and these are subleading with respect to the linear PN initial seeds.

In conclusion, I present {\tt BiGONLight} in my thesis as a reliable numerical tool that can be easily adapted to perform numerical and analytical analyses of light propagation by computing multiple observables within a single calculation. It can be used to perform complex analyses, such as the one presented, with accuracy sufficient to constrain small nonlinear effects, such as those caused by nonlinear PN terms. 

\newpage
 